\documentclass{article}
\usepackage{amssymb}
\usepackage{amsmath}
\usepackage{color}
\numberwithin{equation}{section}

\usepackage{graphicx}

\newtheorem{theorem}{Theorem}[section]

\newtheorem{remark}[theorem]{Remark}

\newcommand{\sgn}{{\rm sgn}}

\def\d{{\mathrm d}}

\renewcommand{\sfdefault}{bch}
\def\barr{\hbox{{\fontfamily{\sfdefault}\selectfont I\hskip -.35ex R}}}

\usepackage{cancel}

\usepackage{xcolor}

\begin{document}
\title{Singularity formation as a wetting mechanism \\ in a dispersionless water wave model
}

\author{R.\ Camassa$^1$, G.\ Falqui$^2$, G.\ Ortenzi$^2$,  M.\ Pedroni$^3$, and G. Pitton$^4$ \\
\smallskip\\
{\small $^1$University of North Carolina at Chapel Hill, Carolina Center for Interdisciplinary}\\ {\small Applied Mathematics,
Department of Mathematics, Chapel Hill, NC 27599, USA }
\\
{\small $^2$Dipartimento di Matematica e Applicazioni, Universit\`a di Milano-Bicocca, 
Milano, Italy}
\\
{\small $^3$Dipartimento di Ingegneria Gestionale, dell'Informazione e della Produzione,}\\ {\small Universit\`a di Bergamo,
Dalmine (BG),  Italy} \\
{\small $^4$SISSA, Via Bonomea 265, 34136 Trieste, Italy}}

\maketitle
\abstract{
The behavior of a class of solutions of the shallow water Airy system originating from initial data with discontinuous derivatives  is considered.
Initial data are obtained by  splicing together self-similar parabolae with a constant background state. These solutions are shown to develop velocity and surface gradient catastrophes in finite time and the inherent persistence of dry spots is shown to be terminated by the collapse of the parabolic core. All details of the evolution can be obtained in closed form until the collapse time, thanks to formation of simple waves that sandwich the evolving self-similar core.   The continuation of solutions asymptotically for short times beyond the collapse is then investigated analytically, in its weak form,  with an approach using stretched coordinates inspired by singular perturbation theory. This approach allows to follow the evolution after collapse by  implementing a spectrally accurate numerical code, which is developed alongside a classical shock-capturing scheme for accuracy comparison. The codes are validated  on special classes of initial data, in increasing order of complexity, to illustrate the evolution of the dry spot initial conditions on longer time scales past collapse.  
}

\section{Introduction}
The interaction of fluid layers with boundaries are arguably among the most challenging problems in the mathematical formulation of fluid dynamics. The Euler system of perfect fluids, while highly idealized,  offers a physically relevant simplification of this dynamics in many setups, however it often presents several obstacles from an analytical standpoint. While a direct investigation of Euler equations could always in principle be carried out numerically, much insight can be gained  
 from the study of reduced  models that can extract the essential elements of confining-boundary/interface interactions, by incorporating boundary conditions and physically relevant properties  either exactly or approximately, e.g., through layer averaging of the horizontal velocity field. 
Such is the case for the Airy (shallow water) system for the water layer long-wave problem which is the focus of this work. Here, the usual free-surface boundary conditions of the Euler equations are enforced  {\em ab~initio} through the process of averaging. The simplest set-up to investigate the interaction of a free surface with a bottom boundary is that of a two-dimensional channel. As can be easily shown (see, e.g.,~\cite{CFOPT}),  contact points between the free surface and the bottom boundary can {\it never} be wiped out as long as the solution of either the Euler or the Airy systems remains  smooth. 

Hyperbolic quasilinear systems can develop shocks, whose study is a classic topic of the literature (see, e.g., \cite{Stoker} in fluid settings, or \cite{Whitham} for a more general esposition).
The general subject of  hyperbolic-parabolic transitions, to which the problem of wetting of dry points/intervals (``vacuum states") belongs in the present context, has also been the subject of many investigations, both in the physical and in the mathematical literature (see, e.g., \cite{bers,CouFried,liusmoller,Lax}). In particular, the authors of~\cite{liusmoller} remark how shocks cannot propagate into vacuum states, and 
how a theory of how  vacuum states can be filled by rarefaction waves may still lack a general, rigorous foundation from the mathematical perspective.
To this end, various methods of regularization in the study of such vacuum states have been proposed, including entropy methods \cite{liusmoller} and dissipative regularizations~\cite{LiLiXin}.  
In this work we confine ourselves to  the purely quasi-linear setting. By using 
specific classes of initial data whose evolution is amenable to analytical methods, we 
demonstrate that the formation of a singularity can circumvent the seemingly unphysical conclusion of the persistence of the dry point. The class of initial data we consider are exact global solutions of the Airy system up to the time of singularity, and in certain sectors provide local solutions even after this time. These solutions evolve from initial conditions obtained by splicing together constant  and time-dependent solutions in the form of local 
parabolic profiles for the free surface elevation, which can be viewed as self-similar in the proper time-like variables. The singularity for these data manifests itself as a ``global" shock, that is, a gradient catastrophe occurs over 
a whole interval in the range of the dependent variables, as opposed to the generic case of an infinite derivative at single point in this range. 

Our main goal is to study the shock problem emanating from the singularity for a suitable class of initial conditions. In this, our approach differs substantially
from previous works on the subject (see, e.g, \cite{Guc,Lyc,Eng,JM,DGK,SpElHo,CFO,KO2}), and, in particular, from the one more recently set forth by Dubrovin and collaborators (see, e.g., \cite{Du08,ElDu,DuGKM}) where the main focus is the conjectured universality behaviour of hyperbolic (and elliptic) shocks, universality being intended both with respect to general dispersive Hamiltonian regularizations and to initial conditions. In this context, 
the Airy system, viewed as the nondispersive version of the defocussing Nonlinear Schr\"odinger equation (NLS), and 
the non-generic nature of gradient catastrophe in the presence of a vacuum point, was investigated using specific initial data in~\cite{MT}; the evolution through singularity was illustrated by numerical simulations the full dispersive NLS, which resolved possible shocks into dispersive wave trains. 
Of course, in our case the evolution of weak (shock) solutions, and more generally the presence of jump discontinuities in the derivatives of the dependent variables, conflicts with   
the derivation of Airy's model from Euler equations by long wave asymptotics.
A full discussion of this point would fall beyond aims of this study, and we leave this to the fairly extensive literature (see, e.g, \cite{Stoker,Whitham}). However, it is worth mentioning here that in  our recent  work \cite{CFOPT}, while testing the validity of non-dispersive models for one and two layer fluids  we found that, at least for the class of initial data we study, hyperbolic shallow water models do provide a reasonably good prediction (both qualitatively and quantitatively) of the dynamics of the parent equations, on short time scales before other physical effects neglected by the model, such as dispersion, become important.

The details of the class of initial data and the parabolic solutions we consider are reported in section~\ref{genpar-sec}, where we focus on the case of an initial dry contact point with the bottom boundary. We then briefly contrast, in~\S\ref{mineq0}, the evolution of the ``dry" case and its singularity formation with that of a nearby class of initial data which maintain a fluid film over the bottom. The shock formation with parabola collapse is examined more closely in~\S\ref{shoulders}, with the aim of bringing forth the analytical obstacles to weak solution continuations. Section~\ref{sec:beyond_shock} introduces new classes of initial data designed to illustrate the main mechanisms of after-shock continuation by eliminating technical obstacles of the original case. These are overcome in~\S\ref{shckfc} by designing ``unfolding" coordinates that magnify space-time region near the singularity at the collapse time, turning the initial value problem on the whole real line into a boundary value problem on a strip whose end points are pinned to the propagating shocks. The new formalism allows for asymptotic solutions to be computed explicitly. Finally, in section~\ref{numerics}, we illustrate the results with numerical algorithms for both unfolding and physical coordinate systems, and validate the numerical simulations with the analytical results, with an eye at isolating possible accuracy shortcomings of the algorithms.

\section{Parabolic solutions for Airy model} \label{genpar-sec}
In a previous paper~\cite{CFOPT} we introduced and studied, in various instances, piecewise smooth initial data for  $1+1$-dimensional models approximating 
the Euler equations, where the density-related variable $\eta$ is obtained by  splicing together constant heights with (self-similar) parabolic profiles. 
In the specific case of the  classical Airy model for shallow water dynamics, in suitable non-dimensional variables, 
\begin{equation}
\eta_t+(u \eta)_x=0\, ,
 \qquad 
 u_t+uu_x+\eta_x=0\, ,
 \label{Airymodel}
\end{equation}
local solutions can be constructed, for $(x,t)$ belonging to certain sectors in space-time, using the simple spatial functions 
\begin{equation}
\eta =\gamma(t) x^2 +\mu(t)   \, ,\qquad u=\nu(t) x\, .
 \label{exact}
 \end{equation}
The evolution described by these functions can be viewed as (generalized) 
self-similar solutions of~(\ref{Airymodel}) since 
(i) they can be cast in the form $u(x,t)=k_u(t) f_1({x}/{l(t)})$, $\eta(x,t)=k_\eta(t)f_2({x}/{l(t)})$, for appropriate functions $k_u$, $k_\eta$, $f_1$, $f_2$ and $l$ of 
one variable, 
and (ii) inserting this prescription in the Airy equations leads to the {\em closed exact} system of ordinary differential equations (ODEs) for the coefficients $(\nu, \gamma, \mu)$
\begin{equation}
 \dot{\nu}+\nu^2+2\gamma=0\,, \qquad \dot{\gamma}+3\nu\gamma=0 \,, \qquad \dot{\mu}+\nu\mu=0\, .
 \label{coeffODEs}
\end{equation}
With initial conditions
\begin{equation}
 \nu(0)=\nu_0\,, \qquad \gamma(0)=\gamma_0\,, \qquad \mu(0)=\mu_0\,, 
\end{equation}
the solutions of system~(\ref{coeffODEs}) can be expressed implicitly by
\begin{equation}
  \nu(\sigma)=\pm \sqrt{4 \gamma_0 \sigma^3+(\nu_0^2-4 \gamma_0)\sigma^2} \, , \qquad \gamma(\sigma)=\gamma_0 \sigma^3\, , \qquad  \mu=\mu_0 \sigma\, ,
  \label{parab-gdep}
 \end{equation}
where  the auxiliary variable $\sigma$ solves the first order nonlinear ODE
  \begin{equation}
  \dot{\sigma}^2= (\nu_0^2-4 \gamma_0)\sigma^4+ 4\gamma_0 \sigma^5\, .
   \label{eq-tau}
   \end{equation}
The parabola curvature sign is invariant in time because the quadrature of system~(\ref{coeffODEs}) implies   
  \begin{equation}
  \gamma=\gamma_0 \exp \left(-3 \int_0^t \nu(t') \, \d t'\right) \, ,
  \end{equation}  
so that $\gamma$ remains positive if it is initially so.

 The study of self-similar solutions~(\ref{exact}), which were independently noted in \cite{Ovs}, has first appeared in \cite{CFOPT}.  Here we focus on the case of solutions
 with positive curvature $\gamma(t)>0$ starting initially with  
a \emph{dry point}, a point where the surface touches the bottom  ($\mu_0=0$). 
For the shallow water model, $\eta=0$  is a line that plays a double role: it is a degenerate characteristic
in the hodograph plane, and it is the ``sonic line,''  i.e. the locus of points where the system 
loses hyperbolicity because the characteristic velocities coincide. Such peculiar feature has some interesting consequences. In particular, from (\ref{parab-gdep}) we have $\mu(t)=0$, and  
the minimum of the parabola (where $\eta=0$) remains dry for as long as the solution maintains its local regularity.  \par
When initial velocities are sufficiently large, $\nu_0 \geq 2 \sqrt{\gamma_0}$, it can be shown that the curvature $\gamma(t)$ tends monotonically to zero and therefore no loss of regularity in the form of shocks can occur.
For small velocities the local parabola can collapse to a vertical segment, a ``global" shock, 
in finite times. This shock effectively reduces a portion of the free surface $\eta$ support containing the dry point to a set of measure zero, which can then be removed from the subsequent dynamics, thus providing a mechanism for the disconnection of the free surface from the bottom.

Next, we derive closed form solutions by quadratures of the first pair of equations in system~(\ref{coeffODEs}),
\begin{equation}
 \dot{\nu}+\nu^2+2\gamma=0\,, \qquad \dot{\gamma}+3\nu\gamma=0 \, .
 \label{coeffODEsr}
\end{equation}
(Since the third equation of~(\ref{coeffODEs}) is slaved to these two, its solution follows by quadratures; for the ``dry spot" case $\mu_0(t)=0$ for all times, and no additional quadrature is required.)
For sake of simplicity  we restrict our study to zero initial velocities, i.e., $\nu_0=0$. 
In this case equation~(\ref{eq-tau}) becomes
 \begin{equation}
  \dot{\sigma}^2= 4 \gamma_0(\sigma^5-\sigma^4)\, ,
   \label{eq-tau-zerovelo}
   \end{equation}
and, since $\gamma(t)>0$ and $\nu_0=0$, for all $t>0$,
\begin{equation}
 \dot{\nu}(t)<0 \quad \Longrightarrow \quad  \nu(t)< 0 \quad \Longrightarrow \quad  \dot{\gamma}(t)> 0.
 \end{equation}
Therefore ${\gamma}(t)$ and $\sigma(t)$ are strictly increasing functions of time,  ${\gamma}(t)\geq \gamma_0$ and $\sigma(t)\geq 1$, and equation~(\ref{eq-tau-zerovelo}) can be rewritten as
\begin{equation}
\dot{\sigma}=2\sqrt{\gamma_0}\, \sigma^2 \sqrt{\sigma-1}\, .
\end{equation}
Its  implicit solution
 \begin{equation}
  t(\sigma)=\frac{\displaystyle \sqrt{\sigma-1}+\sigma\, \arctan\hspace{-2pt}\big(\sqrt{\sigma-1}\hspace{1pt}\big)}{2 \displaystyle \sqrt{\gamma_0 }\, \sigma}\, ,\qquad  
 \gamma=\gamma_0\sigma^3\, ,\quad  \nu =-{\dot{\sigma} \over \sigma}=-2\sqrt{\gamma_0}\, \sigma \sqrt{\sigma-1}\,,
  \label{curvt}
 \end{equation}
 shows that $\sigma$ is a time reparameterization.
As the variable $\sigma\to +\infty$ the curvature of the parabola $\gamma$
blows up at the finite time
\begin{equation}
 t_c=\lim_{\sigma\to +\infty}t(\sigma)=\frac{\pi}{4 \sqrt{\gamma_0 }},
 \label{shocktime}
\end{equation}
while the coefficient $\nu$ diverges, $\nu(t)\to -\infty$ as $t\to  t_c^-$, see (\ref{parab-gdep}). \par

\begin{figure}[t]
\centering
{\includegraphics[width=8cm]{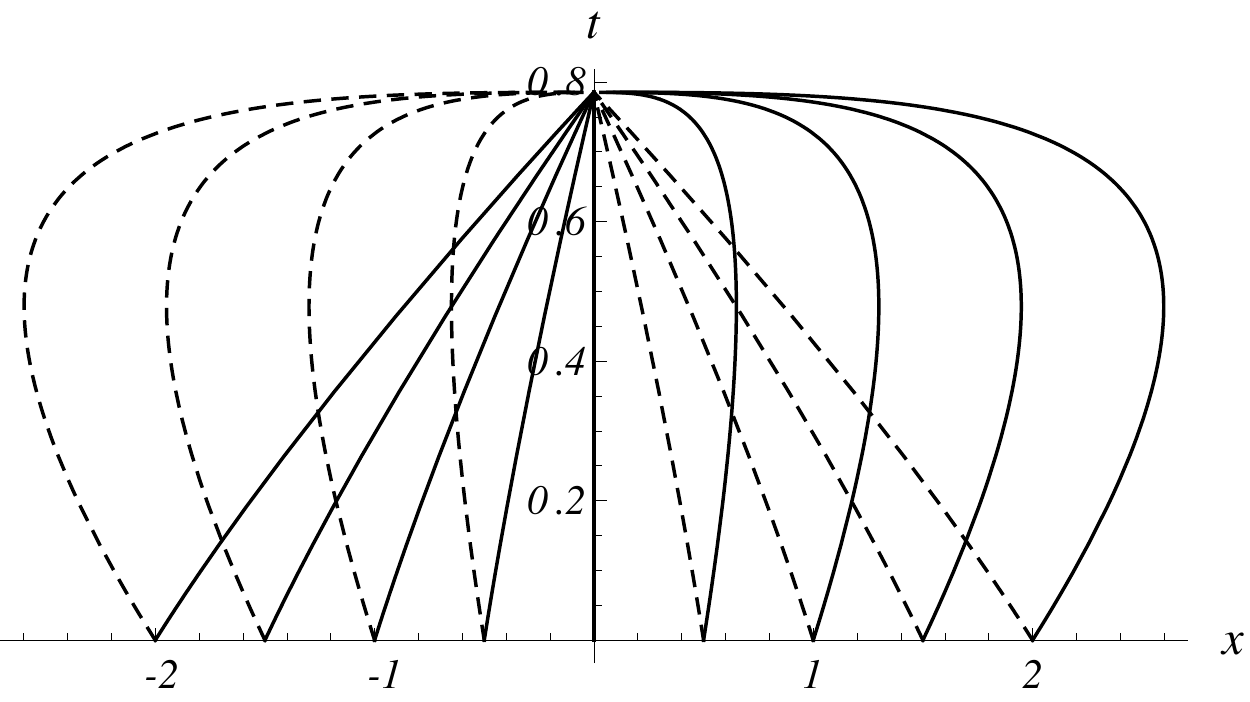}} 
\caption{Characteristics of parabola solutions in $(x,t)$ plane (with 
$\nu_0=0$, $\gamma_0=1$, $\mu_0=0$). The dashed and solid curves are, respectively,  the 
$x_{-}$ and $x_+$ characteristics defined by system~(\ref{paradrychar}) and~(\ref{curvt}).
}
\label{ParChar-plus-fig}
\end{figure}

Next, we look at characteristics corresponding to parabolic initial data. As well known, see e.g.~\cite{Whitham}, the Riemann invariants and characteristic equations for system~(\ref{Airymodel}) are 
\begin{equation}
R_\pm=u\pm2\sqrt{\eta}\, , \qquad \dot{x}_\pm\equiv\lambda_\pm=u\pm \sqrt{\eta}
\label{RI}
\end{equation}
which yield, for the characteristic curves emanating from the initial position $x(0)=x_0$ in the support region of the parabolic initial data, 
\begin{equation} 
\big(\nu(t)\pm2\sqrt{\gamma(t)}\,\big)x=\pm2\sqrt{\gamma_0}\,\, x_0 \,.
\end{equation}
Hence, using solutions~(\ref{parab-gdep}) for $\nu(t)$ and $\gamma(t)$
in terms of the auxiliary time-like variable $\sigma$, the characteristic solutions can be expressed as
\begin{equation}
 x_\pm(\sigma;x_0)
 =x_0 \, \frac{\displaystyle \sqrt{ \sigma}\pm \sqrt{ \sigma-1}}{\sigma } \, .
 \label{paradrychar}
\end{equation}
 The geometry of these characteristic curves is depicted in 
figure~\ref{ParChar-plus-fig}.

\begin{figure}[t]
\centering
{(a) \includegraphics[width=14cm]{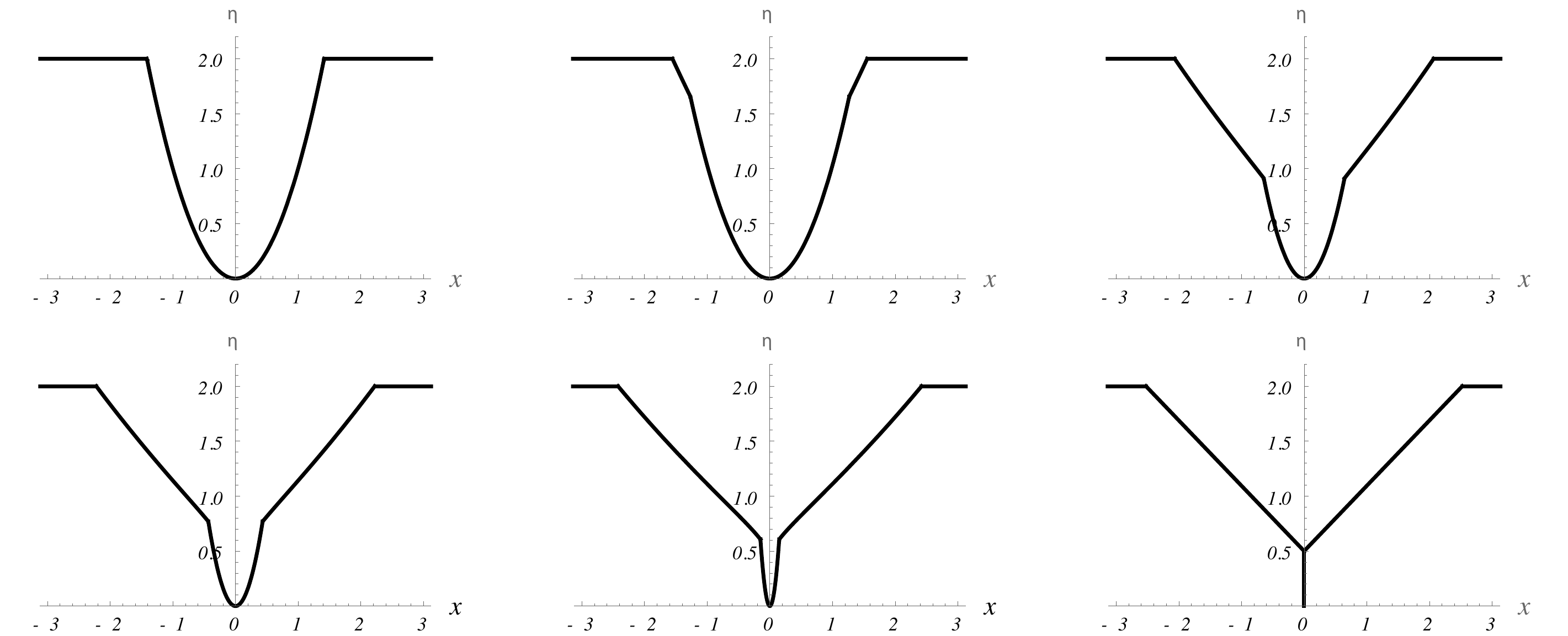}\\
\vspace{1cm}
(b)\includegraphics[width=14cm]{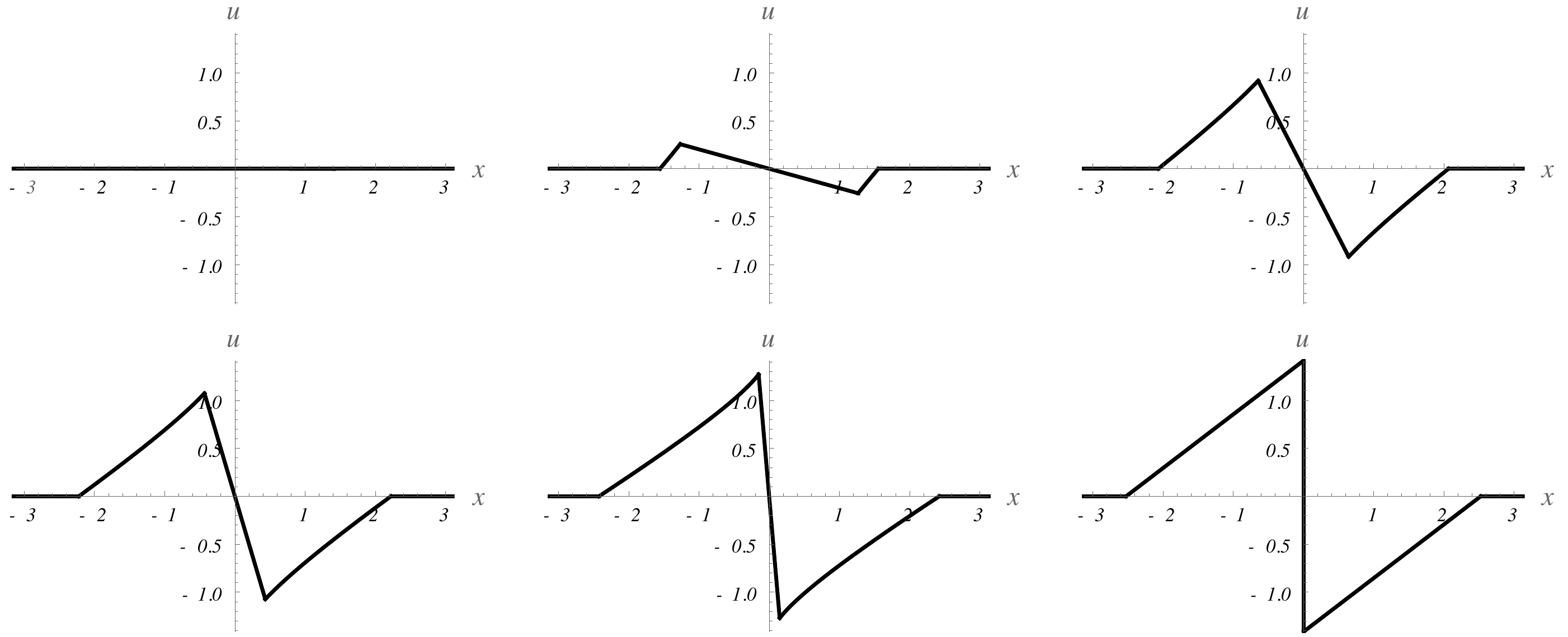}}
\caption{Surface  $\eta$ (a), and velocity $u$ (b) evolution for the Airy model when the initial surface has a dry (contact) point and zero initial velocities.  
The parameters are $Q=2$, $\gamma_0=1$, $\mu_0=0$, yielding a collapse time 
$t_c=\simeq0.7854$. Time snapshots $t=0$, $0.01$, $0.46$, $0.57$, $0.71$, $0.78$.}
\label{para-dry-fig}
\end{figure}

The initial configurations considered in \cite{CFOPT} 
can be called  {\em truncated parabolae}, inasmuch as the profile of the height $\eta$  is obtained by splicing a parabola 
with a constant height $\eta=Q$ for $|x|$ larger than a certain distance, $a_0>0$ say, from the origin, or $x_0\in[-a_0,a_0]$. We can represent these initial conditions as
\begin{equation}\label{inidata-up}
\eta(x,0)=\left\{
\begin{array}{lc} Q,& x< -a_0\\
\gamma_0 x^2,& -a_0 \leq x \leq a_0\\
Q,& x>a_0\end{array}
\right.\,, \qquad u(x,0)=0.
\end{equation}
Continuity requires
 $Q>0$ and $a_0=\sqrt{Q/\gamma_0}$. 
 Time evolution initially preserves  continuity and piecewise differentiability, 
 although the original parabolic section splits into three sections: a shrinking (positively curved) parabola centred at $x=0$, and two convex curves 
 symmetrically placed with respect to the origin, joining the parabola with the constant level $\eta=Q$.
 The behavior of the solution 
for $\mu_0=0$ is shown 
in  figure~\ref{para-dry-fig}. 
In summary, the extrema $\pm a(t)$ of the parabolic section evolve in time from their initial values $\pm a_0$, and as functions of the auxiliary variable $\sigma$ they can be found at
\begin{equation}
 a(t(\sigma))= \frac{\sqrt{Q \sigma}-\sqrt{Q(\sigma-1)}}{\sigma \sqrt{\gamma_0}} \, .
 \label{aevog}
\end{equation}
By comparison with the characteristics (\ref{paradrychar}), as expected by the general theory of singularities (see e.g. \cite{Whitham}), the point $a(t)$, being a point of continuity with discontinuous first derivative of the dependent variables, moves along a characteristic.
When the curve $x=a(t)$ crosses $x=0$ it meets its symmetric counterpart $x=-a(t)$. Let $t=t_s$ denote such crossing time. 
The value $\sigma_s$ of $\sigma$ at  time $t_s$ 
can be obtained from (\ref{aevog}) by setting $a(\sigma_s)=0$, so that $\sigma_s=+\infty$ and
\begin{equation}
 t_s=\frac{\pi}{4 \sqrt{\gamma_0}}\,,
\end{equation}
which coincides with the parabola collapse time $t_c$. This is special to the dry-point case $\mu_0=0$; in general the two times $t_s$ and $t_c$ are different, with $t_s<t_c$ if $\mu_0>0$, as we shall see below. Unless otherwise noted, we shall focus from now on onto the dry case, for which $t_s=t_c$, and use the latter notation to denote the time at which the singularity occurs.
The height of the interface and  the velocity at $x=a(t)$  are given by 
\begin{equation}\label{eq-buone}
\begin{split}
& \eta\big(a(t(\sigma)\big),t(\sigma))=\gamma(t(\sigma)) \, \big[a(t(\sigma))\big]^2  = 
\sqrt{Q}\, \sigma\left(\sqrt{\sigma }-\sqrt{\sigma-1}\,\right)^2 \, , \\
& u\big(a(t(\sigma)),t(\sigma)\big)= \nu(t(\sigma)) \, a(t(\sigma)) =
 -2 \sqrt{Q}\left(\sqrt{\sigma^2-\sigma}-(\sigma-1) \right)\, . 
\end{split}
 \end{equation}
At time $t=t_c$ such quantities  (for the solution with $x>0$ in the limit $t\to t_c^-$, say) are
\begin{equation}
 \eta(a(t_c),t_c)= \frac{Q}{4}\, , \qquad u(a(t_c),t_c)=-\sqrt{Q}\, .
\end{equation}
The support of the parabolic section disappears at $t=t_{s}$, and the solution develops a jump of amplitude~$2\displaystyle\sqrt{Q}$ in the 
velocity. The segment $\eta\in[0,Q/4]$ to which the parabola has reduced can also be viewed as double (from left and right of $x=0$) shock for the surface elevation $\eta$. 
The removal of this segment, i.e., the value $\eta(0,t_c)=0$ and the replacement with $\eta$'s limiting value $\eta(0,t_c)=\lim_{|x|\to 0}\eta(x,t_c) =Q/4$ makes the surface elevation continuous. Hence the gradient catastrophe may be viewed as the mechanism responsible for the detachment of the interface from the bottom (see figure~\ref{para-dry-fig}). Notice that the gradient catastrophe in this exact solution  differs from the more  common case whereby the derivative diverges initially at a single point. This can be viewed as the consequence of characteristics of the same family originating from a finite length segment, such as $x_0\in (0,a_0)$, {\it all} crossing at the same point in the $(x,t)$-plane (see figure~\ref{char-uppara-fig}). For this reason we will often refer to $t_c$ as the ``global shock" time.

\begin{figure}[t]
\centering
{(a)\includegraphics[width=14cm]{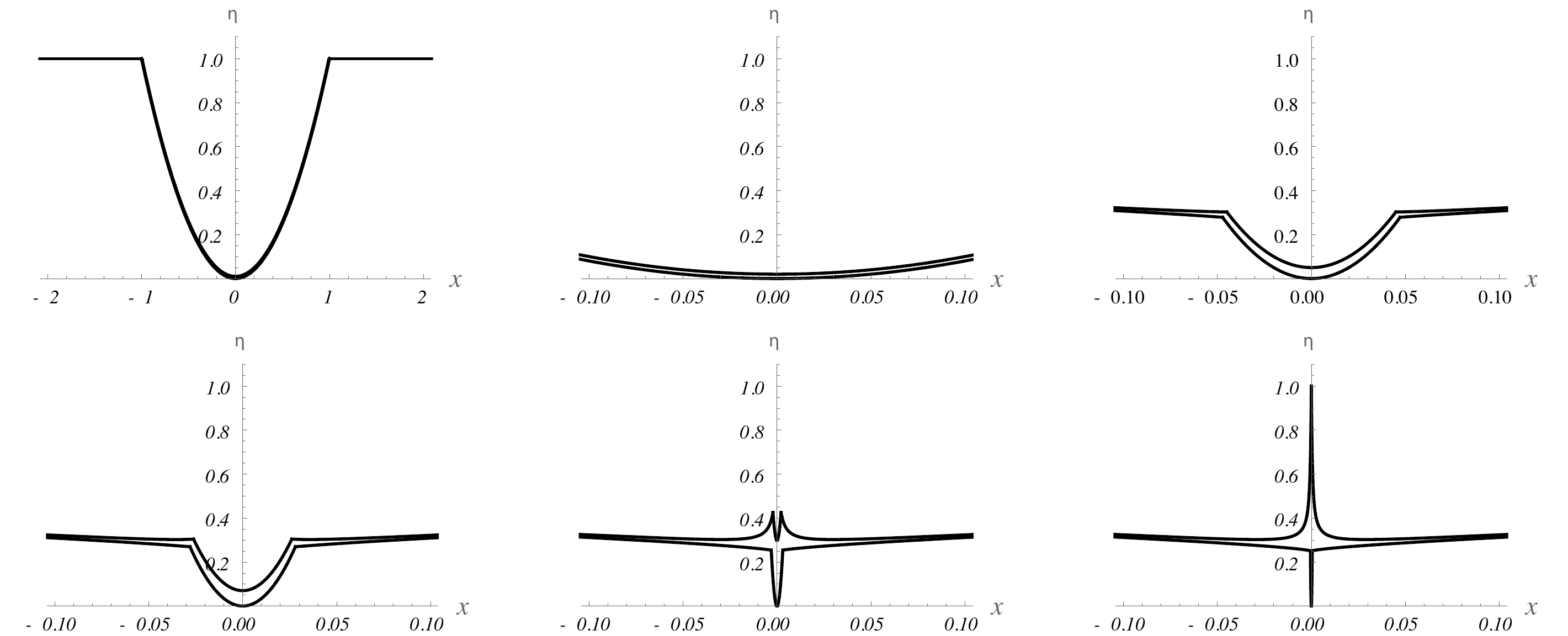}\\
\vspace{1cm}
(b)\includegraphics[width=14cm]{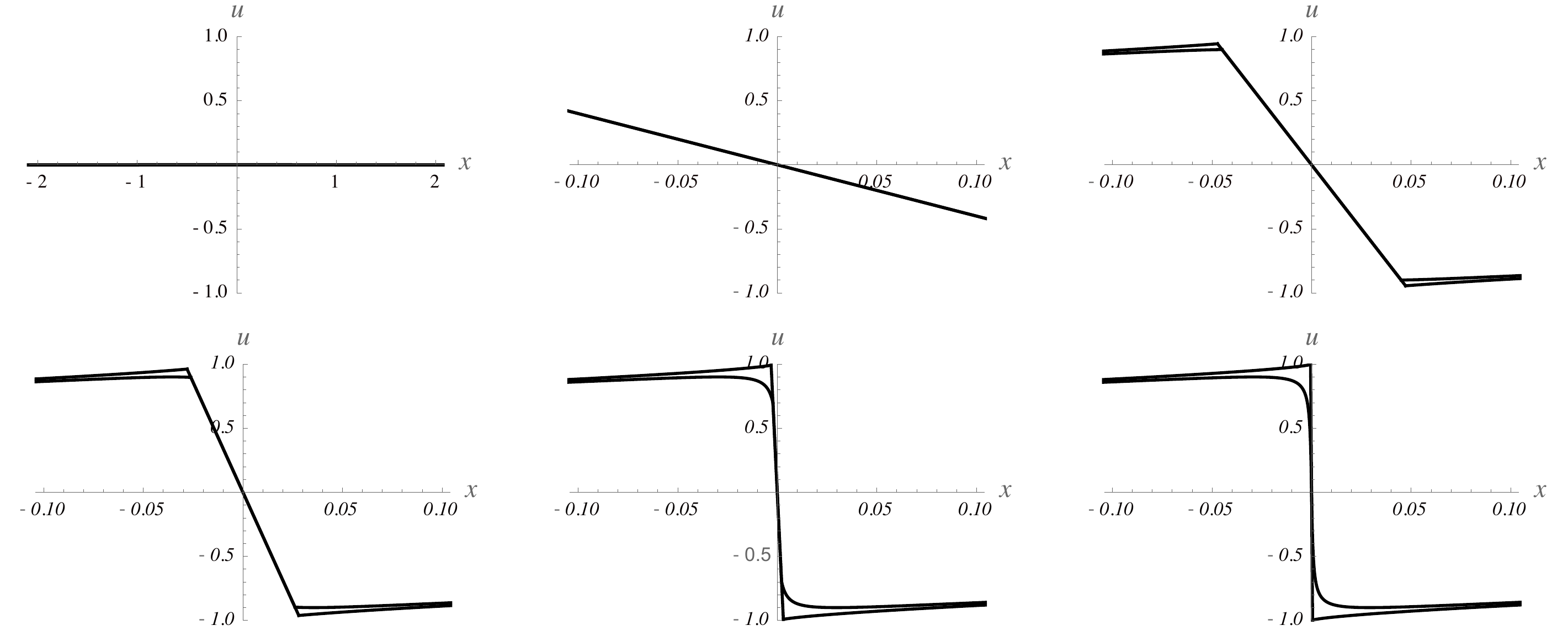}}
\caption{Comparison of Airy model's solutions for (a) the interface $\eta$, and (b) the velocity $u$, evolving from two zero-velocities initial parabolae on a constant background, with and without a dry point.  
Parameters are $Q=1$, $\gamma_0=1$, $\mu_0=0$ and $\mu_0=0.01$.
Snapshots at times $t=0$, $0.64$, $0.75$, $0.77$, $0.783$, $0.785$. The last time is close  to the crossing time $t_s=0.78506$ for $\mu_0=0.01$. The initial conditions are almost indistinguishable (top left panel), and snapshots for $t>0$ are zoomed in the outlined region of initial conditions. 
}
\label{para-eta-comparison-fig}
\end{figure}

\subsection{The ``wet" case $\mu_0>0$} 
\label{mineq0}
When the initial parabola's minimum is above the bottom plate, missing it by a non-zero amount $\mu_0>0$, however small, 
the ($\eta$,$u$) evolution  is substantially different from the contact case above, though the actual analytical 
calculations are similar to those 
in the previous section. We omit all the details (for a full account, see~\cite{CFOPT}) and report here the final results, 
focussing on the differences between the ``near miss" of the surface contact and its contact
counterpart $\mu_0=0$. 
The characteristics for $\mu_0>0$ are 
\begin{equation}
 x_\pm(\sigma;x_0)=\frac{x_0}{\sqrt{\sigma }}\pm \frac{\sqrt{(\sigma -1) \left(x_0^2+\mu_0/\gamma_0\right)}}{\sigma }\, , 
 \label{chars}
\end{equation}
and those for the extrema of the parabola $x_0=\pm a_0$ cross at time $t_s <t_c$, 
\begin{equation}
 t_s=\frac{\sqrt{{Q}/{\mu_0}-1}+({Q}/\mu_0) \arctan\hspace{-2pt}\big(\hspace{-2pt}\sqrt{{Q}/{\mu_0}-1} \hspace{1pt}\big)}{2 \sqrt{\gamma_0 }\,\, {Q}/{\mu_0}}\, ,
\end{equation}
at which the elevation $\eta(x,t)$ recovers its background value,  $\eta(0,t_s)=Q$.

The difference between the two evolutions is illustrated by figure~\ref{para-eta-comparison-fig}. Unlike the case $\mu_0=0$, the surface $\eta(x,t)$ returns to the background elevation $Q$ in a neighborhood of the origin, and both the surface elevation $\eta$ and the fluid velocity $u$ maintain continuity beyond the time $t_c$. This is in contrast with the dry-point case at time $t=t_c$, when the boundary-touching parabola collapses to a segment, forming a global shock $0\leq\eta(0,t_c)\leq Q/4$, with corresponding fluid velocity shock $-\sqrt{Q}\leq u(0,t_c)\leq \sqrt{Q}$.

\subsection{The connecting ``shoulder" states}
\label{shoulders}
\begin{figure}
\centering
{\includegraphics[width=15cm]{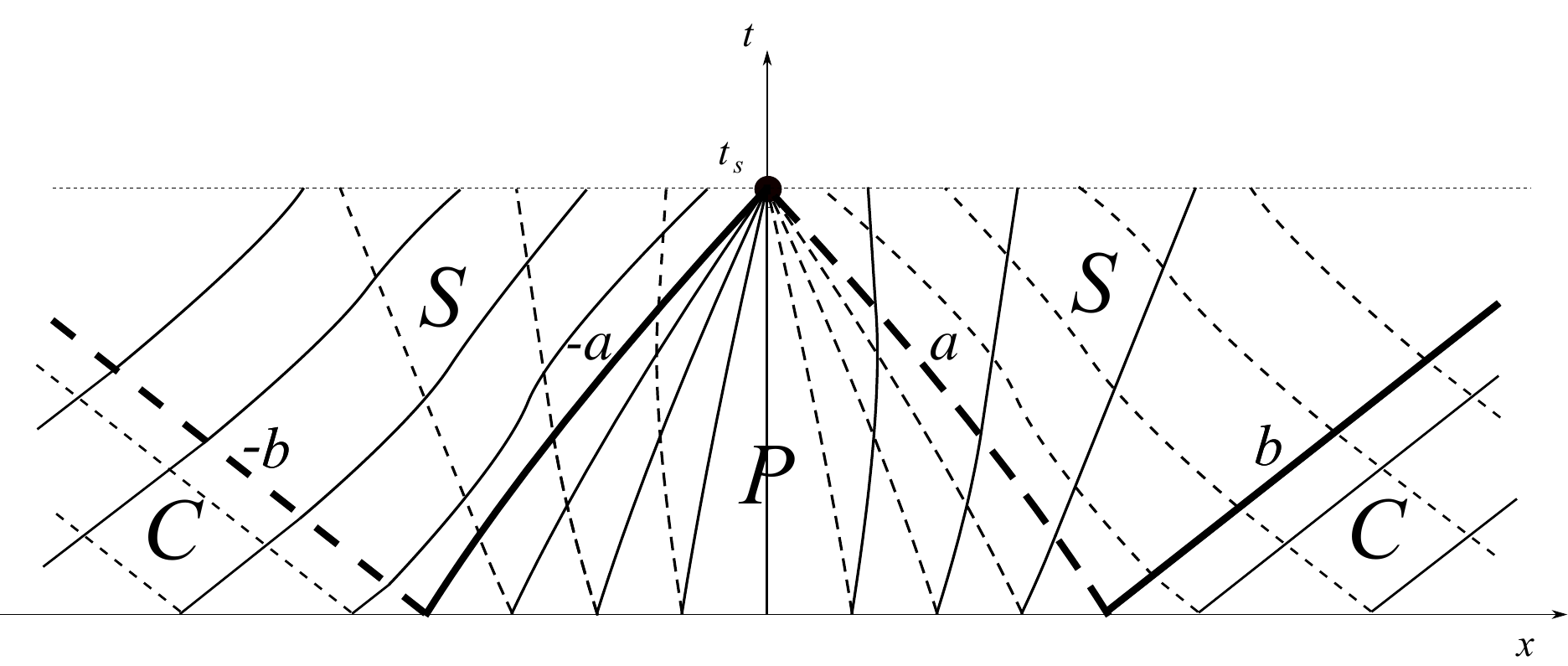}}
\caption{Space-time plane schematic of characteristic for the initial data 
of figure~\ref{para-dry-fig}.
$P$: parabola region; $C$: constant regions; $S$: ``shoulder" simple-wave regions. The merging of characteristics at the single point $(0,t_c)$ is marked by the filled circle, and corresponds to the coalescence of the parabola into the vertical segment $\eta(0,t_c) \in [0,Q/4]$.
}
\label{char-uppara-fig}
\end{figure}

In this section we discuss the evolution of the ``shoulders" that join the parabolic sectors 
with  the constant far-field states $\eta=Q$, by using an approach akin to that of the classical piston  problem (see e.g.,~\cite{Stoker}). 
The behaviour of the solution corresponding to the initial data (\ref{inidata-up}) before the time $t_c$ (global shock or collapse time) is obtained by joining three regions of different analytical behavior. 
Thus, along with a far-field region, $|x|> b(t)$ say, where both $\eta$ and $u$ are constant, and with a region centered around the origin $|x|< a(t)$, where the solution maintains the parabolic form described in the previous section, we have connecting regions~$a(t)<|x| <b(t)$ where the solution is determined by simple waves, which we denote by the pair $(N(x,t), V(x,t))$. With the standard notation for characteristic functions $\chi_I$ of the support interval $I \subseteq \barr$, the whole solution can be  
written explicitly as  
\begin{equation}
\begin{split}
\eta(x,t)=&\gamma \,x^2\,\chi_{(-a,a)}+N(x,t)\chi_{(a,b)}+N(-x,t)\chi_{(-b,-a)}  + 
Q \, \chi_{(-b,b)^c}\\
u(x,t)=&\nu \, x \, \chi_{(-a,a)}+V(x,t)\chi_{(a,b)}-V(-x,t)\chi_{(-b,-a)}\, .
\label{Anssol}
\end{split}
\end{equation}
Here the functions 
$a(t),b(t)$ describe the evolution of the boundaries of the above-mentioned regions where
the solution of system~(\ref{Airymodel}) is continuous but might not be differentiable.
With reference to figure \ref{char-uppara-fig},  these parts of the solution live in the $S$ region, where the pair $(N,V)$ constitutes a simple wave 
(thanks to the symmetry $x\mapsto -x$, we can restrict  to $x\geq 0$).
The $P$ region in figure~\ref{char-uppara-fig} depicts characteristics~(\ref{chars}) and corresponds to the parabola part of the solution. 
In the $C$ regions both Riemann invariants are constant and both families of characteristics are straight lines. In the right $S$ region, the Riemann invariant 
$R_-=u-2 \sqrt{\eta}\equiv V-2 \displaystyle\sqrt{N}$ is constant (independent of~$x_0$), since its value is $R_-(x_0,0)=-2\displaystyle\sqrt{Q}$ at all points $(x_0,0)$ with $x_0\ge a(0)=\sqrt{Q/\gamma_0}$, and this value is conserved along characteristics 
$x_-(t;x_0)$ (dashed curves in figure \ref{char-uppara-fig}). Hence the pair $(N,V)$ is functionally related, i.e., it is a simple wave with 
\begin{equation}
V=2 \sqrt{N}-2\sqrt{Q}\, .
\label{swrel}
\end{equation}
Thus, in this region $S$, the evolution is governed by
\begin{equation}
 N_t+\lambda_+\big(R_+,R_-\big)\,N_x= N_t+\frac14\big(3R_+ +R_-\big)\,N_x =N_t+\big(3 \displaystyle\sqrt{N}-2\displaystyle\sqrt{Q}\big)N_x=0\, ,
 \label{nsmpw}
\end{equation}
so that the non-dashed characteristic curves in the right $S$ region are straight lines starting from $(a(t_0),t_0)$ with (inverse of the) slope given by 
\begin{equation}
3\sqrt{N}-2\sqrt{Q}=3 \sqrt{\gamma(t_0)}a(t_0)-2\sqrt{Q}.
\label{strght}
\end{equation}
Note that the special value $N=4\, Q$/9 makes the characteristic line vertical, which in turn makes this value constant for all times at the corresponding $x$, $x\equiv x_d=\big(\displaystyle\sqrt{3 Q / \gamma_0}\big)/4$, or 
$N(x_d,t)=4 \,Q/9$ for $t>t_0$.
Since the value of $N$ is invariant along on straight lines defined by~(\ref{strght}), we have that  $N(x,t)=\gamma(t_0) a(t_0)^2$, where $t_0$ is given implicitly  by
\begin{equation} 
x=\left(3 \sqrt{\gamma(t_0)}a(t_0)-2\sqrt{Q}\right)(t-t_0)+a(t_0)\, .
\end{equation}
It is convenient to use the parameter $\sigma_0$ corresponding  to $t_0$ defined by  relation~(\ref{curvt}). Using~(\ref{aevog}) and the second equation in~(\ref{parab-gdep}) yields
\begin{equation}
\label{xnv}
\begin{array}{c}
N(x,t)=\gamma(\sigma_0) a(\sigma_0)^2=\sigma_0 Q \left({\sqrt{\sigma_0 }-\sqrt{\sigma_0-1}}\right)^2 \,, 
\vspace*{0.3cm}\\
3\displaystyle\sqrt{N(x,t)}-2 \displaystyle\sqrt{Q}=
\Lambda(\sigma_0)\equiv 3\,\displaystyle\sqrt{Q}\,\sigma_0\left(1-\sqrt{1-{1\over \sigma_0}}-{2\over 3\sigma_0}\right) \,,
\end{array}
\end{equation}
where $\sigma_0=\sigma_0(x,t)$ is implicitly given by
\begin{equation}
\label{xtaut}
 x=\Lambda(\sigma_0)\left(t-\frac{\sqrt{\sigma_0-1}+\sigma_0\, \mathrm{arctan}\left(\sqrt{\sigma_0-1}\right)}{2\, \sigma_0 \,\sqrt{\gamma_0 } } 
 \right)
 +\frac{\displaystyle\sqrt{Q \,\sigma_0 }-\displaystyle\sqrt{Q(\sigma_0-1)}}{\sigma_0 \sqrt{\gamma_0}} \, .
 \end{equation}
From the shoulder solutions' viewpoint, an alternative definition of the global shock time $t_c$ is that of the catastrophe condition, obtained by finding the first time at which the partial derivative $N_x$ becomes infinite.
From (\ref{xnv}) and (\ref{xtaut}) we see that this happens when
\begin{equation}
\frac{\partial x}{\partial\sigma_0}(\sigma_0,t)=0\, .
\label{xinf}
\end{equation}
This shock time for the shoulders coincides with the
shock time for the parabola's collapse to a segment (and with the crossing time $t=t_s$
at which $a(t)=0$), 
\begin{equation}
t_s=t_c=\frac{\pi}{4 \sqrt{\gamma_0}}\, .
\end{equation}
Similarly, the shock position of the shoulders coincides with the shock position by the collapsed parabola, i.e.,  $x(t_c)=0$. Once again, the vertical segment $\eta\in [0,Q/4]$ at $x=0$, $t=t_c$ to which the parabola has collapsed, can be suppressed by setting $\eta(0,t_c)=Q/4$ and $\eta(x,t_c)=N(|x|,t_c)$, thus restoring  continuity at $x=0$ for the surface elevation $\eta$. 

Past the global shock time $t=t_c$, the evolution of solutions of system~(\ref{Airymodel}) continues in weak form. As we shall see below,  weak solutions of~(\ref{Airymodel}) at short time $t>t_c$ are determined by the local behaviour of the simple-wave shoulders around the origin, i.e., before the shocks that emanate from $x=0$ have travelled sufficiently far away from a neighbourhood of the  origin. While of course simple wave solutions are globally known implicitly by the characteristics for~(\ref{nsmpw}), an explicit, albeit asymptotic, form is useful for deriving analytic expressions of $\eta$ and $u$ near $x=0$. This requires knowledge of a new initial condition, $N(x,t_c)\equiv N_0(x)$ say, in the shifted time $\tilde{t}\equiv t-t_c$, for the Hopf-like equation~(\ref{nsmpw}). The form of $N_0$ is given by the joined  left and right shoulders at $\tilde{t}=0$, continued through $x=0$ by taking for $N_0(0)$ its limiting value $Q/4$. Putting $t=t_c$ in~(\ref{xtaut}) and seeking the asymptotic expansion 
$\sigma_0\to \infty$ of the right-hand side yields 
\begin{equation}
x=\frac13\, \sqrt{{Q\over \gamma_0}}\,\sigma_0^{-3/2} +O(\sigma_0^{-5/2})\, , 
\label{xasympt}
\end{equation}
or, inverting, 
\begin{equation}
\sigma_0\sim {1\over 3^{2/3}}\left({Q  \over \gamma_0}\right)^{1/3}\,  x^{-2/3}. 
\end{equation}
This asymptotic solution for the implicit relation~(\ref{xtaut}) gives the leading order behaviour for 
$N_0(x)$, 
\begin{equation}
N_0(x)\equiv N(x,t_c) \sim {Q \over 4} + {3^{2/3}\, \gamma_0^{1/3}\, Q^{2/3} \over 8} \,x^{2/3}+o(x^{2/3})
\qquad \mathrm{as} \quad x \to 0 \, .
\label{twthrds}
\end{equation}
The corresponding asymptotics for the velocity component can of course be obtained from this expression and the simple wave relation~(\ref{swrel}).
Thus, around the origin the spatial behavior  of the left and right shoulders, connected at $x=0$, $\eta=Q/4$,  is that of a branch point (cusp singularity) $O(x^{2/3})$. As we shall see below, this singular behaviour has non-trivial effects on the global evolution of interface $\eta$  and velocity $u$ after the shock time.

A few remarks are now in order. First, while system~(\ref{coeffODEsr}) and its solution~(\ref{curvt}) have been introduced for the case of  parabolic-shape layer thickness at the dry-point, with zero velocity initial data, most of the corresponding results actually apply to {\it any} (symmetric) initial condition for $\eta$ that admits a convergent Taylor series expression in the neighbourhood  of the contact point. As shown in the Appendix, in the presence of dry point $\eta=0$, the evolution of the leading order coefficients of the Taylor series for $\eta$ and $u$  decouples from all their higher order counterparts. 
That is, if by $\eta_{n+1}(t)$ and $u_n(t)$ we denote the $n$-th order pair of Taylor coefficients (cf.~(\ref{Ssolnear0})),  the system of ODE's governing the evolution of $\eta_1(t)$ and $u_0(t)$, equation~(\ref{paraeq}), coincides with system~(\ref{coeffODEsr}). Second, all the higher order coefficients in the Taylor series are determined recursively from the preceding ones through a well defined hierarchy of ODE's. At each stage beyond  the leading order pair~$(\eta_1,u_0)$, these ODE's are {\it linear} in the new unknowns $\eta_{n+1}(t)$ and $u_n(t)$, and hence each 
$n$-th~pair inherits the singularity determined by that of the first pair equations, i.e., system~(\ref{coeffODEsr}),  which as shown above admits a closed form solution. Hence, the catastrophe time at $x=0$ is determined by the local curvature blowup, or $t=t_c$, through relation~(\ref{shocktime}) for any smooth initial data $\eta(x,0)$ and $u=0$  that have the assumed property of nonzero curvature at the contact point.  This generalizes a result reported in~\cite{MT} obtained through a different approach for the case (in our notation) 
$\eta(x,0)=\tanh^2(x)$, $u(x,0)=0$, for the Airy system viewed as the zero dispersion limit of the defocusing Nonlinear Schr\"odinger equation. Further, the singularity time expression~(\ref{shocktime}) signals that as the curvature of $\eta$ at the dry-point vanishes, or $\gamma_0 \to 0$, the singularity might actually never occur in finite time, and that when the contact point at $x=0$ is a zero of higher order for $\eta(x,0)$, a different result might be expected. In fact, it is easy to show from the Taylor series approach, and the ensuing hierarchy of time dependent coefficients, that the first nonzero Taylor coefficient of $\eta$ is time independent, that is, the local shape of $\eta$ in a neighbourhood of the origin is essentially invariant in time. Lastly, note that 
the limiting form of the simple wave~(\ref{twthrds}) suggests a property that was missed
in previous investigations with smooth initial data: the spatial dependence of $\eta$ and $u$ at the singularity is a non-analytic cusp scaling like $x^{2/3}$. Thus, the smoothness loss occurs with a derivative $\eta_x$ that diverges as $x=0$, unlike the case of the divergence of curvature to a corner whereby the derivative would jump between finite values to the left and right of $x=0$.

From the initial condition $N_0(x)$ at time $\tilde{t}=0$ the evolution of the shoulders generates multivaluedness, as the characteristics $x_-(t)$ with $x_0>a_0$ from  right-region $C$  cross their counterpart 
with $-a_0<x_0<0$ from the left-region $S$  (and vice-versa, by symmetry, for $x_+(t)$ characteristics). As well known  for hyperbolic systems, multivaluedness can be remedied  by selecting a shock location, $|x|=x_s(t)$ say, and allowing the evolution of solutions governed by system~(\ref{Airymodel}) to continue in the weak sense.  With respect to the shifted time $\tilde{t}=t-t_c$ (and immediately dropping the tilde as long as this does not generate confusion), the initial condition for the shock position is clearly $x_s(0)=0$; for later purposes it is informative to eliminate, asymptotically for $t\to 0^+$ (or $t\to t_c^+$ in the non-shifted time), the implicitness of the functional dependence $N(x,t)$ at the shock location. This can be achieved directly from~(\ref{nsmpw}) taking  $N_0(x)$ as a new initial condition. In fact, along the characteristic originating from $x=x_0>0$, which for clarity  we simply denote by $x=X(t)$ since we are treating this as a new initial value problem, we have 
$$
\dot{X}(t)=\big(3\displaystyle\sqrt{N(X(t),t)} -2\displaystyle\sqrt{Q}\,\big)\,, \quad X(0)=x_0\,,\qquad N(X(t),t)=N_0(x_0) \, . 
$$     
Hence the characteristics through the point $x=x_s(t)$ at time $t$ are defined implicitly by $X(t)=x_s(t)$, i.e., by an initial value $x_0$ such that 
$$
x_0+t\, \big(3 \displaystyle\sqrt{N_0(x_0)} -2\displaystyle\sqrt{Q}\,\big)=x_s(t) \, . 
$$
For $t \to 0^+$ the asymptotic form of $N_0$~(\ref{twthrds}) can be used in this relation, which results in a cubic equation for $x_0^{1/3}$ 
\begin{equation}
x_0+t\,{3 K \over \displaystyle\sqrt{Q}} \, x_0^{2/3}-\left(x_s(t)+{\displaystyle\sqrt{Q}\over 2 }\, t\right) =0\, , 
\label{cubxs}
\end{equation}
where we have defined the shorthand 
\begin{equation}
K\equiv {3^{2/3}\, \gamma_0^{1/3}\, Q^{2/3} \over 8} \, . 
\label{Kdef}
\end{equation}
Since both $t$ and $x_s(t)$ are small (and expected to be of the same order) as $t\to 0^+$, 
the dominant balance in equation~(\ref{cubxs}) is 
\begin{equation}
x_0\equiv  X_0(t)=x_s(t)+{\displaystyle\sqrt{Q}\over 2 }\, t \, , 
\label{dmbalnc}
\end{equation}
so that the explicit asymptotic form of $N(x,t)$ at the shock location for $t\to 0^+$ is 
\begin{equation}
N(x_s(t),t)=N_0(X_0(t))\sim{Q \over 4} + K\, \big(X_0(t)\big)^{2/3}={Q \over 4} + {3^{2/3}\, \gamma_0^{1/3}\, Q^{2/3} \over 8} \left(x_s(t)+{\displaystyle\sqrt{Q}\over 2 }\, t\right)^{2/3} \, . 
\label{expN0}
\end{equation}

\section{Beyond the collapse time: analytical approach}
 \label{sec:beyond_shock}
The analysis developed in Section~\ref{genpar-sec} describes  the evolution of our class of initial conditions  until the catastrophe time $t_{c}$. In particular, we have seen how the singularity of the velocity field $u(x,t)$ occurring at $t=t_{c}$ for dry-spot initial data leads to a reconnection of the left and right water masses into a single mass with a cuspy ``well" of depth $Q/4$ at $x=0$ (see figure \ref{para-dry-fig}a, last time/panel). Our next task is to discuss the behavior of the system after this collapse and ensuing global (vertical segment) shock. We shall tackle this problem from an analytical perspective first, with a numerical study reported in the following section~\S\ref{numerics}. 

\subsection{The three main examples} 
We begin by briefly recalling some fundamental properties of shock fitting in the context of our model equation~(\ref{Airymodel}). We shall focus here on the salient features of the original problem~(\ref{inidata-up}),  after the parabola collapse and the zero-measure stem connecting the bottom to the elevation $Q/4$ is removed. As discussed above, the general structure giving rise to shocks can hence be taken to be that of simple waves ``shoulders" which have merged at the collapse time. Thus, the solution~(\ref{xnv}) and~(\ref{xtaut}) for the Airy model~(\ref{Airymodel}) is comprised of two simple waves corresponding to the 
Riemann invariants~(\ref{RI}), $R_\pm=\hbox{const.}$ respectively for $x>0$ and $x<0$.  
For both signs  the related characteristic velocities $u\pm \sqrt{\eta}$ increases when 
$u$ increases, and the Lax conditions~\cite{Lax} are satisfied and lead to  shock  jump relations enforcing conservation of mass and momentum, with densities $\eta$   $\eta u$ respectively. 
The shock speed $\dot{x}_s(t)$ for the shock position $x=x_s(t)$  is then given (see, e.g.,~\cite{Whitham}) by
\begin{equation}\label{shockspeed}
\dot{x}_s=\frac{[\eta\, u]}{[\eta]} =\frac{[\eta\, u^2 +\eta^2/2]}{[\eta \, u]} \, , 
\end{equation}
and the consistency between the two expressions can be manipulated to  
\begin{equation}
[u]^2=\frac{[\eta]^2(\eta_++\eta_-)}{2\eta_+\,\eta_-} \, , 
\label{shockcc}
\end{equation}
where, as customary, $[f]=f_+-f_-$ denotes the jump across the shock from right ($+$ subscript), to left ($-$ subscript). Next, we consider three realizations of the structure described above in order of increasing complexity, with the parabola case~(\ref{inidata-up}) at collapse being the most involved due to the loss of smoothness at $t=t_c$ as evidenced by~(\ref{twthrds}). 

\subsubsection{The ``double-Riemann" (toy) model}
\label{riemann}
\begin{figure}[t]
\centering
\begin{center}
(a)\includegraphics[width=.9\textwidth]{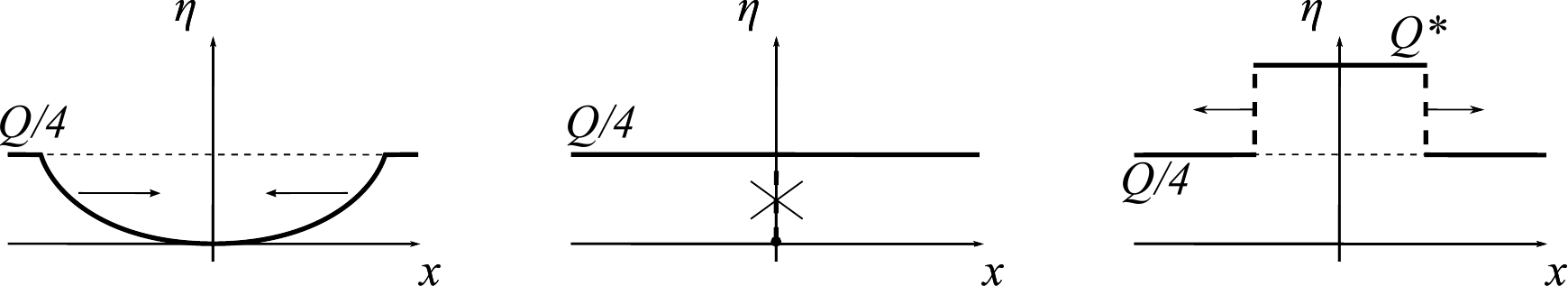}\\
\vspace*{1cm}
(b)\includegraphics[width=.9\textwidth]{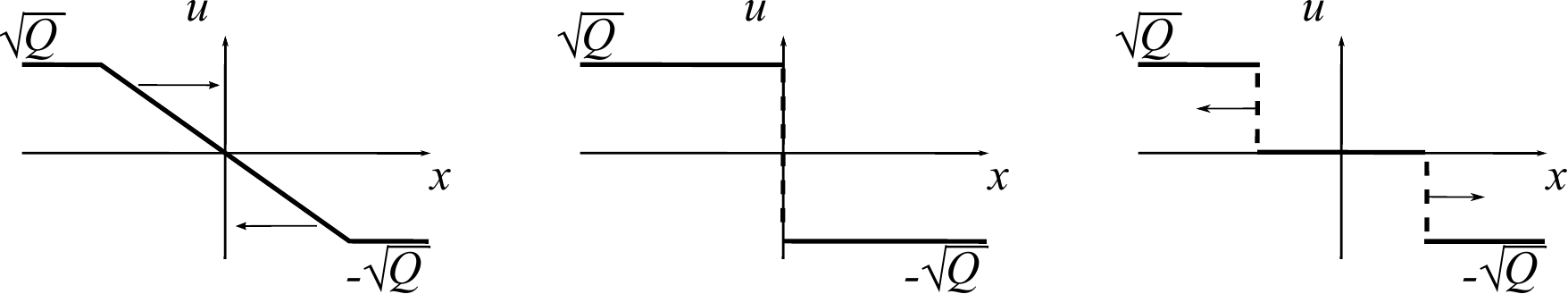}
\end{center}
\caption{Schematics of the evolution of initial data~(\ref{compress}) (with initial time $t_0<t_c$). Panel (a): $\eta(x,t)$; panel (b): $u(x,t)$. }
\label{toposhock-fig}
\end{figure} 
The simplest ``cartoon" of evolution after the collapse of the dry-spot parabola of \S2 is arguably that of  the Riemann problem obtained by taking an initial condition determined solely by the 
asymptotic value $\eta(0, t_{c})=Q/4$ and a jump at $x=0$ of the velocity $u(0, t_{c})$ 
of amplitude $2\displaystyle\sqrt{Q}$,  
\begin{equation}
\eta(x,t_c)={Q \over 4}\,, \qquad  u(x,t_c)=-\sqrt{Q}\, \sgn(x) \, .
\end{equation}
Such an initial condition for the Airy system~(\ref{Airymodel}) can be viewed as mimicking the local behavior around $x=0$ at the global shock time $t=t_c$ created by the merging of the shoulder simple waves. This Riemann problem can itself be considered as that emerging at the catastrophe time from the evolution of  a piecewise continuous initial condition obtained by reverse evolution of the classical expansion wave emanating from a step-like initial condition, as well known (see, e.g.,~\cite{GK}).
Thus, prior to the catastrophe, we consider a
solution of the form, which we dub ``double-Riemann,"   
\begin{equation}
 \eta(x,t)=\left\{ \begin{array}{c}
               {Q/4},    \vspace*{0.2cm}\cr
               \displaystyle\frac{1}{9} \left( \displaystyle\frac{x}{t_c-t}\right)^2, \vspace*{0.2cm}\cr
               {Q/4} ,
              \end{array}
\right. \quad
u(x,t)=\left\{ \begin{array}{cr}
               \displaystyle\sqrt{Q}, & x<-a(t) \vspace*{0.2cm}\cr
               -\displaystyle\frac{2}{3} \left( \displaystyle\frac{x}{t_c-t}\right), & \hspace{1cm} -a(t)<x<a(t)\vspace*{0.2cm}\cr
                             -\displaystyle\sqrt{Q}, & x>a(t) 
\end{array}
\right. \,,
\label{compress}
\end{equation}
with
\begin{equation}
a(t)= {3 \over 2} \sqrt{Q}\, (t_c-t)\, .
\end{equation}
When $t=t_c$ a global shock forms, with the elevation becoming constant, $\eta=Q/4$, except at $x=0$ where $\eta=0$, and with the velocity jumping at $x=0$, 
$ u=-\displaystyle\sqrt{Q}\, \sgn(x)$.
The evolution is not affected by the value of the new initial data at one point, so we can assume that $\eta={Q/4}$ everywhere.
Thus, in our case for $x>0$, $f_+$ corresponds to the constant ``background" solution 
$\eta=Q/4$ and $u=-\displaystyle \sqrt{Q}$, while $f_-$ is determined by the solution to the cubic equation for $\eta_-$ which follows from~(\ref{shockcc}) when $u_-=0$.
We obtain
\begin{equation}
 \eta=\left\{ \begin{array}{c}
               {Q/4}  \\
               Q^* \\
               {Q/4} 
              \end{array}
\right.\,, \qquad
u=\left\{ \begin{array}{c}
               \sqrt{Q}  \\
               0  \\
               -\sqrt{Q} 
              \end{array}
              \right.\,, \qquad
 \begin{array}{r}
                \ x<-x_s \\
               -x_s<x<x_s \\
                x>x_s 
              \end{array}
 \,,
\label{spit}
\end{equation}
where
\begin{equation}
 x_s(t)= \frac{Q^{\frac32}}{4Q^*-{Q}}\, (t-t_c) \, .
\end{equation}
Here $Q^*$ is given by the unique solution of the cubic equation for $\eta_-$ which follows from the compatibility condition~(\ref{shockcc})
\begin{equation}
\eta_-^3 -{Q\over 4}\eta_-^2-{9 Q^2 \over 16}\eta_-+{Q^3 \over 64}=0
\end{equation}
subject to $\eta_->Q/4$. The decimal value of this root  is approximately
\begin{equation}
Q^* \simeq 3.49396\, {Q \over 4}=0.87349\, Q \, , 
\label{cub}
\end{equation} 
which leads to the (constant) shock speed   
\begin{equation}
 \dot{x}_s \equiv s_0=\frac{Q^{\frac32}}{4Q^*-{Q}}\simeq 0.4009689\, \sqrt{Q} \, . 
 \label{sriem}
 \end{equation}
The evolution is depicted in figure~\ref{toposhock-fig}.

\begin{figure}
\includegraphics[width=.9\textwidth]{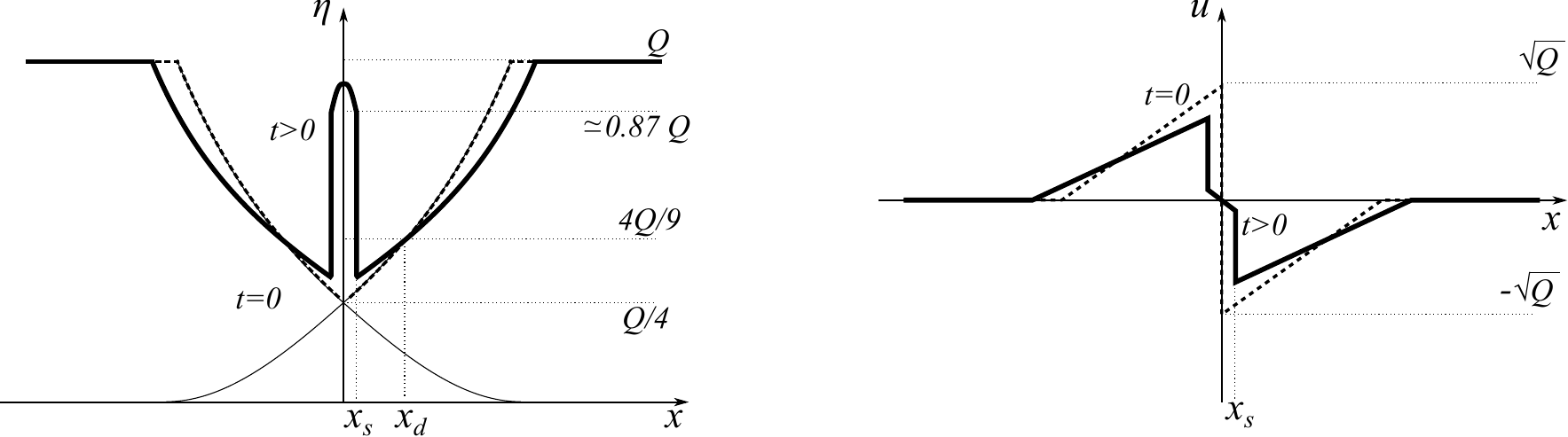}
\caption{Schematics of the initial condition (dash) and evolution (solid) with shock development at short times $t>0$ for the Airy solutions~(\ref{2stok}). The thin segments of parabolae are removed from the initial data and the shocks develops from the initial discontinuity in the velocity (right panel).
}
\label{screwtex}
\end{figure}

\subsubsection{The ``double-Stoker" model}
\label{doubleSt}
In order to capture the time evolution and spatial dependence of the initial conditions  that emerge from the parabola collapse, the toy double-Riemann problem model described above is not adequate and 
can be modified as follows.
We consider the evolution from an initial condition obtained by ``cutting" and ``splicing" together two Stoker (parabolic) rarefaction wavescrossing at $x=0$, $\eta(0, 0)=Q/4$   (for relevant definitions, see e.g., \cite{Stoker}, \S10.8), 
\begin{equation}
\vspace*{0cm}
N_S(x,t)=\left\{\begin{array}{l} Q,\cr
\displaystyle\frac{1}{9}\, \left( {\frac {x-x_{{d}}}{t+t_{{d}}}}+2\,\sqrt {Q} \right)^{2}\vspace*{0.1cm},\cr
\displaystyle\frac{1}{9}\, \left( {-\frac {x+x_{{d}}}{t+t_{{d}}}}+2\,\sqrt {Q} \right)^{2}\vspace*{0.1cm},\cr
Q,
\end{array}\, 
\right.\quad  
V_S(x,t)=\left\{\displaystyle \begin{array}{lr}0,& x\geq x_Q\vspace*{0.2cm}\cr
\displaystyle\frac23\,{\frac {x-x_{{d}}}{t+t_{{d}}}}-\frac23\,\sqrt {Q},&0<x<x_Q\vspace*{0.2cm}\cr
\displaystyle\frac23\,{\frac {x+x_{{d}}}{t+t_{{d}}}}+\frac23\,\sqrt {Q},&\hspace{0.5cm}-x_Q<x<0\vspace*{0.2cm}\cr
0, & x\leq -x_Q
\end{array}
\right.
\label{2stok}
\end{equation}
where 
\begin{equation}
x_{{d}}=
\,{1\over 4}\sqrt{3 Q \over  g_0},\quad 
t_d={1\over 2}\sqrt {3 \over g_0}\,, \quad x_Q=\sqrt{Q}\, t+\displaystyle{3\over 4}\sqrt{3Q\over g_0}
 \, .
\label{stparam}
\end{equation}
This choice of initial data mimics the state reached at the collapse time for the ``full" case $N(x,t_c),V(x,t_c)$, when the global shock at the origin can be removed from  the simple wave shoulders in contact at $x=0$, $\eta(0^\pm,t_c)=Q/4$. Just as in that case, the evolution from initial data~(\ref{2stok}) maintains the elevation $\eta(x_d,t)=4Q/9$ and 
$u(x_d,t)=\mp 2\displaystyle\sqrt{Q}/3$ at all times (for as long as shocks do not reach the ``hinge" position $x=\pm x_d$). 
The location
$x_Q$ corresponds to the moving point where the rarefaction waves reaches the background elevation  $Q$ and the fluid velocity vanishes. The analogous point for the full case moves at the same 
characteristic speed $\sqrt{Q}$, but at the initial time $t=t_c$ putting $g_0=\gamma_0$ yields 
a different initial position for this point. We dub this initial condition the ``double-Stoker" case. 

The evolution out of these initial data forms, at time $t=0^+$, two shocks which move away from the origin, 
just as in the double-Riemann case of section~\ref{riemann}. However, unlike that case, a time-dependent shock speed and a non-trivial spatial and temporal dependence for $\eta$ and $u$ in the region between shocks can now be expected.  For $t>t_c$, $|x|>x_s(t)$, the functions $N_S$, $V_S$ continue to be solutions of  Airy's system  outside the region  $|x| < x_s(t)$ between shocks. 

\subsubsection{The ``full" case}
\label{fullsec}
This is the case of initial data (in the shifted time $t-t_c$) 
\begin{equation}
\eta(x,0)=N_0(|x|) \, , \qquad  u(x,0)=2\,\sgn(x)\left(\displaystyle\sqrt{N_0(|x|)}-\displaystyle\sqrt{Q}\right) \, ,  
\label{icN0}
\end{equation}
analyzed at the end of \S\ref{shoulders}. The surface elevation $\eta$ is continuous with a cusp point $O(x^{2/3})$ at the origin, while the corresponding velocity $u(x,0)$ has a  jump at the origin of amplitude $2\displaystyle\sqrt{Q}$. The local behavior at early times $t \to 0^+$ in a neighbourhood of the origin is summarized by the asymptotic 
behaviour~(\ref{expN0}). The schematic of the initial condition and subsequent short time evolution is qualitatively indistinguishable from that sketched in figure~\ref{screwtex} for the double-Stoker case (though for the velocity $u(x,t)$ there are substantial differences between the two cases, as we shall see below).

\subsection{Shock-fitted coordinates: unfolding the collapse singularity}
\label{shckfc}
Motivated by the tools of singular perturbation theory (see, e.g.,~\cite{cole}), in order to zoom onto the region $|x|<x_s(t)$ it is convenient to define ``unfolding" coordinates $(\xi,\tau)$ to set  the shock positions at fixed locations in time, e.g., $\xi=\pm 1$, which can be achieved by the new independent variables
\begin{equation}
\xi\equiv{x \over x_s(t)} \, ,  \qquad \tau\equiv\log (x_s(t)) \, , 
\label{chngcrdnts}
\end{equation}
so that 
\begin{equation}
\partial_x={1\over x_s}\partial_\xi
 \, ,  \qquad \partial_t= {\dot{x}_s\over x_s}\left(\partial_\tau -\xi \, \partial_\xi\right)\, . 
\label{chngcpartl}
\end{equation}
With this mapping, the Airy's system~(\ref{Airymodel}) assumes the form (with a little abuse of notation by maintaining the same symbols for dependent variables) 
\begin{equation}
{\partial \eta \over \partial \tau}- \, \xi \, {\partial \eta \over \partial \xi} +
\displaystyle{ 1\over \dot{x}_s}{\partial \over \partial \xi} \big(\eta \, u \big)=0 \, , 
\qquad 
{\partial u \over \partial \tau}- \, \xi \, {\partial u \over \partial \xi} + { 1\over \dot{x}_s} 
{\partial \over \partial \xi}\Big({u^2 \over 2}
+\eta\Big) =0 \, . 
\label{unfold}
\end{equation}
Here $\dot{x}_s$ is a placeholder for the expression that couples the evolution equation for the physical time $t$ to the new evolution variable $\tau$. The shock position evolves according to the 
equation that defines $\dot{x}_s$    in terms of the jump amplitudes $[\eta]$ and $[\eta u]$:
\begin{equation}
{d (e^{ \tau}) \over d t}=\dot{x}_s={[\eta u]\over [\eta]}=\displaystyle{N(x_s(t),t)\, V(x_s(t),t)-\eta(1,\tau(t))\,u(1,\tau(t))
\over N(x_s(t),t)-\eta(1,\tau(t))} \,, \qquad x_s(0)=0, 
\end{equation}
so that, in the new variables, 
\begin{equation}
{d t \over d \tau}=\displaystyle{e^{\tau} \,  \big(N(e^{\tau},t(\tau))-\eta(1,\tau)\big)
\over N(e^{\tau},t(\tau))\, V(e^{ \tau},t(\tau))-\eta(1,\tau)\,u(1,\tau)} \,, \qquad t(\tau)\to 0 \, \quad \hbox{as} \,\,\, \tau \to -\infty\,.  
\label{time}
\end{equation}
Thus, the system governing the evolution of the ``inner" solution between shocks consists of this ordinary differential equation   together with the partial differential equations~(\ref{unfold}), to be solved within the strip $\xi \in [0,1]$ (by symmetry only half the $\xi$ domain 
$[-1,1]$ can be used) subject to the boundary conditions 
\begin{equation}
u(0,\tau)=0\, , \qquad u(1,\tau)=V(e^{\tau},t(\tau))
+\displaystyle\sqrt{{\big(N(e^{\tau},t(\tau))-\eta(1,\tau)\big)^2 \big(N(e^{\tau},t(\tau))+\eta(1,\tau)\big) \over 2 N(e^{ \tau},t(\tau))\,\eta(1,\tau)}} \, . 
\label{bcunf}
\end{equation}
The first equality is a consequence of the antisymmetry of the velocity, $u(\xi,\tau)=-u(-\xi,\tau)$. The second relation expresses the compatibility condition~(\ref{shockcc}) in terms of the new independent variables. Here and in what follows, we suppress subscript labels that differentiate between the different cases of simple wave solutions which bracket the region between shocks, so long as it does not generate confusion; further, recall that for simple waves $V$ and $N$ are functionally related so that for all our cases $V$ can be thought of as a placeholder for the quadratic relation $V=2\displaystyle\sqrt{N}-2\displaystyle\sqrt{Q}$.

With the shorthand notations 
\begin{equation}
\phi\big(e^{\tau},t(\tau);\eta(1,\tau),u(1,\tau)\big)\equiv \displaystyle{ N(e^{\tau},t(\tau))-\eta(1,\tau)
\over N(e^{\tau},t(\tau))\, V(e^{ \tau},t(\tau))-\eta(1,\tau)\,u(1,\tau)}-{1\over s_0} \, , 
\label{phidef}
\end{equation}
and 
\begin{equation}
\psi\big(e^{\tau},t(\tau);\eta(1,\tau)\big)\equiv V(e^{\tau},t(\tau))
+\displaystyle\sqrt{{\big(N(e^{\tau},t(\tau))-\eta(1,\tau)\big)^2 \big(N(e^{\tau},t(\tau))+\eta(1,\tau)\big) \over 2 N(e^{ \tau},t(\tau))\,\eta(1,\tau)}} \, ,
\label{psidef}
\end{equation}
where $s_0$ is the initial shock speed defined by~(\ref{sriem}),  the full system of evolution equations for the inner shock region $(\xi,\tau)\in [0,1]\times \barr$  
can be written compactly as 
\begin{equation}
{\partial \eta \over \partial \tau}- \, \xi \, {\partial \eta \over \partial \xi} +
\left(s_0^{-1}+\phi\right)
{\partial \over \partial \xi} \big(\eta \, u \big)=0 \, , 
\quad 
{\partial u \over \partial \tau}- \, \xi \, {\partial u \over \partial \xi} + 
\left(s_0^{-1}+\phi\right)
{\partial \over \partial \xi}\Big({u^2 \over 2}
+\eta\Big) =0 \, , \quad
{d t\over d\tau}= e^{\tau} \left({1\over s_0}+\phi\right)\, , 
\label{unfoldbis}
\end{equation}
with boundary conditions 
\begin{equation}
u(0,\tau)=0\,, \qquad u(1,\tau)=\psi \, . 
\label{bc_zetau}
\end{equation}

As $\tau\to -\infty$ ``initial data" for $u$ and $\eta$ can be assigned so  that they are compatible with the boundary conditions~(\ref{bcunf}). This can be done by taking
 $u(\xi,\tau) \to 0$ and $\eta(\xi,\tau)\to Q^*\simeq 3.49396\, N(0,0)$ as $\tau \to -\infty$, because of the definition~(\ref{cub}) of $Q^*$ as the solution  of the equation 
\begin{equation}
-2\sqrt{Q}+2\sqrt{N(0,0)}
+\displaystyle\sqrt{{\big(N(0,0)-Q^*\big)^2 \big(N(0,0)+Q^*\big) \over 2 N(0,0)\,Q^*}}=0 \, , 
\label{q*def}
\end{equation}
constrained by 
the condition $Q^*>N(0,0)$ (with $N(0,0)=Q/4$ for all cases considered here).

We remark that system~(\ref{unfoldbis}) shares the same Riemann invariants as those of the original model, although the corresponding characteristic eigenvalues now depend explicitly on the unfolding  coordinates $\xi$ and $\tau$ as well. The Riemann invariant form of system~(\ref{unfoldbis}) is 
\begin{equation}
R_{\pm}= u\pm 2\sqrt{\eta}\,, \qquad {d R_{\pm} \over d \tau}=0\, , \qquad 
{d \xi_{\pm} \over d \tau}=-\xi_\pm+\big(s_0^{-1}+\phi\big) \big(u\pm \sqrt{\eta}\big)\,.
\label{unfriemann}
\end{equation}

We do not attempt here to find solutions in closed form  of the initial boundary value problem for system~(\ref{unfoldbis}); this is most efficiently approached numerically. Analytical progress can be made however by focussing  on the short time behavior after shock formation, i.e., as $\tau \to -\infty$. In this limit, the initial evolution is governed by the perturbation of system~(\ref{unfoldbis}) around the initial data $\eta=Q^*$, $u=0$ and $t=0$. Denoting by $\tilde{\eta}$ the difference $\tilde{\eta}=\eta-Q^*$, and abusing notation a little by immediately dropping the tilde while maintaining  the same symbols for the dependent variables, unless necessary to avoid confusion, the governing equations for short times after shock formation are
\begin{eqnarray}
\label{unfoldbispert}
&&{\partial \eta \over \partial \tau}- \xi \,{\partial \eta \over \partial \xi} + Q^*\, (s_0^{-1}+\phi_0) 
{\partial u \over \partial \xi}=0 \, , 
\\
\nonumber
&&{\partial u \over \partial \tau}- \xi \,{\partial u\over \partial \xi} +  (s_0^{-1}+\phi_0) 
{\partial \eta \over \partial \xi}=0 \, , 
\\
\nonumber
&&{d t \over d \tau}=
e^\tau \big(s_0^{-1}+\phi_0+
\phi_\eta\, \eta(1,\tau)+\phi_u\, u(1,\tau)\big)\, , 
\end{eqnarray}
with boundary conditions 
\begin{equation}
u(0,\tau)=0\, ,\quad  u(1,\tau)=\psi_0+
\psi_\eta\,\eta(1,\tau)\, ,
\quad t(\tau)\, ,\eta(\xi,\tau)\, , u(\xi,\tau)\to 0 \quad \hbox{as}\,\,\, \tau \to -\infty \, . 
\label{psibclin}
\end{equation}
Here we have defined 
\begin{eqnarray}
\label{phipsi0}
&&\hspace{-0.5cm}
\phi_0(\tau)\equiv\phi(x_s,t;Q^*,0)=\phi(e^\tau,t(\tau);Q^*,0)=\displaystyle{ N(e^\tau,t(\tau))-Q^*\over N(e^\tau,t(\tau))\, V(e^\tau,t(\tau))}-{1\over s_0}\, , 
\\
\nonumber
&&\hspace{-0.5cm}
\psi_0(\tau)\equiv\psi(x_s,t;Q^*)=\psi(e^\tau,t(\tau);Q^*)=V(e^{\tau},t(\tau))
+\displaystyle\sqrt{{\big(N(e^{\tau},t(\tau))-Q^*\big)^2 \big(N(e^{\tau},t(\tau))+Q^*\big) \over 2  N(e^{ \tau},t(\tau))\, Q^*}}\, ,
\end{eqnarray}
and the gradient components of $\phi$ and $\psi$ with respect to the fields $\eta$ and $u$ are, respectively,
\begin{eqnarray}
\label{phipsi1}
&&\left.{\partial \phi \over \partial \eta}\right|_{\eta=u=0}=-\displaystyle{1\over N(e^\tau,t(\tau))\, V(e^\tau,t(\tau))}\,, \quad  
\left.{\partial \phi \over \partial u}\right|_{\eta=u=0}=\displaystyle{Q^* \big(N(e^\tau,t(\tau))-Q^*\big)\over \big(N(e^\tau,t(\tau))\, V(e^\tau,t(\tau))\big)^2}\, ,
\nonumber
\\
&&\left.{\partial \psi\over \partial \eta}\right|_{\eta=0}=\displaystyle{ 2{Q^*}^3-N(e^\tau,t(\tau))\, {Q^*}^2
-\big(N(e^\tau,t(\tau))\big)^3 \over 
4 \, N(e^\tau,t(\tau))\, {Q^*}^2 \,
\big(\psi_0(\tau)-V(e^\tau,t(\tau))\big)} \, .
\end{eqnarray}
Note that, as $\tau\to -\infty$, from the definitions~(\ref{cub}) and~(\ref{sriem}) of the elevation~$Q^*$ 
and shock speed $s_0$ at $t=0^+$, respectively, the boundary ``forcing" terms  $\phi_0(\tau)$ and $\psi_0(\tau)$ 
both vanish, with asymptotic rate dictated by the analytic feature of simple-wave shoulder solutions~$N(e^\tau,t(\tau))$ in this limit.  
Thus, for instance, for the double-Stoker case $N$ (and $V=2\displaystyle\sqrt{N}-2\displaystyle\sqrt{Q}$) are analytic functions of $e^\tau$ and $t(\tau)$, so that both $\phi_0(\tau)$ and $\psi_0(\tau)$ are of order~$O(e^\tau)$ as $\tau \to -\infty$, while for the full case the branch-point singularity evidenced by~(\ref{twthrds}) leads to order~$O\big(e^{{2\over 3} \tau}\big)$ for these quantities in the same limit. Conversely, in general all gradient 
components~$\phi_\eta$, $\phi_u$ and $\psi_\eta$ in~(\ref{phipsi1}) have finite non-zero limits as $\tau \to -\infty$.

The general solution of the perturbation system~(\ref{unfoldbispert}) (which is essentially the classical wave-equation in a 
space- and time-varying medium) satisfying the boundary condition at $\xi=0$ is 
\begin{equation}
\eta(\xi,\tau)=\frac12\big(F(e^\tau\xi-\Phi(\tau))+F(-e^\tau\xi-\Phi(\tau))\big)  \, , 
 \quad 
u(\xi,\tau)={1\over 2\displaystyle{\sqrt{Q^*}}}\big(F(e^\tau\xi-\Phi(\tau))-F(-e^\tau\xi-\Phi(\tau))\big) \, ,\label{uetaF}
\end{equation}
for any function $F(\cdot)$ of sufficient regularity, with the function $\Phi$ defined by
\begin{equation}
\Phi(\tau)\equiv \displaystyle\sqrt{Q^*}
\int_{-\infty}^\tau e^{\tau'}\left({1\over s_0}+\phi_0(\tau') \right)d \tau' 
 \, . 
\label{caphi}
\end{equation}
The asymptotic conditions on the system's solution~$(\eta,u)$ as $\tau \to -\infty$ requires $F(0)=0$, and substitution of these expressions evaluated at $\xi=1$ into the boundary condition~(\ref{psibclin})  leads to a functional equation for $F$, coupled to the evolution equation for $t(\tau)$ in system~(\ref{unfoldbispert}). 

From this general formulation of the perturbed unfolding problem, it is useful to  derive identities that can be used for validation purposes. The initial  time evolution of the curvature of $\eta$ and the slope of $u$ in a neighbourhood of the origin can be explicitly computed directly from the perturbed evolution equation~(\ref{unfoldbispert}). 
In fact, at $\xi=0$, the first two equations of system~(\ref{unfoldbispert}) yield
\begin{equation}
\eta_{\xi\xi}(0,t)=-{s_0 \over 1+s_0\, \phi_0}\left.\big(u_{\xi\tau}-u_\xi\big)\right|_{\xi=0}, 
\label{etaxxi}
\end{equation}
and
\begin{equation}
u_\xi(0,t)=-{s_0 \over Q^*(1+s_0\, \phi_0)}\, \left. \eta_\tau \right|_{\xi=0},
\qquad u_{\xi\tau}(0,t)=-{s_0 \over Q^*(1+s_0\, \phi_0)}\, 
\left.\big(\eta_{\tau \tau} -Q^* {\phi_0}_\tau u_\xi\big)\right|_{\xi=0} \, . 
\label{uxitau}
\end{equation}

Further progress towards closed form expressions depends on the particular asymptotic behaviour of the 
simple-wave shoulder 
solution~$N(e^\tau,t(\tau))$. We next examine each of our three representative cases above, the ``double-Riemann," ``double-Stoker" and ``full" cases,
highlighting their respective differences in the unfolding coordinate formulation.

\subsubsection{Shock unfolding for double-Riemann initial data}
The weak solution and shock fitting after the collapse of the initial parabola for this case has already been achieved in \S\ref{riemann} following classical results for the Riemann problem. In fact, it is easy to see that in the unfolding formulation above the inner problem between shocks is time independent and consistent with the initial conditions as $\tau \to -\infty$, since  the fields $N$ and $V$ are constant, $N=Q/4$ and 
$V=-\displaystyle\sqrt{Q}$,  and compatible with these initial data. Thus, the second boundary condition in~(\ref{bc_zetau}) is constant, 
and the evolution equation~(\ref{unfoldbis}) admits the trivial solution  
$u(\xi,\tau)=0$ and $\eta(\xi,\tau)=Q^*\simeq0.87349\, Q$ for all $\tau \in \barr$.

\subsubsection{Shock unfolding for double-Stoker initial data}
\label{2stoksec}
Next, and in order of increasing complexity, we look at the double-Stoker case~(\ref{2stok}), where the~$(x,t)$-dependence of the boundary forcing terms $N$ and $V$  is known explicitly, and hence so is the functional form of $N(e^{\tau},t(\tau))$ and $V(e^{\tau},t(\tau))$.

We use  asymptotics for $\tau \to -\infty$ in equations~(\ref{phipsi0})-(\ref{caphi}). Expanding the function $F$ in Taylor series as its argument goes to zero, or 
$e^\tau-\Phi(\tau)=O(e^\tau)$ as $\tau \to -\infty$,  yields at leading order
\begin{equation}
{1\over \displaystyle\sqrt{Q^*}}\, F'(0) \, e^\tau\sim A \, e^\tau- B\, F'(0) \, \Phi(\tau) \, , 
\label{fp1}
\end{equation}
where the constants $A$ and $B$ follow from the asymptotic limits as $\tau\to-\infty$ 
$$
A=\lim_{\tau\to -\infty} e^{-\tau}\psi_0(\tau)\simeq -0.22302 \left({\displaystyle\sqrt{Q} \over s_0} +2\right) \sqrt{g_0} \, ,  \qquad B=\lim_{\tau\to -\infty} \psi_\eta(\tau)\simeq {1.4765 \over \displaystyle\sqrt{Q}} \, .
$$
Here we have used the asymptotic expression 
$$
t(\tau)\sim {1\over s_0}\,e^\tau \,,
$$
which follows from the leading order of the third equation in system~(\ref{unfoldbispert}).
The same leading order evaluation of this equation yields  the asymptotics for $\Phi$   
\begin{equation}
\Phi(\tau) \sim {\displaystyle\sqrt{Q^*} \over s_0}\, e^\tau \equiv \Phi_0 \, e^\tau\,, 
\label{phigrdasy}
\end{equation}
so that equation~(\ref{fp1}) provides a closed form relation  to the parameters of the double-Stoker data for the derivative 
\begin{equation}
F'(0)={\displaystyle\sqrt{Q^*}}\, \, {A\, s_0  \over s_0+ B\,Q^* } \simeq -0.22215 \sqrt{g_0 \,Q}\, . 
\label{fpfin}
\end{equation}
With this formula at hand, the asymptotic leading order behaviour of the boundary values  
$\eta(1,\tau)$ and $u(1,\tau)$ can be readily found,  
\begin{equation}
\eta(1,\tau)\sim -F'(0)\,{\displaystyle\sqrt{Q^*}\over s_0}\,\,  e^\tau 
, \qquad 
u(1,\tau)\sim {F'(0) \over \displaystyle\sqrt{Q^*}} \, e^\tau \, . 
\label{eta1u1}
\end{equation}
Note that the spatial dependence of $u(x,t)$ in the inner region between shocks $x\in\big(-x_s(t),x_s(t)\big)$ at leading order as time $t\to 0^+$ can already be read off from 
the second equation in~(\ref{eta1u1}), since
\begin{equation}
u(x,t)=u(\xi,\tau)\sim {1\over \displaystyle\sqrt{Q^*}}\, F'(0)\,  \xi \, e^\tau = {F'(0)\over \displaystyle\sqrt{Q^*}}\, { x \over x_s(t)}\,\cdot x_s(t) =
{F'(0) \over \displaystyle\sqrt{Q^*}}\,  x \equiv \nu^{(0)} x \simeq -0.237649 \sqrt{g_0}\, x\,,
\label{2stokslopu}
\end{equation}
that is, the velocity field  between shocks jumps from a vertical step of amplitude $2\displaystyle\sqrt{Q}$ at $t=0$ to a linear behaviour with finite (negative) slope  
\begin{equation}
\nu^{(0)} \equiv {A\, s_0\over  s_0+B\, Q^*} \simeq -0.23769 \sqrt{g_0}\, . 
\label{alpha0}
\end{equation}
The first equation in~(\ref{eta1u1}) shows that from the shock value 
$Q^*$ the free surface  $\eta(0,t)$ initially evolves in the inner region (restoring the tilde notation to 
return to physical variables)  with an $x$-independent base-point that increases linearly in time according to 
\begin{equation}
\eta(x,t)=Q^*+\tilde{\eta}(\xi,\tau)\sim Q^*-F'(0)\Phi(\tau) \sim 
{Q^*}\, (1-{\nu^{(0)}})\,  t
\equiv {Q^*}+\mu^{(1)} t \, ;
\label{mu1}
\end{equation}
the decimal value of the slope $\mu^{(1)}$ for this linear increase in time is 
$ \mu^{(1)}=-\nu^{(0)} Q^* \simeq 0.20762\, \sqrt{g_0} \, Q \, $. These scaling behaviours with respect to time  and parameters are consistent with the general relations~(\ref{etaxxi}) and~(\ref{uxitau}) that follow directly from the perturbed evolution equations~(\ref{unfoldbispert}). The choice of notation $\mu^{(1)}$ and $\nu^{(0)}$ in the above formulae is deliberate, since these parameters play a similar role as their counterparts for the self-similar solutions~(\ref{exact}); however, we stress that in this case~(\ref{2stokslopu}) and~(\ref{mu1}) provide asymptotic approximations, rather than exact solutions,  for the evolution at short times $t>0$.  

The above expansions can be continued to order $O(e^{2\tau})$, so that the asymptotic functional dependence 
on $\xi$ for $\eta$ and on $\tau$ for $u$, respectively, can be computed.  A glance at the general solution~(\ref{uetaF}) shows that the respective structure of these higher order corrections
is that of quadratic and linear dependence on $\xi$. With the quadratic corrections the asymptotic form of $\tilde\eta$ and $ u$  as $\tau \to -\infty$ becomes 
\begin{eqnarray}
&&\tilde{\eta}(\xi,\tau)\sim -F'(0) \Phi_0\, e^\tau+
\left(-F'(0) \Phi_1 + \frac12 \,F''(0) \big(\xi^2+\Phi_0^2\big) \right)e^{2\tau}\, ,
\label{et2u2eta}
\\
&&{u}(\xi,\tau) \sim {F'(0)\over \displaystyle\sqrt{Q^*}}\, \xi \, e^\tau -{F''(0)\over \displaystyle\sqrt{Q^*}}\, \Phi_0 \,  \xi \, e^{2\tau} \, ,
\label{et2u2u}
\end{eqnarray}
where the Taylor coefficient $F''(0)$ and that for the asymptotic expansion of $\Phi(\tau)$ (extending~(\ref{phigrdasy}), where $\Phi_0=\displaystyle\sqrt{Q^*}/s_0$),
\begin{equation}
\Phi(\tau)\sim \Phi_0 \, e^{\tau}+\Phi_1 \, e^{2\tau}\,,
\label{Phiasy2}
\end{equation}
at second order  are
\begin{equation}
\Phi_1=
-{\displaystyle\sqrt{g_0\, Q^*}\over 3\sqrt{3}}
{(1+4 Q^*/Q)\big(\displaystyle\sqrt{Q}+2 s_0\big) \over Q\, s_0}\simeq 
-3.6325 \displaystyle\sqrt{{g_0\over Q}}\,, 
\qquad 
F''(0)\simeq -0.58487 \, g_0 \, . 
\label{F2p}
\end{equation}
The second order correction also enters the asympotics for the shock location $x_s$;  from system~(\ref{unfoldbispert}), it can be shown that 
\begin{equation}
x_s(t)\sim s_0 t + s_1 t^2\, ,\qquad  s_1\simeq 0.11703 \sqrt{g_0 Q} \, , 
\label{s1}
\end{equation} 
as $\tau \to -\infty$ (and hence $t\to 0^+$).
We omit the somewhat lengthy details of the calculation for these expressions, but use these results to validate the numerical approaches in section~\ref{numerics} below. An alternative construction of the shock-continuation solution for this case is also sketched in the Appendix.
In the original $\eta(x,t)$, $u(x,t)$ variables, the asymptotic expressions~(\ref{et2u2eta}),(\ref{et2u2u}) lead to
\begin{eqnarray}
&&\hspace{-1.5cm}{\eta}(x,t)\sim Q^*-F'(0) \Phi_0 \,s_0 \,t -\left(F'(0)\Phi_0\, s_1+\Big(F'(0)\Phi_1-\frac12 \,F''(0)\Phi_0^2\Big) s_0^2\right)t^2 +\frac12 \,F''(0) x^2\, ,
\label{et2u2etaxt}
\\
&&\hspace{-1.5cm}{u}(x,t) \sim {x \over \displaystyle\sqrt{Q^*}}\left( F'(0) -F''(0)\, \Phi_0 \,  (s_0\,t +s_1\,t^2) \right)\, .
\label{et2u2uxt}
\end{eqnarray}

We remark that, similarly to what happens at leading $O(e^{\tau})$-order for the slope $u_x(0,t)$ for this choice of initial data, the curvature $\eta_{xx}$ for the solution between shocks has a finite initial value at $t=0^+$, in this case set by the $O(e^{2\tau})$-order. That both the $u$-slope and the $\eta$-curvature turn out to be time-independent at their respective orders is a peculiarity of this double-Stoker problem, and their time evolution can be expected to be defined by higher order terms and later time evolution.

The construction of the solution outlined above is schematically illustrated  
in figure~\ref{screwtex}. As we shall see next, most of the qualitative features of this solution's continuation past the shock time
 persist for the full case of initial data~(\ref{inidata-up}).

\subsubsection{Shock unfolding for the full case}
The result~(\ref{expN0}) of section~\ref{shoulders} for the full case shows that the asymptotic behaviour as $t\to t_c^+$, which yields the corresponding asymptotics as $\tau \to -\infty$ of $N(e^\tau,t(\tau))$, is 
\begin{equation}
N(e^\tau,t(\tau))\sim 
{Q \over 4} +K \left(e^\tau+{\displaystyle\sqrt{Q}\over 2 }\,\displaystyle{e^{\tau}\over s_0}\right)^{2\over 3}=
{Q \over 4} +\frac12  \left({3 \,\sqrt{\gamma_0}\, Q\, 
(\displaystyle\sqrt{Q}+2 s_0)\over 16 \, s_0}\right)^{2\over3}e^{{2\over3}\tau}\, ,
\label{Nasy23}
\end{equation}
where we have used the leading order for $t(\tau)$ from system~(\ref{unfoldbispert}), i.e., $t \sim x_s(t)/s_0 $, in~(\ref{expN0}). 
The asymptotics for $\tau \to -\infty$ of equations~(\ref{phipsi0})-(\ref{caphi}) imply that the dependence of the function $F$ on its argument as this limits to zero 
must be through a functional composition with a $2/3$-power. 
Hence, the counterpart of~(\ref{fp1}) in the full case becomes the asymptotic relation
\begin{equation}
{1 \over 2 \displaystyle\sqrt{Q^*}}F'(0)\left( 
\big(e^\tau -\Phi(\tau)\big)^{2\over 3}-\big(e^\tau +\Phi(\tau)\big)^{2\over 3}\right)
\sim e^{{2\over3}\tau}A+ {B \over 2 }F'(0)
\left(
\big(e^\tau -\Phi(\tau)\big)^{2\over 3}+\big(e^\tau +\Phi(\tau)\big)^{2\over 3}\right) \,, 
\end{equation}
where the coefficient $A$  for the full case, $A_f$ say, is now
\begin{equation}
A_f=\lim_{\tau\to -\infty} e^{-{2\over 3}\tau}\psi_0(\tau)\simeq -{0.51691 \, \gamma_0^{1/3} \,Q^{1/6}} \, .
\label{af}
\end{equation}
The leading order asymptotic for $\Phi$~(\ref{phigrdasy}) applies to the full case as well, hence the counterpart of equation~(\ref{fpfin}) 
becomes 
\begin{equation}
F'(0)={2 s_0^{2/3}\displaystyle\sqrt{Q^*} A_f \over 
(-s_0 + \displaystyle\sqrt{Q^*})^{2/3} (1 - \displaystyle\sqrt{Q^*} B) 
-(s_0 + \displaystyle\sqrt{Q^*})^{2/3} (1 + \displaystyle\sqrt{Q^*} B) }
\simeq   0.16752 \,  \gamma_0^{1/3}\,  Q^{2/3} \, .
\label{fp23}
\end{equation}
The leading order asymptotics as $\tau \to -\infty$ determines the functional form of the perturbing fields $\tilde{\eta}(\xi,\tau)$ and $u(\xi, \tau)$,
\begin{eqnarray}
\label{etauzeta23a}
&&\tilde{\eta}(\xi,\tau)\sim{F'(0) \over 2 } 
\left(
  \left(\sqrt{Q^*}/s_0-\xi\right)^{2/3} 
+ \left(\sqrt{Q^*}/s_0+\xi\right)^{2/3} 
\right) e^{{2\over 3}\tau}
\\
\label{etauzeta23}
&&u(\xi,\tau)\sim{F'(0) \over 2 \displaystyle\sqrt{Q^*}} 
\left(
  \left(\sqrt{Q^*}/s_0-\xi\right)^{2/3} 
- \left(\sqrt{Q^*}/s_0+\xi\right)^{2/3} 
\right) e^{{2\over 3}\tau} \, ,   
\end{eqnarray}
or, in the original variables and in the asymptotic limit~$t\to t_c$, the solution between shocks is explicitly
\begin{eqnarray}
\label{etauzeta23xt}
&&\eta(x,t)\sim Q^*+{F'(0) \over 2 } 
\left(
  \left(\sqrt{Q^*}\, (t-t_c)-x\right)^{2/3} 
+ \left(\sqrt{Q^*}\, (t-t_c)+x\right)^{2/3} 
\right) \chi_{(-x_s,x_s)}\,,
\nonumber
\\
&&u(x,t)\sim{F'(0) \over 2 \displaystyle\sqrt{Q^*}} 
\left(
  \left(\sqrt{Q^*}\,(t-t_c)-x \right)^{2/3} 
- \left(\sqrt{Q^*}\,(t-t_c)+x\right)^{2/3} 
\right)\chi_{(-x_s,x_s)}  \, ,  
\nonumber
\\ 
&&
x_s(t)\sim s_0\, (t-t_c)\, .
\end{eqnarray}
Here, we have restored the original (unshifted) time~$t$ and elevation~$\eta$ to further highlight the dynamics that develops shortly after the collapse time $t=t_c$. 
In particular, the velocity's slope  $u_x(0,t)$ and elevation's curvature $\eta_{xx}(0,t)$ relax in time from their  infinite values at $t=t_c$ ($u$ jumps from $u(0^-,t_c)=\displaystyle\sqrt{Q}$ to $u(0^+,t_c)=-\displaystyle\sqrt{Q}$, and $\eta(x,t_c)$ has cusp singularity at $x=0$)  with a monotonic increase from $-\infty$ which initially  scales  according to, respectively,   
\begin{eqnarray}
&&u_x(0,t)=-{2 F'(0)\over 3 \displaystyle{{Q^*}^{2/3}}(t-t_c)^{1/3}}
\simeq-{0.122216\,  \gamma_0^{1/3} \over (t-t_c)^{1/3}} \, , 
\label{asyful13a}
\\
&&\eta_{xx}(0,t)=-{2 F'(0)\over 9 \,\displaystyle{{Q^*}^{2/3}}(t-t_c)^{4/3}}
\simeq-{0.040739  \,\gamma_0^{1/3} \over (t-t_c)^{4/3}} \, . 
\label{asyful13}
\end{eqnarray} 
Of course, these relations are consistent with those computed directly from the perturbed evolution equations~(\ref{unfoldbispert}) through~(\ref{etaxxi}) and~(\ref{uxitau}), as can be easily verified. 

The  higher order correction to the shock position, which at leading order is 
$x_s\sim s_0 \, t$, can now be determined. From the time advancing equation~(\ref{unfoldbispert}), the $O\big(e^{{5 \over 3}\tau}\big)$ terms are
$t(\tau)\sim x_s/s_0+ \hbox{const.}\, \, x_s^{5/3}$
so that
\begin{equation}
x_s\sim s_0 \, t + s_1 \,  t^{5/3}\,, \qquad s_1\simeq 0.18006 \, \gamma_0^{1/3}\displaystyle\sqrt{Q} \, ,
\label{s0acc}
\end{equation}
i.e., the shock accelerates, as a function of the parameters $\gamma_0$ and $Q$,  from the initial Riemann shock speed $s_0$ at a rate $O(t^{-1/3})$ as $t \to 0^+$. 
Note that unlike the double-Stoker case, the slope $u_x$ does not jump from  (negative) infinity  to a finite value at $t=t_c^+$, as the slope increase in the full case diverges with scaling $O(t-t_c)^{-1/3}$. 

\section{Numerics}
\label{numerics}
The theoretical results described in sections~\ref{genpar-sec} and~\ref{sec:beyond_shock} can be illustrated by simulating numerically the evolution governed by the Airy model~(\ref{Airymodel}) for
the class of initial data we have considered.  In turn, 
the closed form solutions together with  the asymptotic approximations we have derived can be used to monitor the accuracy of numerical algorithms after shocks develop.

We first consider the numerical approximation of the Airy's system in the shock-fitted, unfolding  coordinates~(\ref{unfold}). The resulting nonlinear wave equation is expected to always have regular solutions, thus avoiding the need for shock-capturing schemes.
However, the numerical task for devising an algorithm is by no means trivial given the highly nonlinear nature of the coupled equations for the time advancement~(\ref{time}) and for the boundary condition~\eqref{bcunf} at $\xi=1$, which both drive and are determined by the solution of the wave-system~(\ref{unfoldbis}). This structure is reminiscent of the classical water wave problem (see, e.g.,~\cite{Whitham}), in that the solution of a PDE in an interior domain is required to determine its boundary conditions, whose evolution in turn depends on the PDE solution. 

Conversely, over the whole physical domain the original system of PDEs~\eqref{Airymodel} with the initial data considered in Section~\ref{sec:beyond_shock} is expected to develop shocks even for continuous initial data.
Implementing a numerical algorithm for a hyperbolic system of conservation laws up and beyond the shock formation while maintaining  accuracy can also be a challenging  task, especially when high precision is required, and it is worth describing our approach for this problem in some detail.

The first two parts of this section are devoted to the description of two approximation schemes, respectively for the shock-fitted variables $(\xi,\tau)$ and for the original system in (physical) variables $(x,t)$; the third part consists of a panoramic view of the numerical results, including a discussion on the validation of the numerical computations by the theoretical predictions.

\subsection{Shock-fitted scheme}
The Airy system in shock-fitted coordinates~\eqref{unfold} consists of a couple of nonlinear wave equations set on the interval $[0,1]$.
For a proper numerical treatment of the boundary conditions, it is convenient to rewrite the system in terms of the Riemann variables~\eqref{unfriemann}, which leads to the coupled equations
\begin{equation}
\label{eq:fitted_riemann}
\begin{array}{c}
{R_+}_\tau - \xi {R_+}_\xi +(s_0^{-1}+\phi) 
\left(\frac34 R_{+} + \frac14 R_{-}\right) {R_+}_\xi = 0  \,,
\vspace*{0.3cm}
\\
{R_-}_\tau - \xi {R_-}_\xi +(s_0^{-1}+\phi) 
\left(\frac34 R_{-} + \frac14 R_{+}\right) {R_-}_\xi = 0  \,.
\end{array}
\end{equation}
The original variables can be expressed in terms of the Riemann invariants, 
\begin{equation}
  \eta = \frac{\left(R_+ - R_-\right)^2}{16} \, ,  \qquad 
  u = \frac{R_+ + R_-}{2}\, ,
\label{eq:orig_riemann}
\end{equation}
and with this change of variables the boundary conditions~\eqref{bcunf} can be rewritten as
\begin{gather}
  R_+(0,\tau) + R_-(0,\tau) = 0
\label{eq:riemann_bc_0} \\
  R_+(1,\tau) + R_-(1,\tau) = 2 V(e^\tau,t(\tau)) + \sqrt{\frac{(16N(e^\tau,t(\tau)) - R_d(1,\tau)^2)^2(16N(e^\tau,t(\tau)) + R_d(1,\tau)^2)}{128 N(e^\tau,t(\tau))R_d(1,\tau)^2}},
\label{eq:riemann_bc_1}
\end{gather}
where for simplicity we introduced the notation (and new variable) $R_d=R_+ - R_-$.
To ensure the well-posedness of problem~\eqref{eq:fitted_riemann}, the boundary conditions must be imposed in the form of inflow conditions for the appropriate Riemann invariant.
Since the variable $R_+$ has a positive velocity, the condition~\eqref{eq:riemann_bc_0} at $\xi=0$ must be imposed on $R_+$; conversely, since $R_-$ has a negative propagation velocity, the condition~\eqref{eq:riemann_bc_1} for $\xi=1$ must be imposed on $R_-$.

For the numerical approximation of system~\eqref{eq:fitted_riemann}, we choose the Chebyshev Collocation Method (CCM) (briefly sketched below, see e.g.,~\cite{CHQZ2, CHQZ3} for a comprehensive exposition).
The CCM is a special instance of a spectral method, meaning that the solution to the original PDE is expanded in a suitable set of basis functions (with an approximate solution then obtained by truncating the expansion). The truncated equations are the approximate analogue of the original PDE, so that the approximation error decreases exponentially fast with the number $M$ of basis functions retained, at least as long as the solution is sufficiently regular.

To set up the CCM, we introduce a set of \emph{collocation points} $\{\xi_i\}_{i=1}^M\subset [0,1]$ that will be used both for the definition of the basis functions and in the determination of the expansion coefficients,
\begin{equation}
  \xi_i = \frac12 + \frac12\cos\left(\pi + \frac{i-1}{M-1}\pi\right) \qquad \text{for }\, i=1,\dots,M \,.
\label{eq:def_GLC}
\end{equation}
The points $\{\xi_i\}$ defined above are often termed the \emph{Gauss--Lobatto--Chebyshev (GLC)} points, to emphasize that the endpoints of the interval $[0,1]$ are included in the set of collocation points.
It is now possible to introduce the set of Lagrange interpolant polynomials on the GLC points, as
\begin{equation}
  T_j(\xi) = \prod_{\substack{i=1 \\ i\ne j}}^M \frac{\xi - \xi_i}{\xi_j - \xi_i}\, .
\label{eq:def_Tj}
\end{equation}

The CCM is based on writing the approximate solutions $R_+^M$, $R_-^M$ as a linear combination of the polynomials $T_j$,
\begin{align}
  R_+^M(\xi, \tau) = \sum_{j=1}^M r_j(\tau) T_j(\xi) \,, \qquad &&
  R_-^M(\xi, \tau) = \sum_{j=1}^M s_j(\tau) T_j(\xi)\,,
\end{align}
where the unknown coefficients $r_j, s_j$ are determined by imposing that the original PDE~\eqref{eq:fitted_riemann} holds pointwise at the set of collocation points $\{\xi_i\}_{i=1}^M$
\begin{align}
\sum_{j=1}^M\left[\left(r_j'(\tau) - \xi_i r_j(\tau)\right)T_j(\xi_i) + e^{-\tau}t'(\tau)\left(\frac34 \sum_{k=1}^M r_k(\tau)T_k(\xi_i) + \frac14 \sum_{l=1}^M s_l(\tau) T_l(\xi_i)\right) r_j(\tau)T_j'(\xi_i)\right] = 0 \,,
\label{eq:cheb_r}
\\
\sum_{j=1}^M\left[\left(s_j'(\tau) - \xi_i s_j(\tau)\right)T_j(\xi_i) + e^{-\tau}t'(\tau)\left(\frac14 \sum_{k=1}^M r_k(\tau)T_k(\xi_i) + \frac34 \sum_{l=1}^M s_l(\tau) T_l(\xi_i)\right) s_j(\tau)T_j'(\xi_i)\right] = 0\,,
\label{eq:cheb_s}
\end{align}
for all $i=1,\dots,M$. Here primes denote derivatives with respect to either of the independent variables $\xi$ or $\tau$, as appropriate for each functional argument. 
Note that by virtue of~\eqref{eq:def_Tj}, $T_j(\xi_i) = \delta_{ji}$, and the first derivatives $T_j'(\xi_i)$ are known in closed form, see e.g.~\cite[p.~89]{CHQZ2}.
Furthermore, due to the interpolant nature of the polynomials $T_j(\xi)$, the expansion coefficients $r_j, s_j$ can be interpreted as the pointwise 
values $R_+(\xi_j)$ and $R_-(\xi_j)$, respectively. Note also that the nonlinearity of these equations is not confined to the quadratic terms, but enters through the coupling to the boundary conditions in the $\tau$ dependent factor $t'(\tau)\, e^{-\tau}$, which depends on the Riemann invariants evaluated at $\xi=1$. 

Equations~\eqref{eq:cheb_r} and~\eqref{eq:cheb_s}
are a set of first order, nonlinear, coupled ordinary differential equations that govern the time evolution of the coefficients $\{r_j\}$ and $\{s_j\}$.
In this work, we approach equations~\eqref{eq:cheb_r} and~\eqref{eq:cheb_s} by means of  explicit, either Euler or fourth-order Runge--Kutta (RK4),  methods.
Thus, for instance, the Euler scheme time-advancing leads to the difference equations
\begin{align}
r_j^{n+1} &= r_j^n - h\left[-\delta_{ij}\xi_jr_j^n + (s_0^{-1}+\phi^n)\left(\frac34 r_i^n + \frac14 s_i^n\right)r_j^n \, T_j'(\xi_i)\right]\, , \qquad &\forall \, i=1,\dots,M
\label{eq:euler_cheb_r} \\
s_j^{n+1} &= s_j^n - h\left[-\delta_{ij}\xi_js_j^n + (s_0^{-1}+\phi^n)\left(\frac14 r_i^n + \frac34 s_i^n\right)s_j^n \, T_j'(\xi_i)\right]\, , \qquad &\forall \, i=1,\dots,M
\label{eq:euler_cheb_s} \\
t_{n+1} &= t_n + h\, e^{\tau_n}(s_0^{-1} + \phi^n)\, ,
\label{eq:euler_cheb_t}
\end{align}
where we have adopted the shorthands $r_j^n = r_j(\tau_n)$ (and similarly for $s_j$),  
$t_n=t(\tau_n)$ and $\phi^n=\phi(e^{\tau_n}, t_n, \eta^n_M, u^n_M)$ for the values of the numerical approximations on the equispaced, constant  $h$-duration subintervals $[0,\tau_1,\dots,\tau_n,\tau_{n+1},\dots]$.

Since both the forward Euler and the RK4 methods are explicit, a Courant-Friedrichs-Levy (CFL) condition holds on the time step size $h$.
It is known from theoretical considerations that the velocity will achieve its maximum (in modulus) at $\xi=1$, and the collocation grid size achieves its minimum precisely at the same point, hence it is sufficient to check that the CFL condition is satisfied at the last grid point, $\xi_M=1$.

After advancing in time with the RK4 analogue of equations~\eqref{eq:euler_cheb_r} and~\eqref{eq:euler_cheb_s}, the boundary condition~\eqref{eq:riemann_bc_0} at $\xi=0$ can be imposed as follows:
\begin{equation}
r_1^{n+1} = -s_1^n,
\label{eq:bc_discrete}
\end{equation}
thus making sure that the relative inflow boundary condition is enforced.
For the boundary condition at $\xi=1$, we found by trial and error that the accuracy in the approximation of the nonlinear relation~\eqref{eq:riemann_bc_1} has a strong influence on the stability of the approximation method.
To address this, we set up a Newton iterative scheme to make sure that, after propagating $r_M^{n+1}$ according to the RK4 analogue of equation~\eqref{eq:euler_cheb_r}, $s_M^{n+1}$ is chosen so that the nonlinear relation~\eqref{eq:riemann_bc_1} holds up to machine precision accuracy.

Lastly, we mention that the unavoidable aliasing errors are controlled by guaranteeing a sufficiently high resolution in space, see e.g.~\cite[\S3.10]{CHQZ2}. This approach towards de-aliasing, usually deemed impractical in two- or three-dimensional computations, is not unreachable in one-dimensional settings.

\subsection{Shock-capturing scheme}
The shock-capturing approximation scheme adopted for this work is a high-order, finite difference method 
commonly referred to as Weighted Essentially Non-Oscillatory (WENO).
For an overview of the WENO method we refer to~\cite{ShuReview}, and to~\cite{Shu97} for a discussion on the details of a practical implementation.

Suppose that the Airy system~\eqref{Airymodel} is set on the interval $I_a=[-c,c]$, and consider the set of $M$ ordered, equispaced, \emph{grid points} $\{x_i\}_{i=0}^{M-1}\subset I_c$, chosen so that $x_0=-c$, $x_{M-1}=c$.
For convenience, the distance between two consecutive grid points is assumed to be the same for all the grid points, and say $\delta=x_{i+1} - x_i$.
Here we are concerned with finding an approximation for the pointwise values of $\eta$ and $u$ at the grid points.

For this purpose, it is convenient to rewrite equation~\eqref{Airymodel} in conservative form,
\begin{equation}
\eta_t + m_x = 0 \,,   \qquad  m_t + \left(\frac12 {m^2 \over \eta} + \eta\right)_x = 0\, ,
  \label{eq:Airy_conservative}
\end{equation}
where, respectively, the conserved quantities $m=\eta\, u$ and $\eta$ can be interpreted  physically as mass and momentum densities, and 
\begin{equation}
 G(m,\eta) = m\,,  \qquad  L(m, \eta) = \frac 12 \frac{m^2}{\eta} + \eta\, ,
  \label{eq:FG_flux}
\end{equation}
are the corresponding fluxes.
In the following, we denote  with $x_{i+1/2}$ the arithmetical average between the consecutive grid points $x_i$ and $x_{i+1}$, and with $x_{x-1/2}$ the arithmetical average between the consecutive grid points $x_{i-1}$ and $x_i$.
Similarly, we denote with $m_i$, $\eta_i$ the pointwise values of $m$ and $\eta$ at the grid point $x_i$, and with $G_{i\pm1/2}$, $L_{i\pm1/2}$ the pointwise values of the fluxes $G$ and $L$ at $x\pm1/2$.

The finite difference WENO scheme  provides a numerical approximation to the solutions of system~\eqref{eq:Airy_conservative} through the ODE system
\begin{equation}
 \frac{\d \eta_i}{\d t}  = - \frac{1}{\delta}\left(\widehat{G}_{i+1/2} - \widehat{G}_{i-1/2}\right)\,,
\qquad
    \frac{\d m_i }{\d t}    = - \frac{1}{\delta}\left(\widehat{L}_{i+1/2} - \widehat{L}_{i-1/2}\right)\, ,
  \label{eq:weno_fd}
\end{equation}
where the space derivative of the fluxes is expressed formally as the difference quotient of some approximations $\widehat{G}$, $\widehat{L}$ of $G$ and $L$, evaluated at the two half-coordinate grid points centered in $x_i$.
The time derivatives in equation~\eqref{eq:weno_fd} will be discretised with a finite difference method based on the knowledge of the values of $m$ and $\eta$ at the grid point $x_i$ at the previous time steps.
The time advancement  
is done by means of a third-order, strong stability preserving, Runge--Kutta method, for the details of which we refer to~\cite{GottliebSSP}.

In this work, we consider a WENO scheme with fifth-order accuracy in space, meaning that the flux approximation $\widehat{G}$ fulfills
\begin{equation}
  \frac{1}{\delta}\left(\widehat{G}_{i+1/2} - \widehat{G}_{i-1/2}\right) = G(u_{i+1/2},\eta_{i+1/2}) + O(\delta^5),
  \label{eq:F_fifth}
\end{equation}
and a similar property holds for $\widehat{L}$.
WENO schemes construct the approximations $\widehat{G}$, $\widehat{L}$ from the pointwise values of $G$ and $L$ on the grid $\{x_i\}_{i=0}^{M-1}$.
For the Airy model, the expression for the fluxes is particularly simple, since it only involves quadratic nonlinearities, and it does not depend on the derivatives of $u$ and $\eta$.

\begin{figure}[t]
\centering
\includegraphics[width=0.4\textwidth]{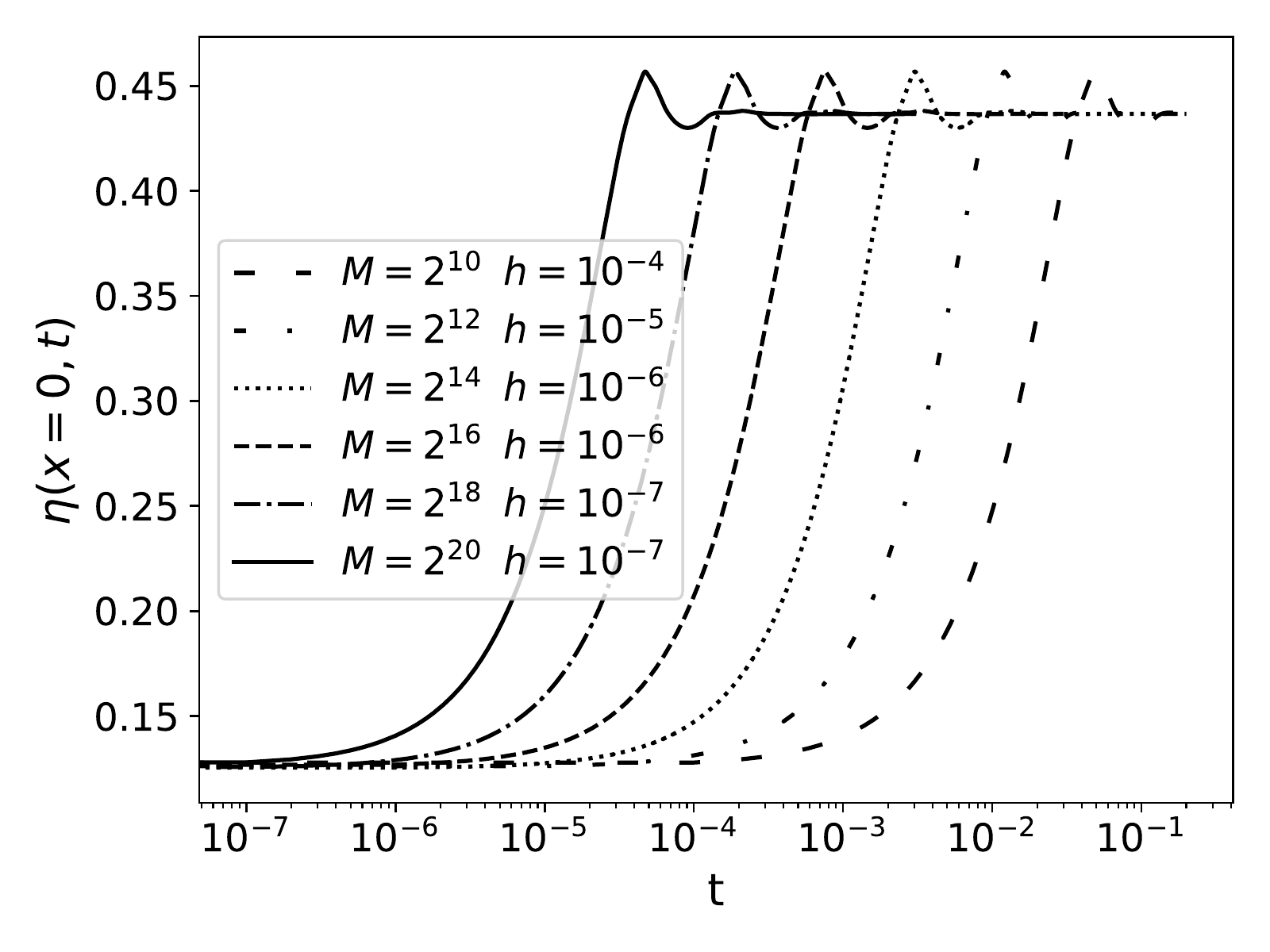}
\caption{Evolution of $\eta(0,t)$ from double-Riemann initial data,  for several refinement levels of spatial grid $M$ and time step $h$. All results are obtained with the WENO shock-capturing algorithm.}
\label{fig:riemann_refinement}
\end{figure}

To avoid spurious oscillations for the approximate fluxes, $\widehat{G}_{i\pm1/2}$  and $\widehat{L}_{i\pm1/2}$ are computed by means of Lagrange interpolation formulas based on three sets of points, namely:
\begin{equation}
  S_{i,1} = \{x_{i-1},x_{i},x_{i+1}\}\, , \qquad S_{i,2} = \{x_i,x_{i+1},x_{i+2}\}\, , \qquad S_{i,3} = \{x_{i+1},x_{i+2},x_{i+3}\}.
  \label{eq:stencils_plus}
\end{equation}
Then, by means of a local regularity estimator for $u$ and $\eta$, the approximate fluxes can be reconstructed using all of the three stencils in 
\eqref{eq:stencils_plus} if the solution does not have a discontinuity in the interval $[x_{i-1},x_{i+3}]$, thus giving a fifth-order approximation; otherwise, the approximate fluxes are reconstructed by neglecting the stencils in which the solution presents a shock, leading to an approximation which is locally of third order.
For stability issues, it is convenient to split the fluxes as
\begin{equation}
  G(m,\eta) = G^+(m,\eta) + G^-(m,\eta)\, , \qquad
  L(m,\eta) = L^+(m,\eta) + L^-(m,\eta),
  \label{eq:F_G_split}
\end{equation}
with the scheme adopted here being the Lax--Friedrichs splitting,
\begin{equation}
  G^{\pm}(m,\eta) = \frac12\left(G(m,\eta) \pm \alpha m\right), \qquad
  L^{\pm}(m,\eta) = \frac12\left(L(m,\eta) \pm \alpha \eta\right),
  \label{eq:F_G_Fried}
\end{equation}
where $\alpha$ is the maximum modulus among the eigenvalues of the Jacobian of the system of conservation laws.
For the Airy system, this value can be computed explicitly as
\begin{equation}
  \alpha = \max\{|u + \sqrt \eta|, |u - \sqrt \eta|\}.
  \label{eq:alpha_Fried}
\end{equation}
\begin{figure}[t]
\centering
\includegraphics[width=0.325\textwidth]{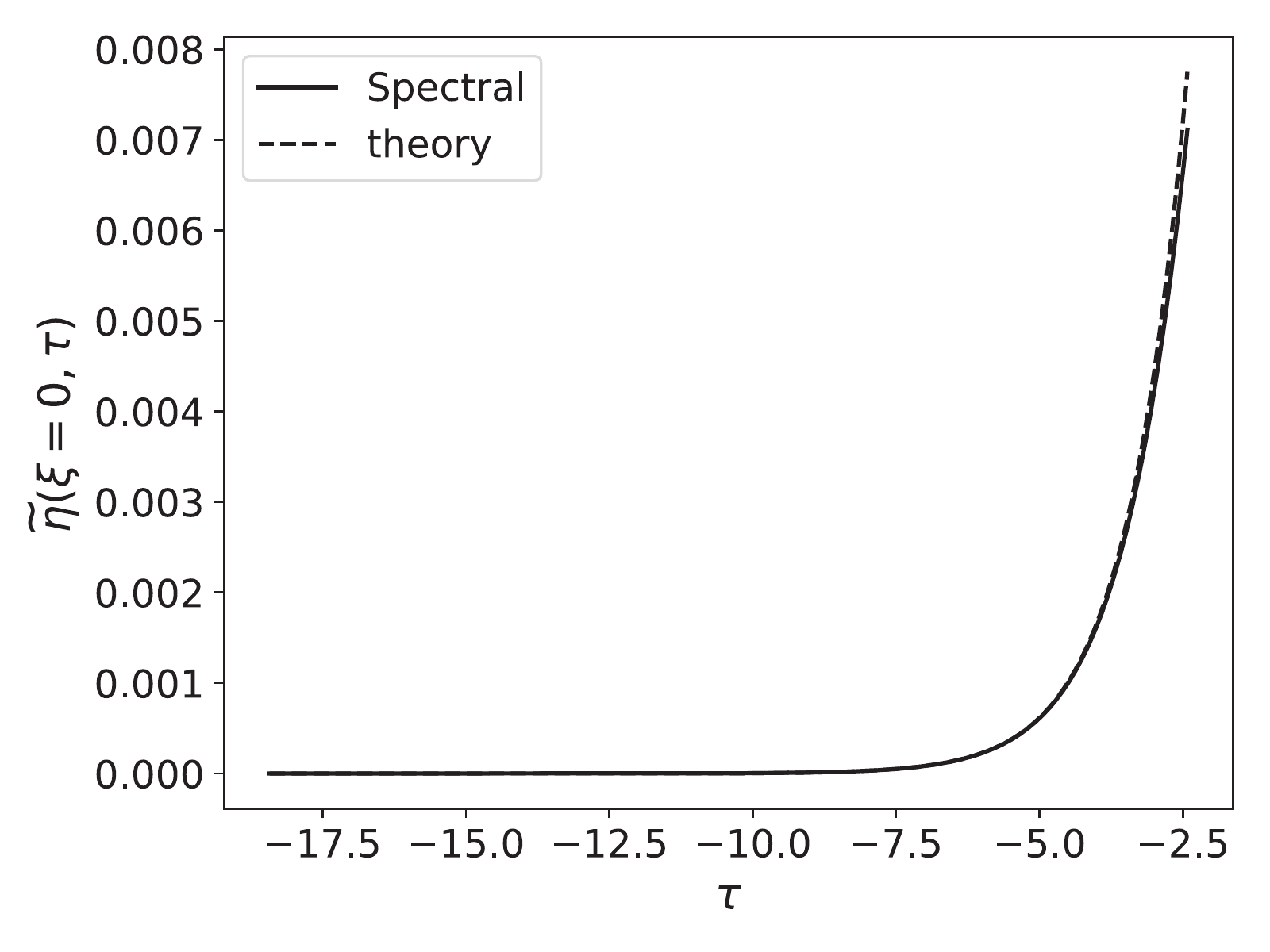}
\includegraphics[width=0.327\textwidth]{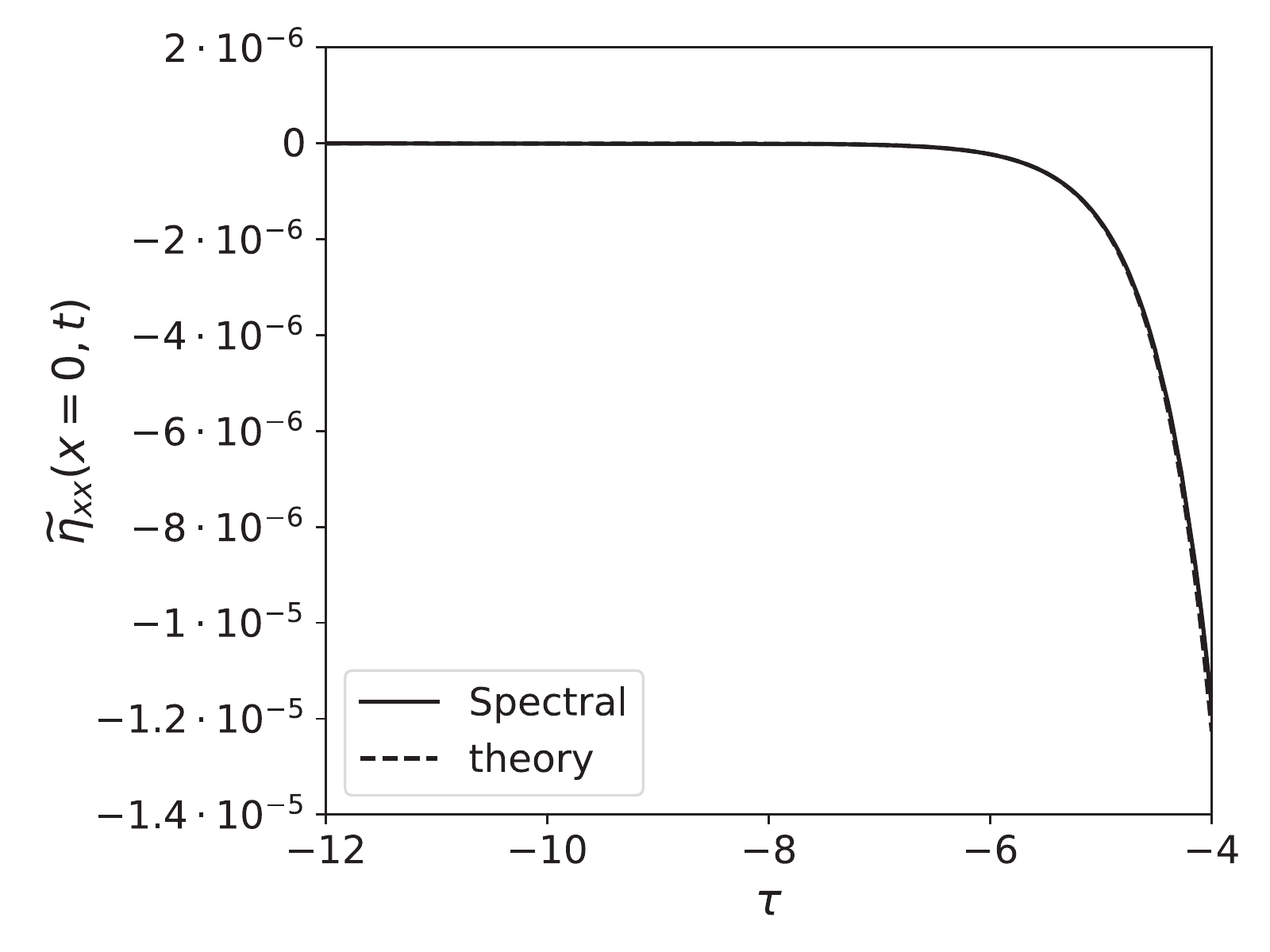} 
\includegraphics[width=0.325\textwidth]{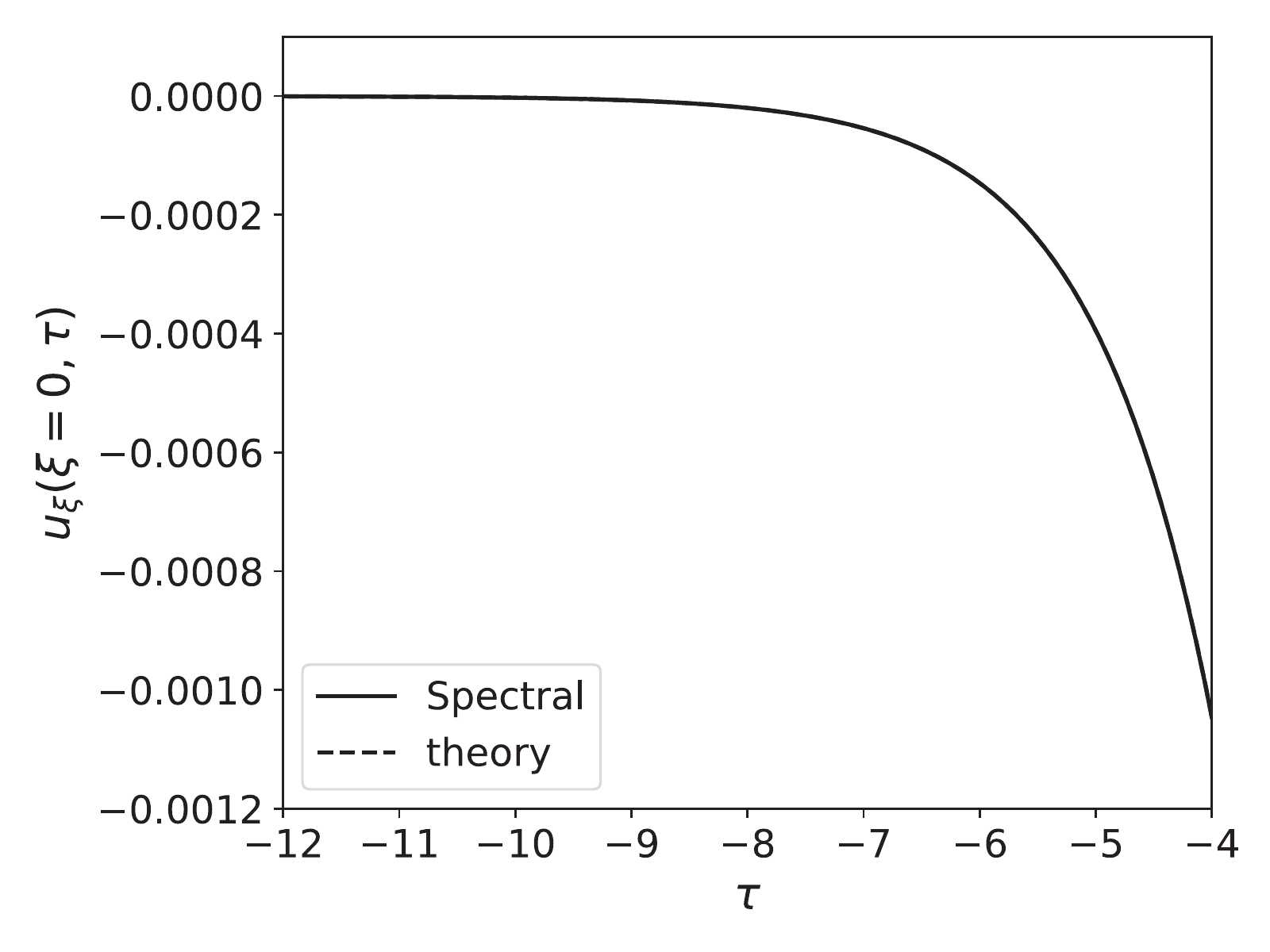}\\
\includegraphics[width=0.325\textwidth]{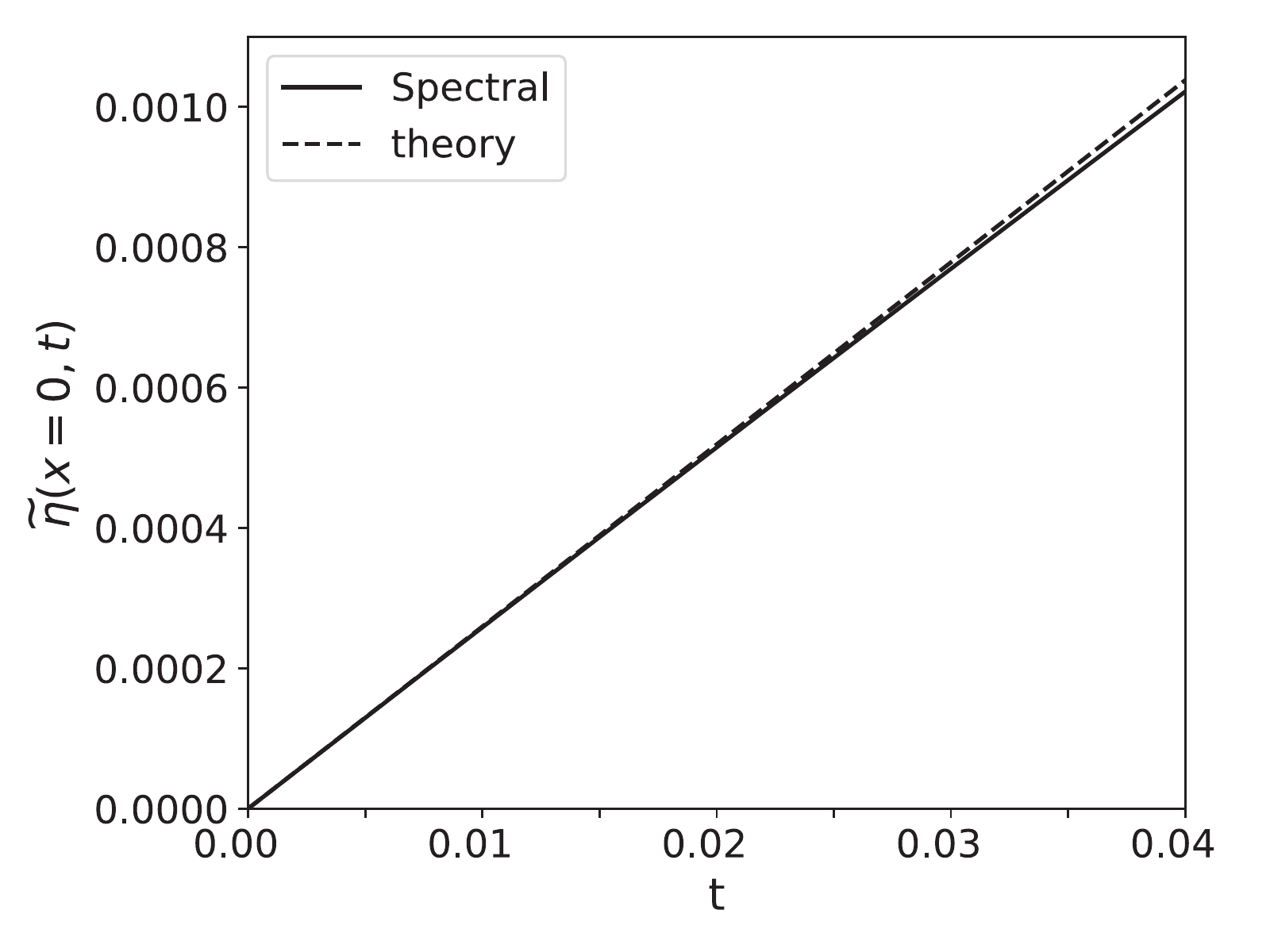}
\includegraphics[width=0.325\textwidth]{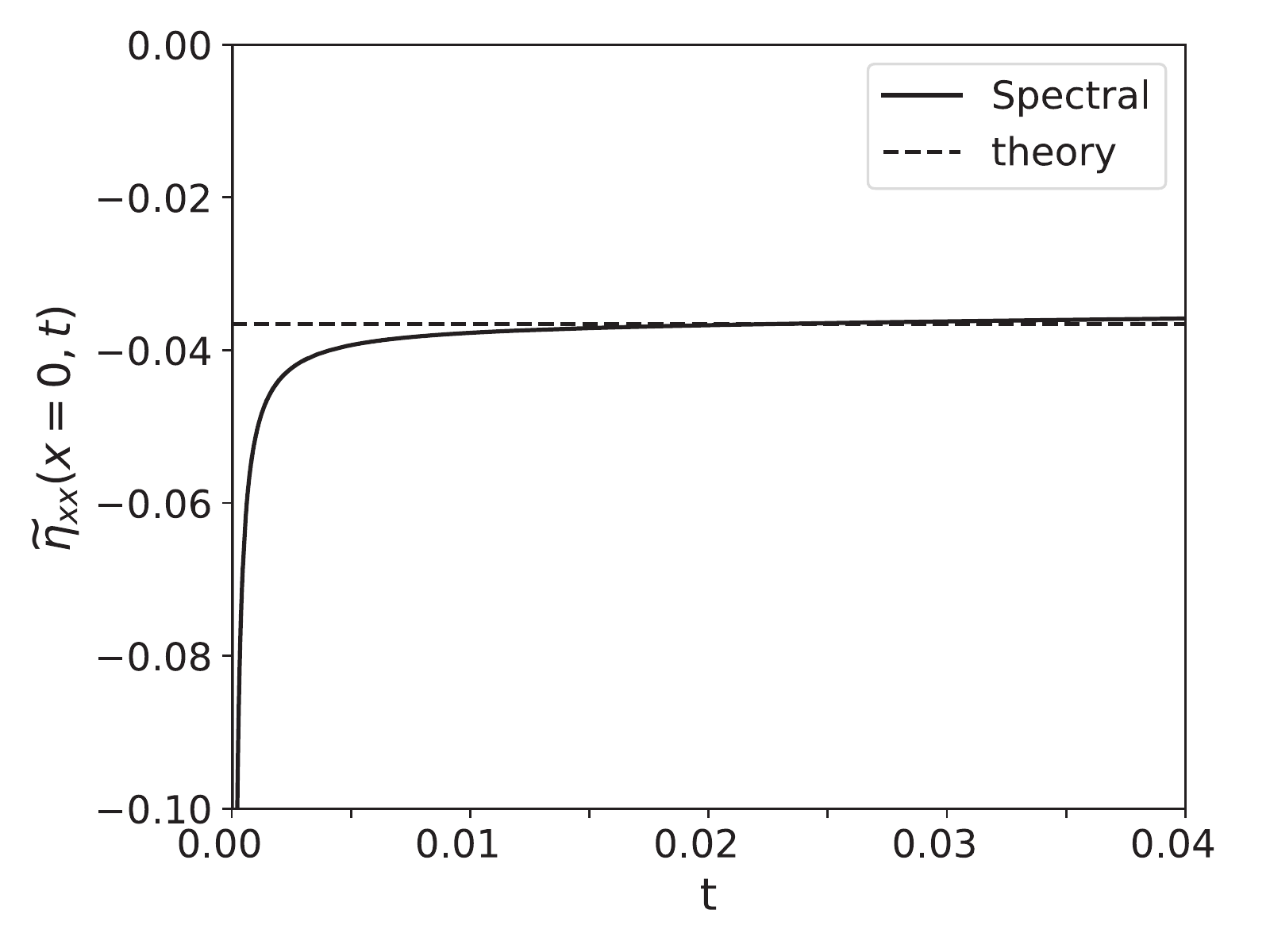} 
\includegraphics[width=0.325\textwidth]{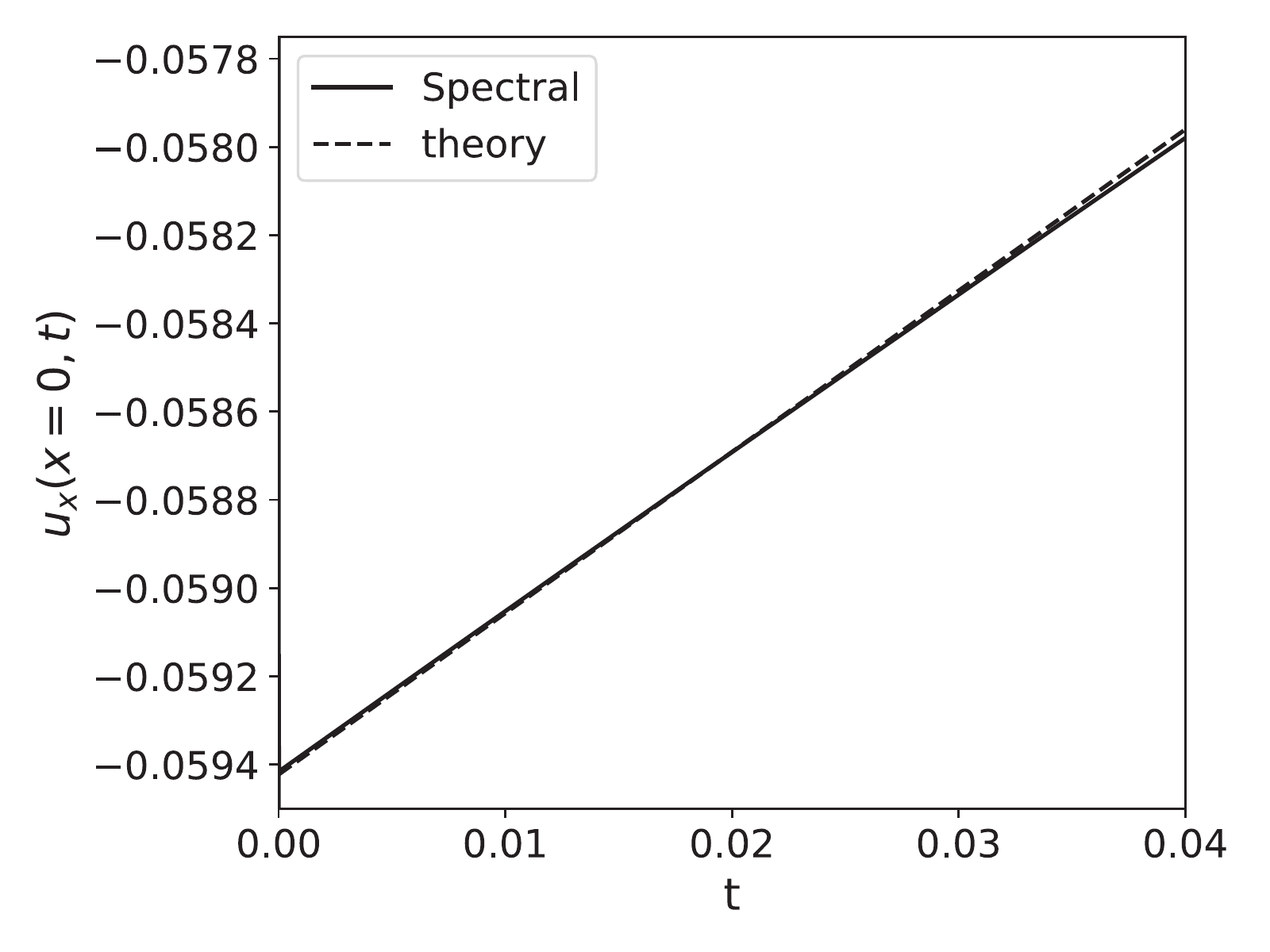}
\caption{Top row: comparison of the theoretical results (dashed line) in equations~\eqref{et2u2eta} (top row) and~\eqref{et2u2u} for the double-Stoker case with the numerical computations (continuous line) obtained with the Chebyshev Collocation Method on shock-fitted coordinates, in the original variables $(\xi,\tau)$.
Bottom row: same as above, but after the change of variables to return to physical coordinates $(x,t)$}
\label{fig:stoker_tau_xi}
\end{figure}

\subsection{Numerical results}
Among various alternatives for illustration and validation of the numerical schemes' performance, it is particularly interesting to focus on the full case and monitor quantities along the $x=0$ symmetry line,  such as the short time evolution of $\eta$, $u_x$, $\eta_{xx}$, and consider pointwise differences between solutions by algorithms and their closed form analytical counterparts when available.
We focus mainly on the ``full'' case both because of its interest in the theoretical question of the wetting of a dry spot, and because of its challenging non-analytic behaviour for the continuation past the collapse point.
Nonetheless, the analogous results about the ``double-Riemann" and the ``double-Stoker'' problems  will be discussed briefly to illustrate how the numerics performs with respect to the increasing level of complexity at the collapse point.
Unless otherwise specified, all spectral computations are performed with~$M=64$ modes and  timestep~$h=10^{-4}$, while for the shock-capturing scheme the grid will vary, depending on a target accuracy, from $M=2^{14}$ to $M=2^{18}$ with timesteps from $10^{-8}$ to $10^{-6}$.

\begin{figure}[t]
\centering
\includegraphics[width=0.4\textwidth]{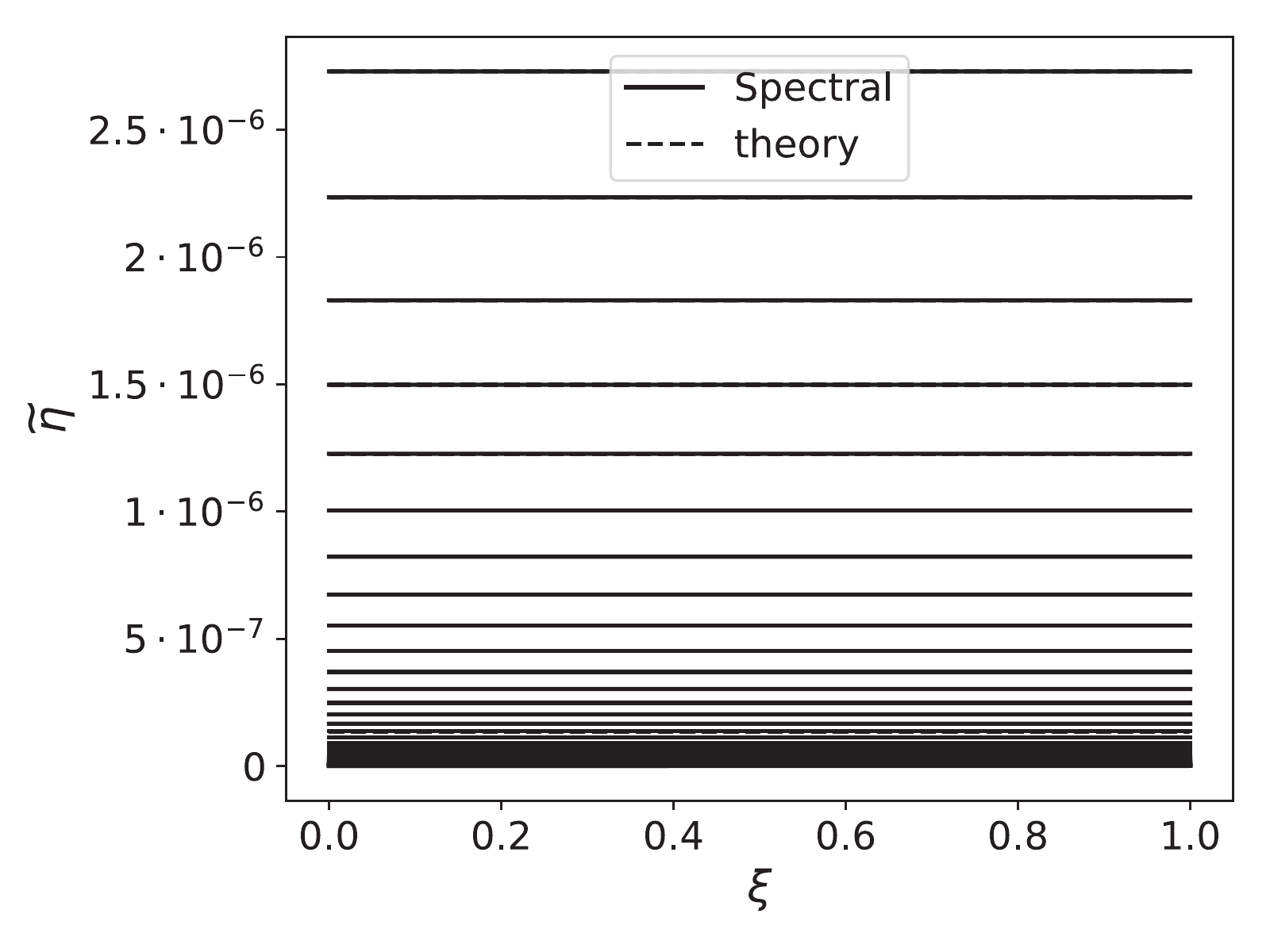}
\includegraphics[width=0.4\textwidth]{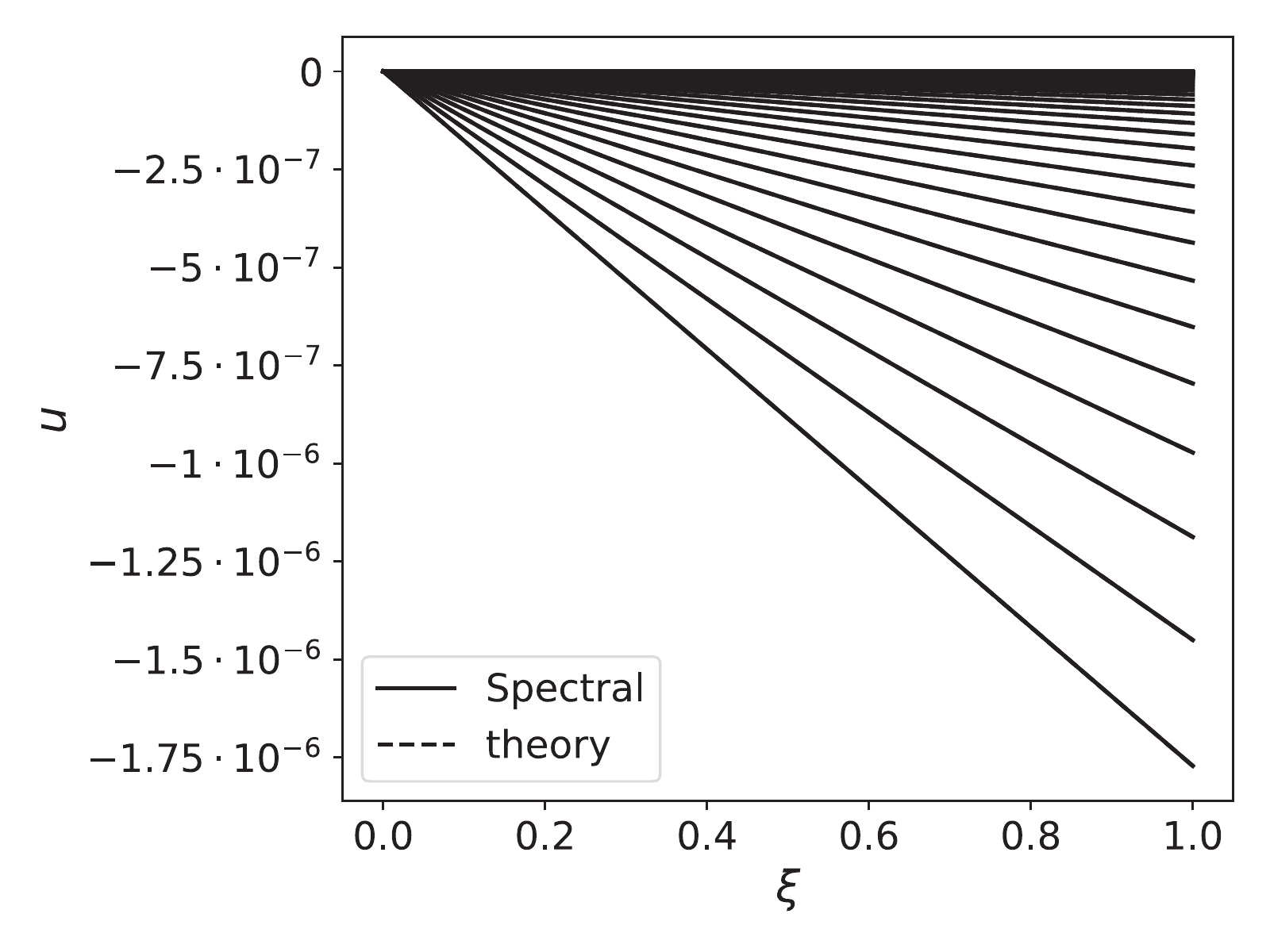}
\caption{Snapshots of  evolution in the half domain $\xi\in[0,1]$ for $\eta$ (left) and $u$ (right) for the double-Stoker problem in the shock-fitted variables $(\xi,\tau)$. The curves in the plot are equispaced in $\tau$ with a time step of $2\cdot10^{-3}$ units. The elevation curvature from~(\ref{et2u2eta}) is barely noticeable in the left panel at these scales, but the overlap of the theoretical and numerical curves documents the validity of the numerical scheme.}
\label{fig:stoker_profiles}
\end{figure}

\subsubsection{Double-Riemann simulations}
As already remarked, the dynamics for this case in shock fitting coordinates is trivial and does not pose a challenge to the spectral code implementation. On the other hand, the WENO algorithm is designed to deal with problem like these, and so this is an excellent test for this scheme, allowing to check pointwise accuracy and sharpness in resolving discontinuities in both space and time.
In figure~\ref{fig:riemann_refinement}, we show the results of a Riemann problem with constant initial density $\eta_0=Q/2$ and discontinuous initial momentum $m=-\sgn(x)\displaystyle\sqrt{Q}(Q/2)$.
For the numerical test we take $Q=1/2$, and run the code on the interval $[-1,1]$ for several space and time resolutions. The inflow conditions are the same as the initial conditions, namely $\eta=Q/2$ and $m=\mp \displaystyle\sqrt{Q}(Q/2)$ are imposed respectively at $x=\pm 1$.
The results show that the code is indeed high order at a sufficient distance from the discontinuity, as the height of $\eta$ after the collision matches the theoretical prediction to 5 digits even for a modest resolution of $2^{10}$ grid points and $h=10^{-4}$.
However, by focusing on the plot for $\eta(x=0,t)$, the numerical solution after shock formation shows a significant overshoot followed by a region of damped oscillations. While the temporal support of these oscillations decreases with increasing resolution, their amplitude does not. We can safely interpret these oscillations as numerical artifacts, as their frequency is significantly smaller than the Nyquist frequency for all the timesteps considered. 
These oscillations also make estimating the shock position and velocity  inherently imprecise,  as the discontinuity's support jumps erratically from one timestep to the next over a set of a few spatial grid points.

\subsubsection{Double-Stoker simulations}

The purpose of this section is to check the consistency of the theoretical results of section~\ref{doubleSt} with the numerical computations.
More precisely, we consider the initial datum~\eqref{2stok} with $Q=1/2$ and $g_0=1/16$, and aim at checking the asymptotics of equations~\eqref{et2u2eta} and~\eqref{et2u2u}.

To avoid the proliferation of figures, we consider only the shock-fitted spectral method, referring to Section~\ref{sec:numerics_full} for a complete comparison between the shock-capturing and the shock-fitted schemes for the full case.
The results that we report here are the values of $\tilde\eta$, $\tilde{\eta}_{\xi\xi}$ and ${u}_{\xi}$ evaluated at $\xi=0$ as a function of $\tau$ (figure~\ref{fig:stoker_tau_xi}, first row); the analogous values of $\tilde\eta$, $\tilde{\eta}_{xx}$ and ${u}_{x}$, evaluated at $x=0$ as a function of $t$ (figure~\ref{fig:stoker_tau_xi}, bottom row); the profiles of $\tilde\eta$ and $u$ as a function of $\xi$, for a few representative values of $\tau$.

It can be seen that in all the cases the agreement with the theoretical predictions is quite reasonable.
The agreement between asymptotic results and numerical computations is to some extent less clear for the curvature of $\eta$ seen in the original variables $x,t$.
We attribute this effect to the fact that for computing $\tilde{\eta}_{xx}$ starting from $\tilde{\eta}_{\xi\xi}$, it is necessary to compute a ratio between $\tilde{\eta}_{\xi\xi}$ and $e^{2\tau}$.
Both these quantities are very small for small times, making the resulting ratio an ill-conditioned problem.
Such cancellation errors fade away as $\tau$ and $\tilde{\eta}_{\xi\xi}$ grow, and this is confirmed by the good fit for times $t$ between $0.01$ and $0.04$.

Finally, we check the spectral code with the analytic expression in equations~\eqref{et2u2eta} and~\eqref{et2u2u} for the profiles of $\eta$ and $u$ in shock-fitted coordinates.
The results are summarized by figure~\ref{fig:stoker_profiles}, which shows how the theoretical curves are indistinguishable from the numerical computed solutions within plotting resolution.

\begin{figure}[t]
\centering
\includegraphics[width=0.4\textwidth]{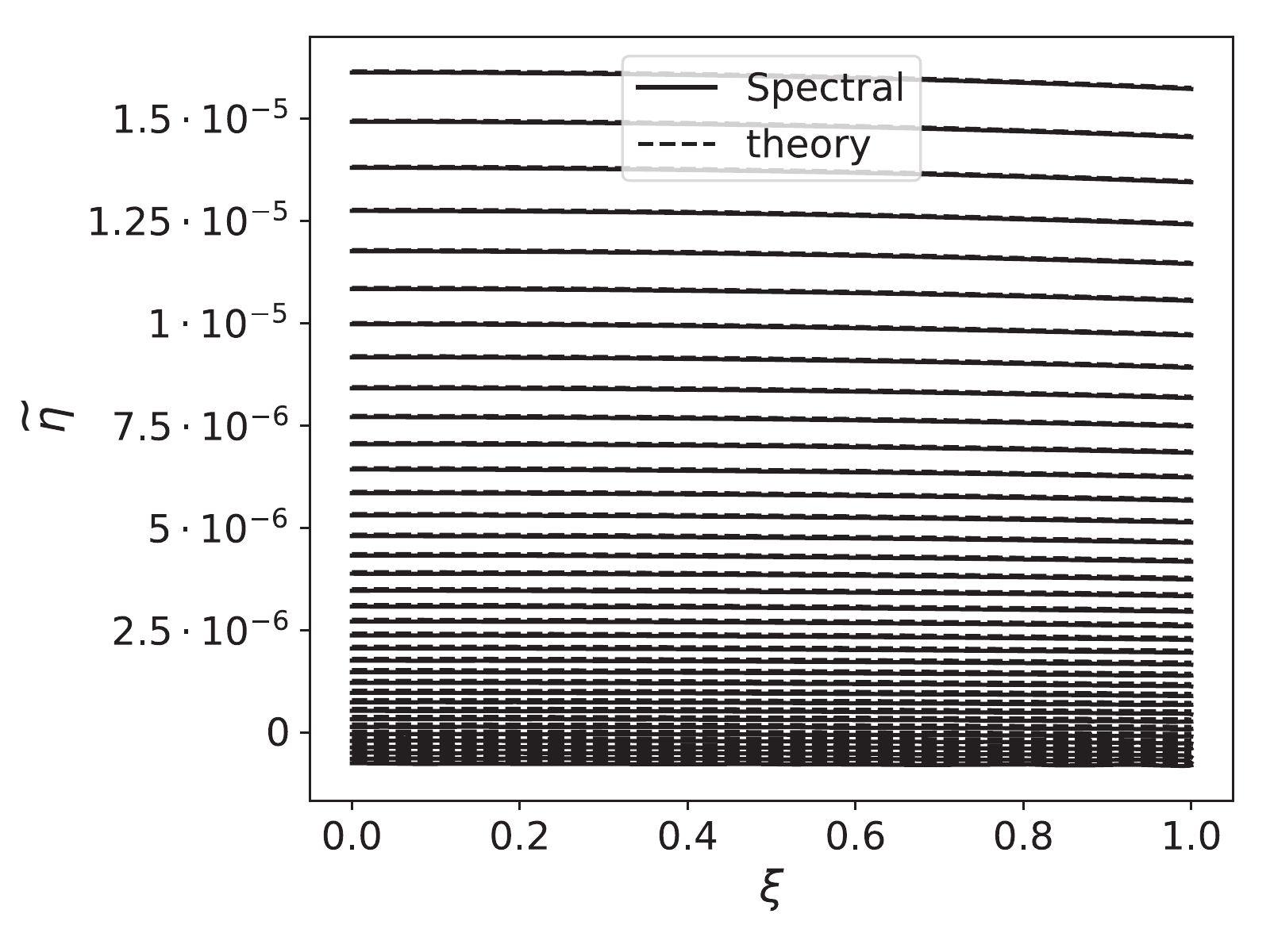}
\includegraphics[width=0.4\textwidth]{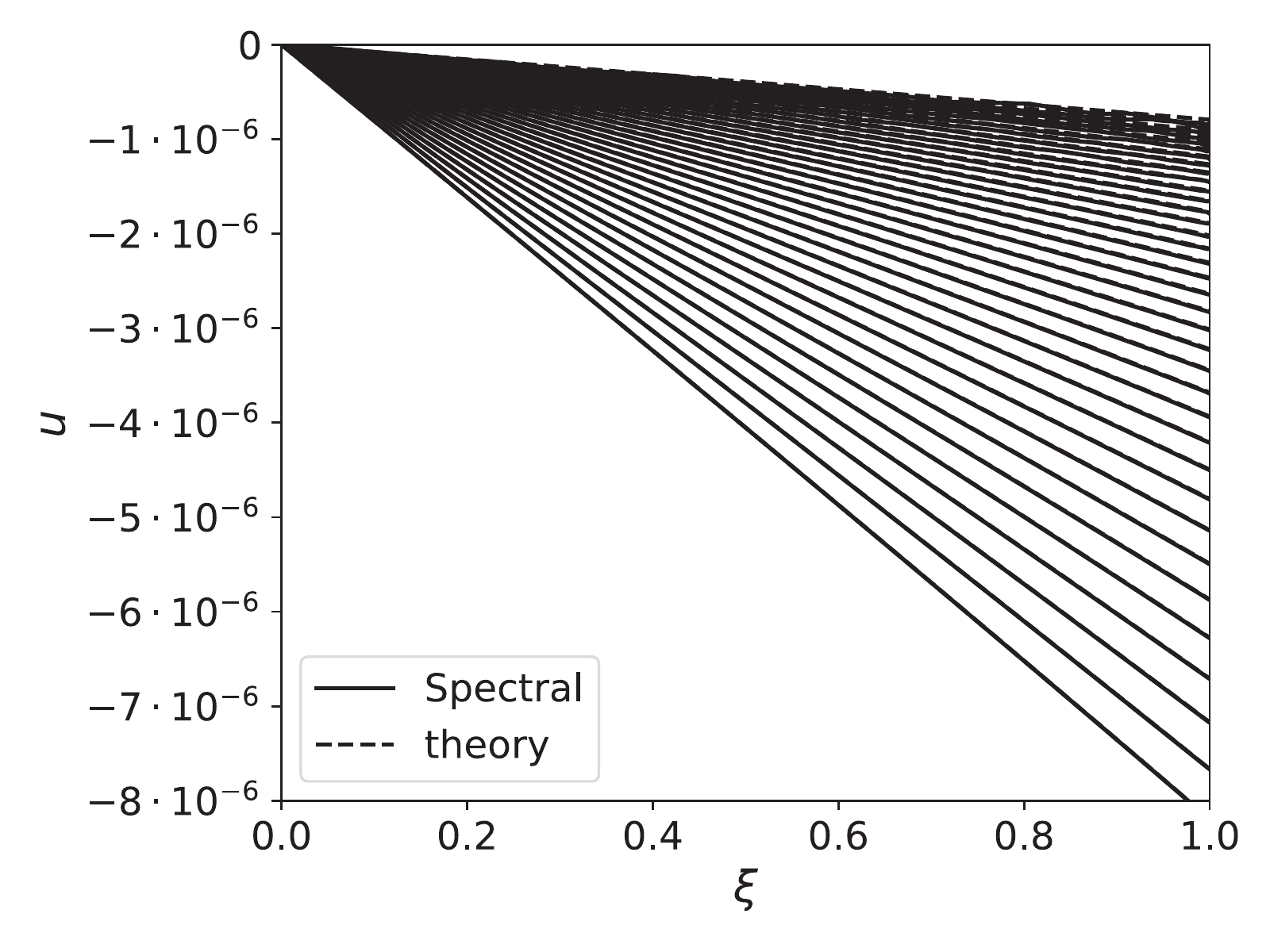}
\caption{Same as figure~\ref{fig:stoker_profiles} but for the full case~\S\ref{fullsec}. Here the curves in the plot are equispaced $0.1$ units apart in $\tau$, from $\tau=-16$ to $\tau=-12.5$. The spectral results are compared with the theoretical predictions of equations~\eqref{etauzeta23a} and~\eqref{etauzeta23}.
}
\label{fullxitauspectr}
\end{figure}

\subsubsection{Full case simulations}
\label{sec:numerics_full}

For the shock-fitted spectral method, the initial datum corresponds to an asymptotic condition for $\tau\to-\infty$.
We approximate the asymptotic condition by taking $x_s=10^{-8}$, and a corresponding initial value of $\tau=\log x_s\simeq -18.4207$.
Initially, the Riemann variables are set to their asymptotic values of $R_{\pm} = \pm\displaystyle\sqrt{Q^*}$.
There is clearly an inconsistency arising from the use of the asymptotic values for the Riemann variables while the initial value for $\tau$ is a finite number.
A result of this inconsistency is the generation of spurious, small amplitude, waves in the initial instants of the numerical computations. These ``acoustic" waves propagate in the computational domain and are reflected at the boundaries.
Due to the absence of numerical diffusivity in Chebyshev spectral methods, the acoustic waves are dissipated only by interaction with the shock wave at $\xi=1$.
A side-effect of these spurious waves is that the numerical results can be trusted only after a few units of $\tau$. However we verified that the dissipation of the acoustic waves is sufficiently fast that the small time asymptotics can still be easily detected.
Being interested in the short time asymptotics, the boundary conditions at $\xi=1$ are further simplified by adopting the expansion~\eqref{expN0} for $N(e^\tau,t(\tau))$.

We first test the spectral code against the asymptotic expressions in equations~\eqref{etauzeta23a} and~\eqref{etauzeta23} for the profile of $\eta$ and $u$ with respect to the shock-fitted coordinates $(\xi,\tau)$.
Figure~\ref{fullxitauspectr} shows time snapshots of the profiles, and the good agreement between theory and numerics can be clearly seen.

\begin{figure}[h]
\centering
\includegraphics[width=0.4\textwidth]{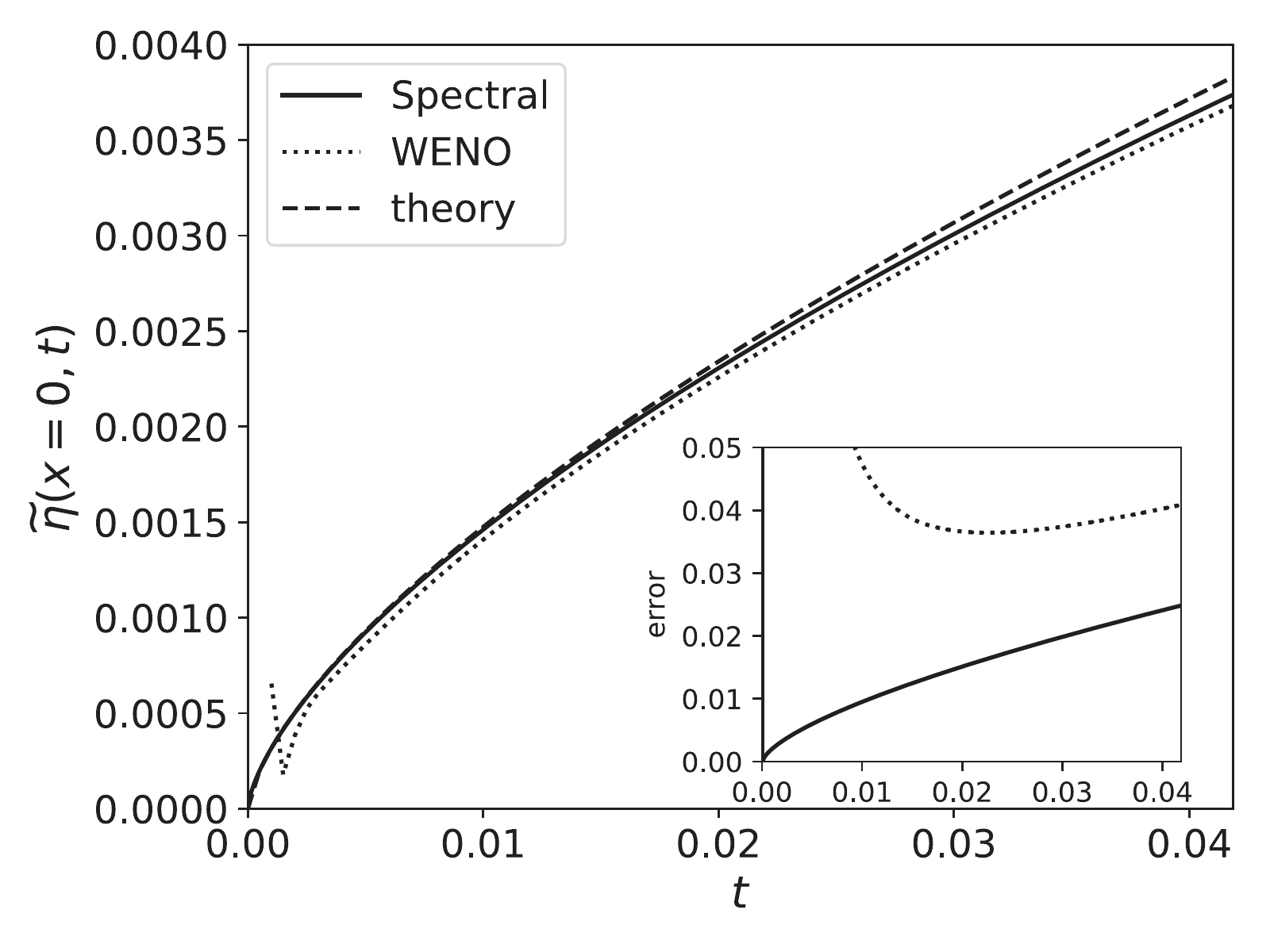}
\includegraphics[width=0.4\textwidth]{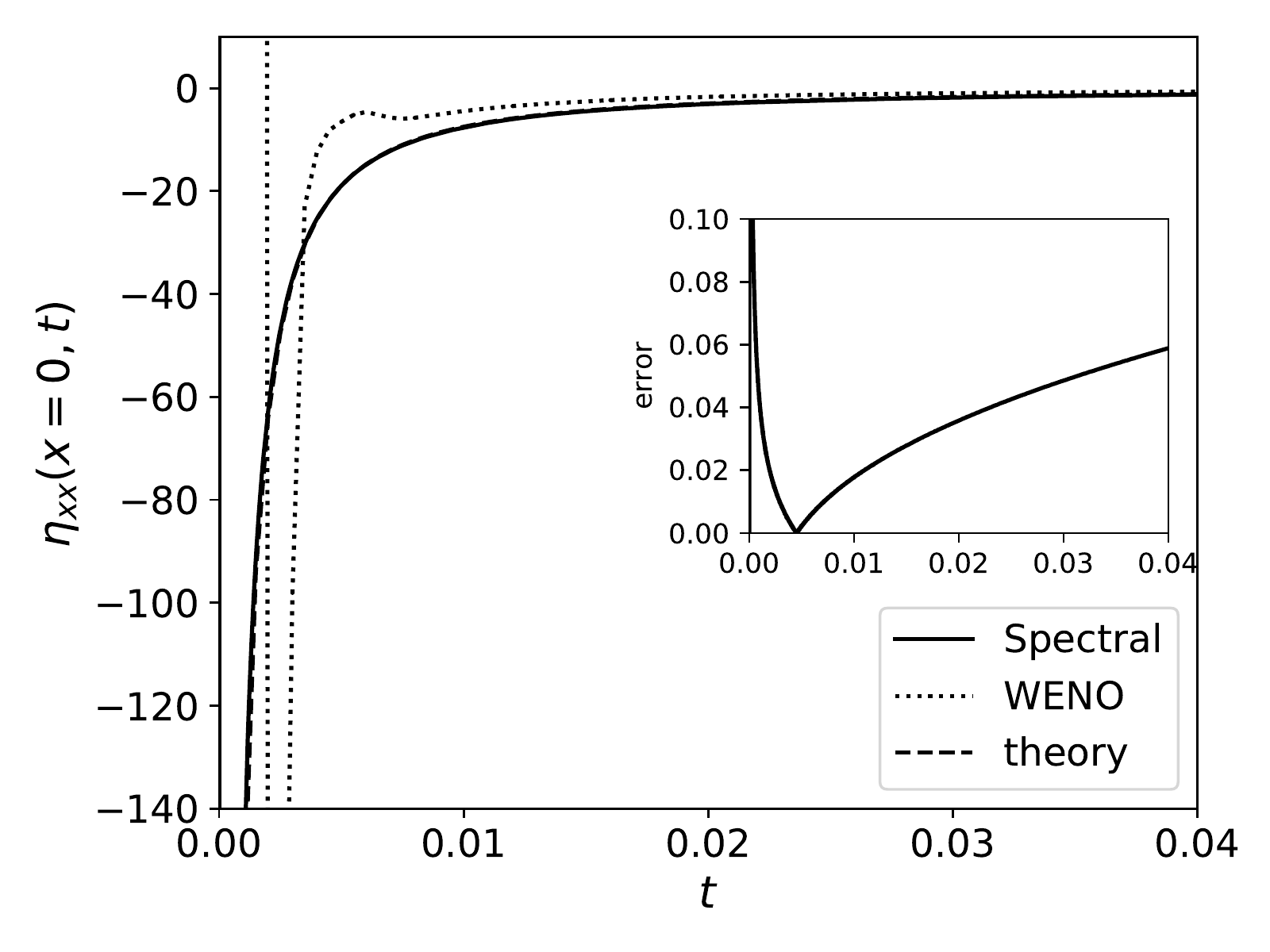} \\
\includegraphics[width=0.4\textwidth]{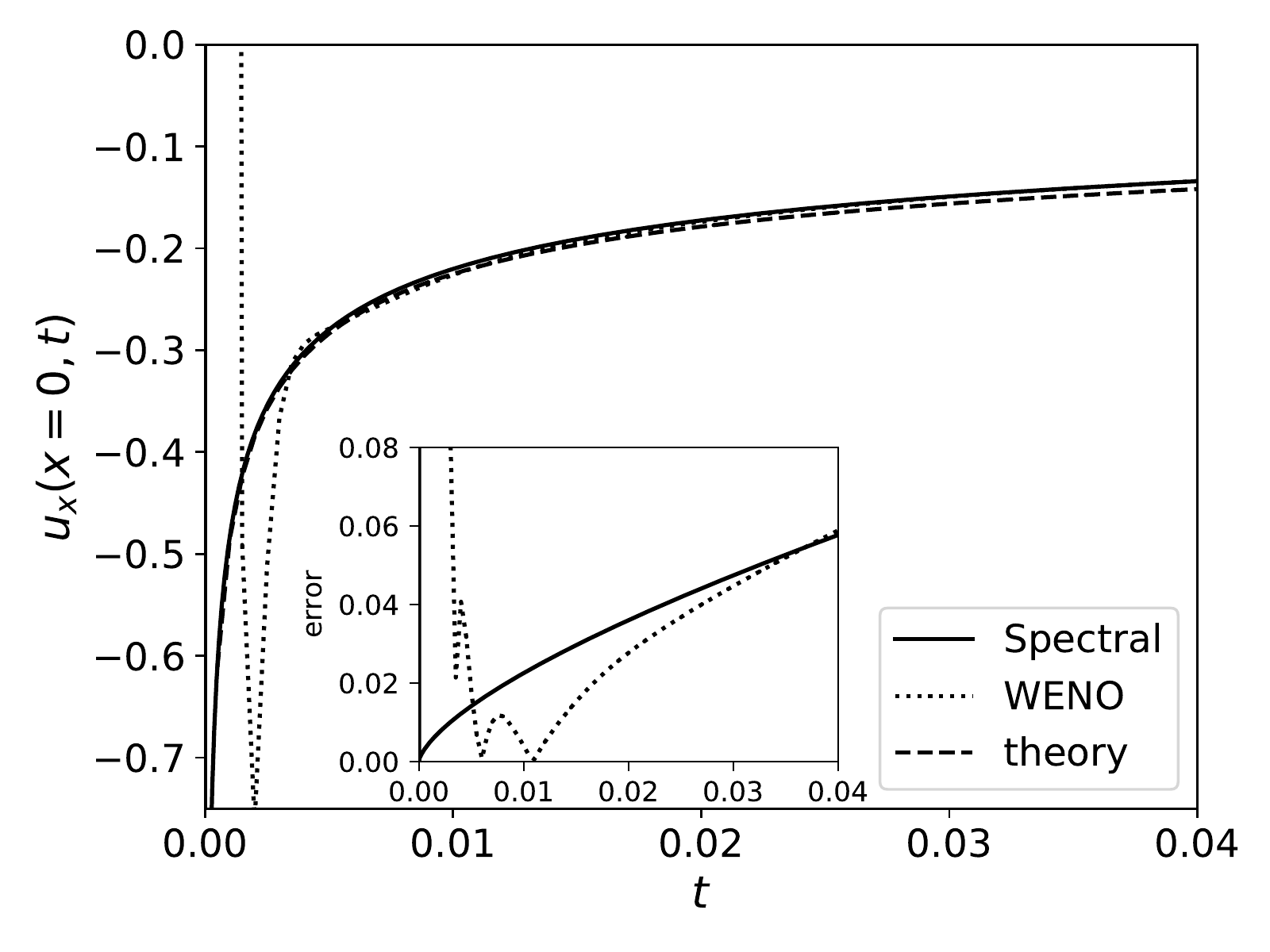}
\caption{Comparison of the numerical results for the full case obtained with the Chebyshev Collocation Method (continuous curve), the WENO Finite Difference Method (dotted) and the theoretical predictions of equations~\eqref{etauzeta23a},~\eqref{asyful13a} and~\eqref{asyful13} (dashed). The upper left plot shows $\tilde{\eta}(x=0,t)$, the upper right plot $\eta_{xx}(x=0,t)$ and the bottom plot $u_x(x=0,t)$. The insets on each graph show the relative error of the two numerical methods as a function of time. Notice how the error for the WENO scheme is off the scale  in the inset of the top-right panel for $\eta_{xx}$.}
\label{fig:cmp_weno_spec}
\end{figure}

Next, the two numerical methods considered here are compared with the theoretical results in figure~\ref{fig:cmp_weno_spec}. The purpose of this plot is to check the predicted growth rates of $\eta$, $u_x$ and $\eta_{xx}$ along the centerline for small times (here we perform the comparison up to $t=0.04$).
The plots generally show good agreement of both numerical schemes with the theoretical rates for $\eta$ and $u_x$, with relative errors always below 10\%; for $\eta_{xx}$ the method based on shock-fitted variables still provides accurate results, while for the shock-capturing method the error is substantially higher, being on average about 80\%.
A further drawback of the shock-capturing algorithm shows up as oscillations appearing just after the shock formation (see figure~\ref{fig:riemann_refinement}).
These numerical artifacts are to be contrasted with the small amplitude acoustic waves of the shock-fitted spectral method, which perturb the numerical results for a much smaller (in fact,  exponentially smaller) period of time.

\begin{figure}[htbp]
\centering
\includegraphics[width=0.4\textwidth]{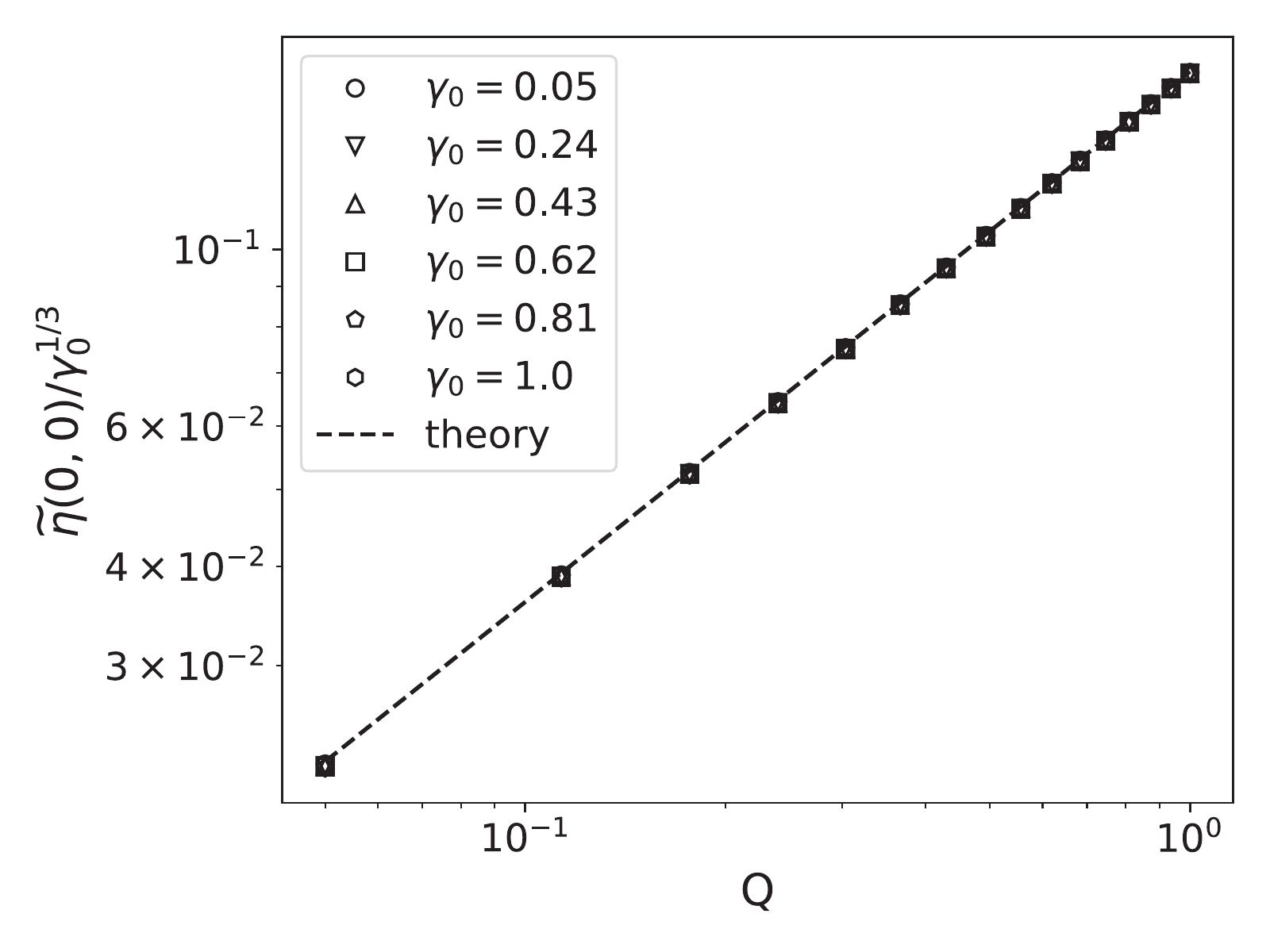}
\includegraphics[width=0.4\textwidth]{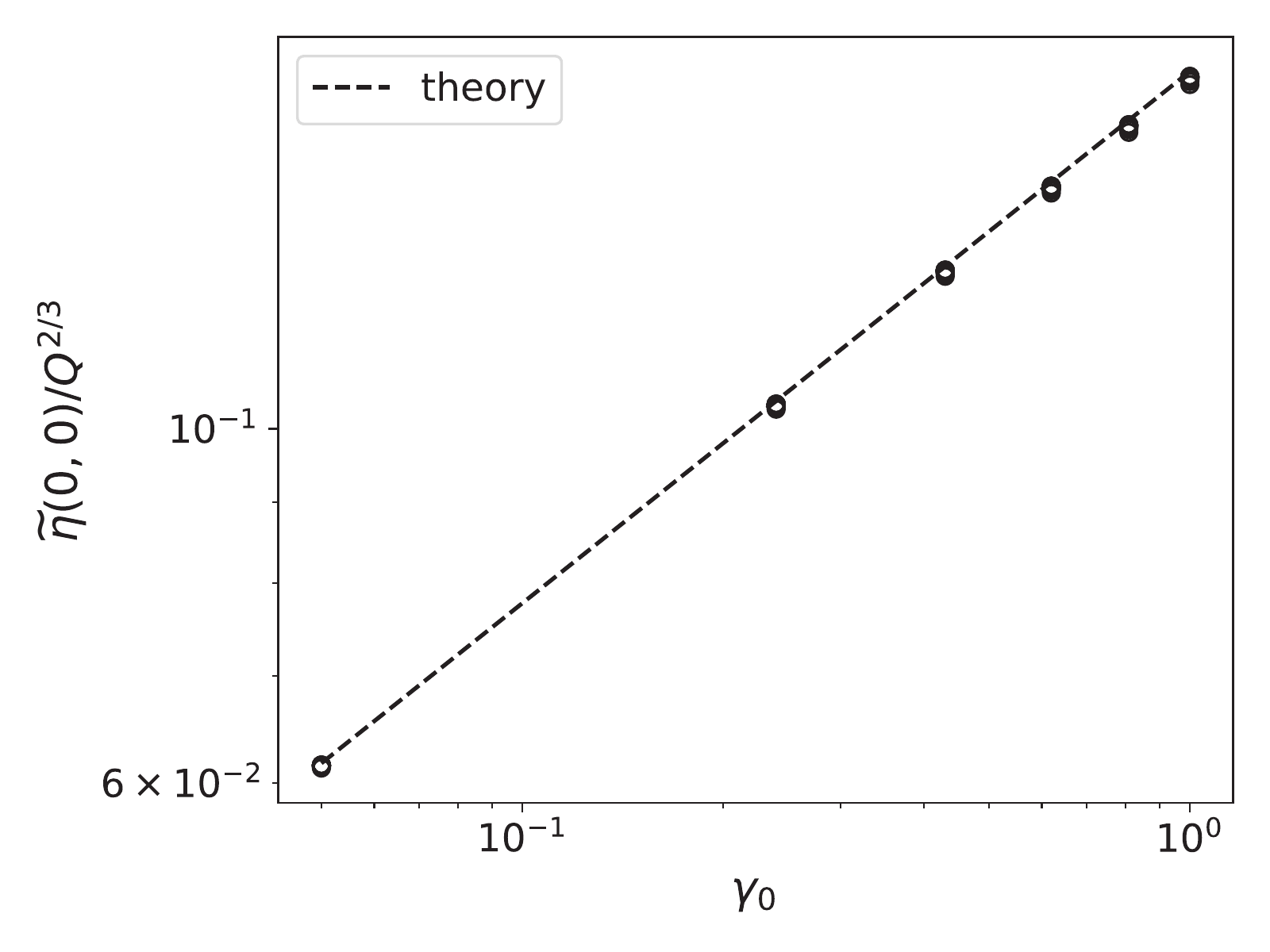}
\caption{Results from the full case initial data simulations. Left: comparison of the numerical coefficient $\tilde{\eta}(0,0)\gamma_0^{-1/3}$ for the values of $\gamma_0$ shown in the legend, with the theoretical prediction of equation~\eqref{fp23}, as a function of $Q$. The dashed line has slope $2/3$, and the markers are so close that there is a significant overlap between them. Right: for the same values of $\tilde{\eta}(0,0)$ as in the left panel, the scaling with respect to $\gamma_0$ is shown after a renormalization with $Q^{2/3}$ (in this case we do not attempt to resolve the individual values of $Q$, since this would clutter the plot beyond readability). On the right panel, the dashed line has slope $1/3$.}
\label{fig:plot_Fp0}
\end{figure}

Lastly, we check the prediction of equation~\eqref{fp23} for the scaling of $\eta$ and 
$u$ with the initial condition parameters $\gamma_0$ and $Q$.
This comparison is more conveniently done with the numerical scheme based on shock-fitted coordinates and the corresponding spectral code.
The value of $\eta(\xi=0,\tau=0)$ is computed by extrapolating to $\tau=0$ the value $\tilde{\eta}(\xi=0,\tau)$ obtained numerically, in this case with $\tau=-6$. This provides the following approximation for $\tilde{\eta}(0,0)$,
\begin{equation}
  \tilde{\eta}(0,0) \simeq \exp\left(\log\big(\eta(0,\tau)\big) - \frac23 \tau\right) \, . 
\label{eq:approx_eta_00}
\end{equation}
Next, evaluating equation~\eqref{etauzeta23a} at $\xi=0, \tau=0$ gives
\begin{equation}
  \tilde{\eta}(0,0) \simeq \left({\displaystyle\sqrt{Q^*}\over s_0}\right)^{2/3}F'(0),
\label{eq:approx_eta_00_th}
\end{equation}
and replacing in equation~\eqref{eq:approx_eta_00_th} the value computed 
numerically for $\eta$, it is possible to obtain an estimate for $F'(0)$.
The computations described above are repeated for several values of $\gamma_0$ and $Q$ ranging between 0.05 and 1.
For a clearer comparison, we show in figure~\ref{fig:plot_Fp0} (left) the values of $\tilde{\eta}(0,0)\gamma_0^{-1/3}$, so that the resulting data should depend only on $Q$, and not on $\gamma_0$.
For analogous reasons, on the right plot of figure~\ref{fig:plot_Fp0} we show the values of $\tilde{\eta}(0,0)Q^{-2/3}$ as a function of $\gamma_0$.
The numerical results agree with the theoretical predictions, confirming the numerical coefficient of $F'(0)$ to within three decimal digits.

\begin{figure}[t]
\centering
\includegraphics[width=0.4\textwidth]{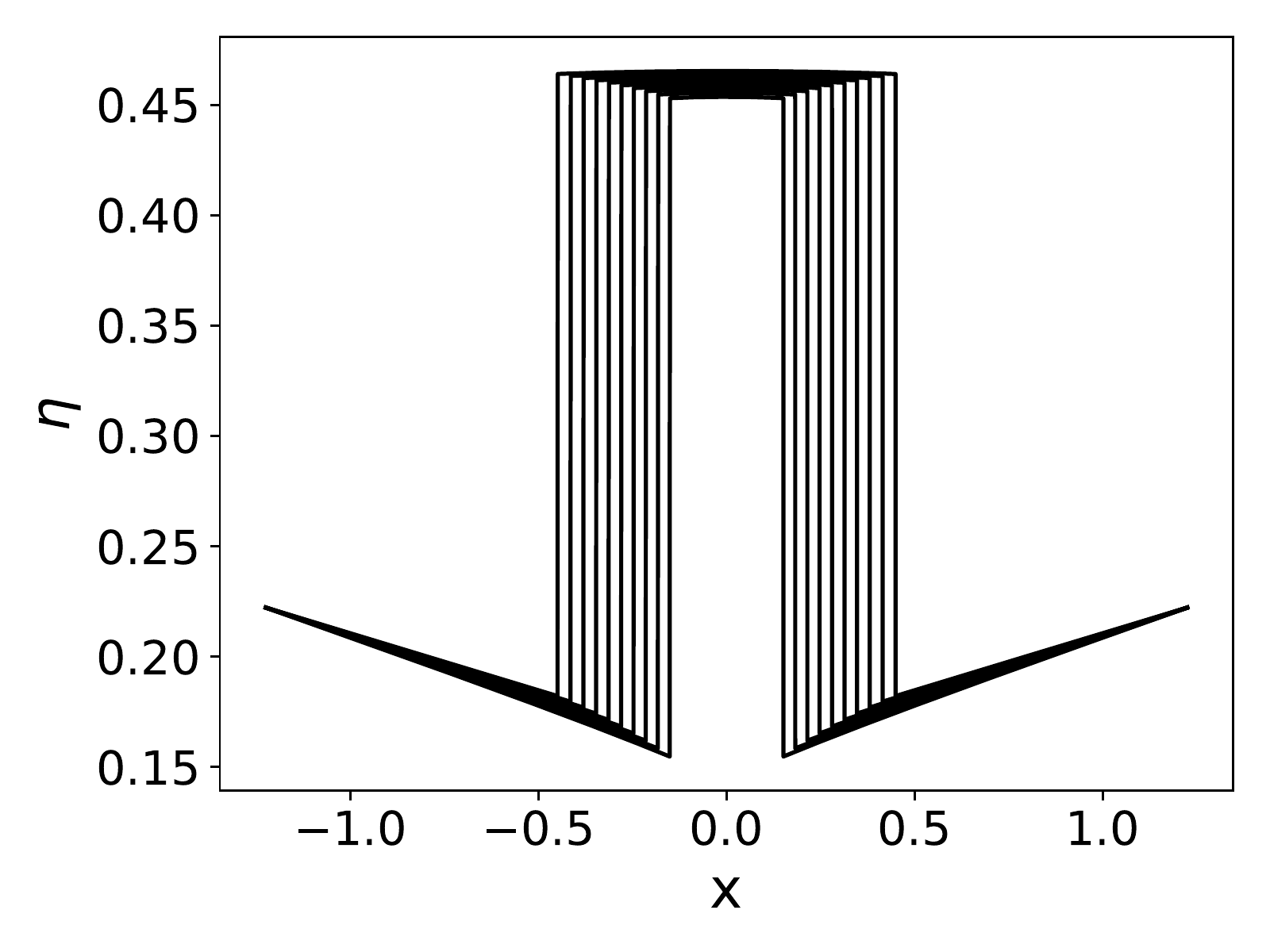}
\includegraphics[width=0.4\textwidth]{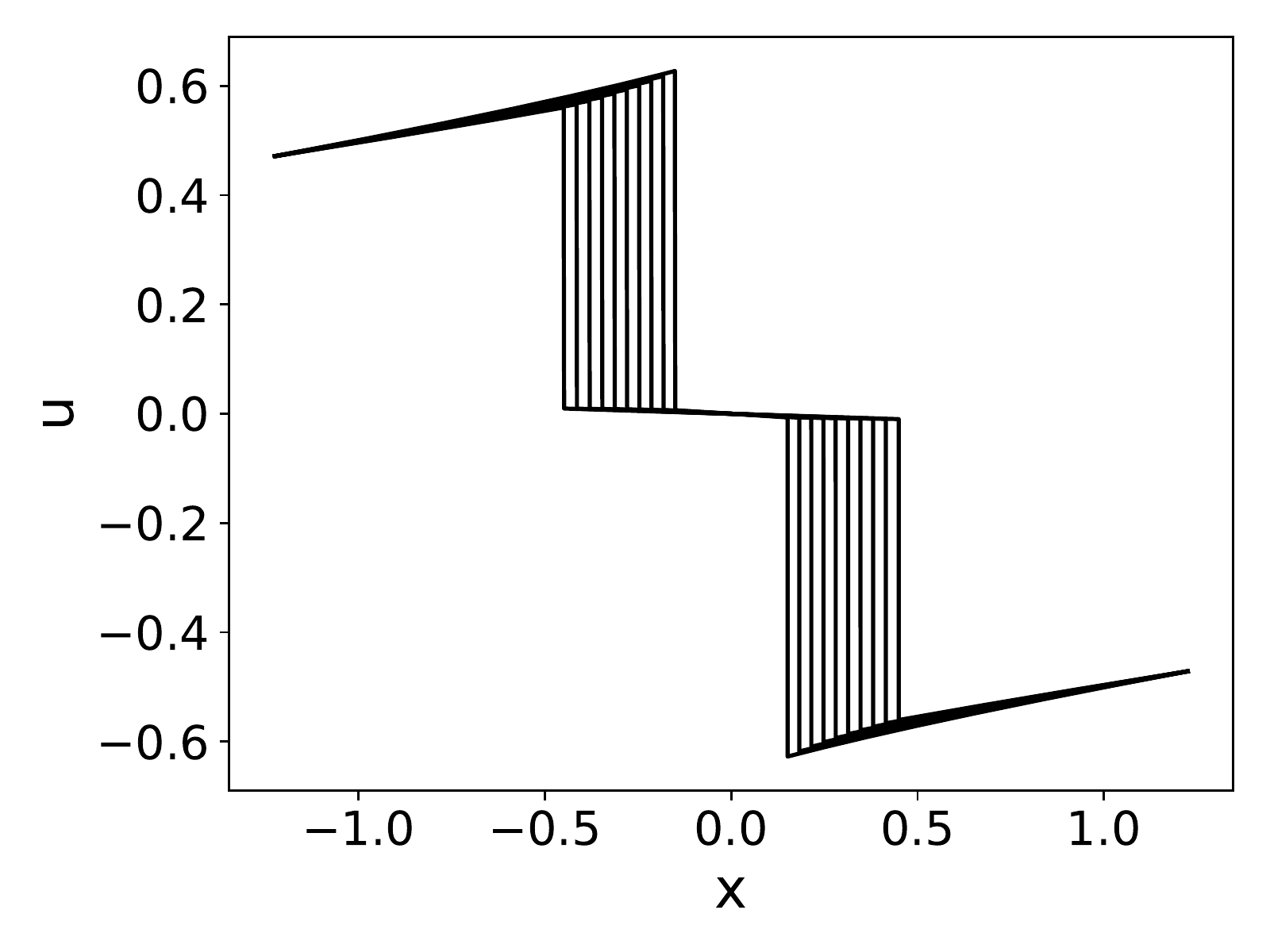}
\includegraphics[width=0.4\textwidth]{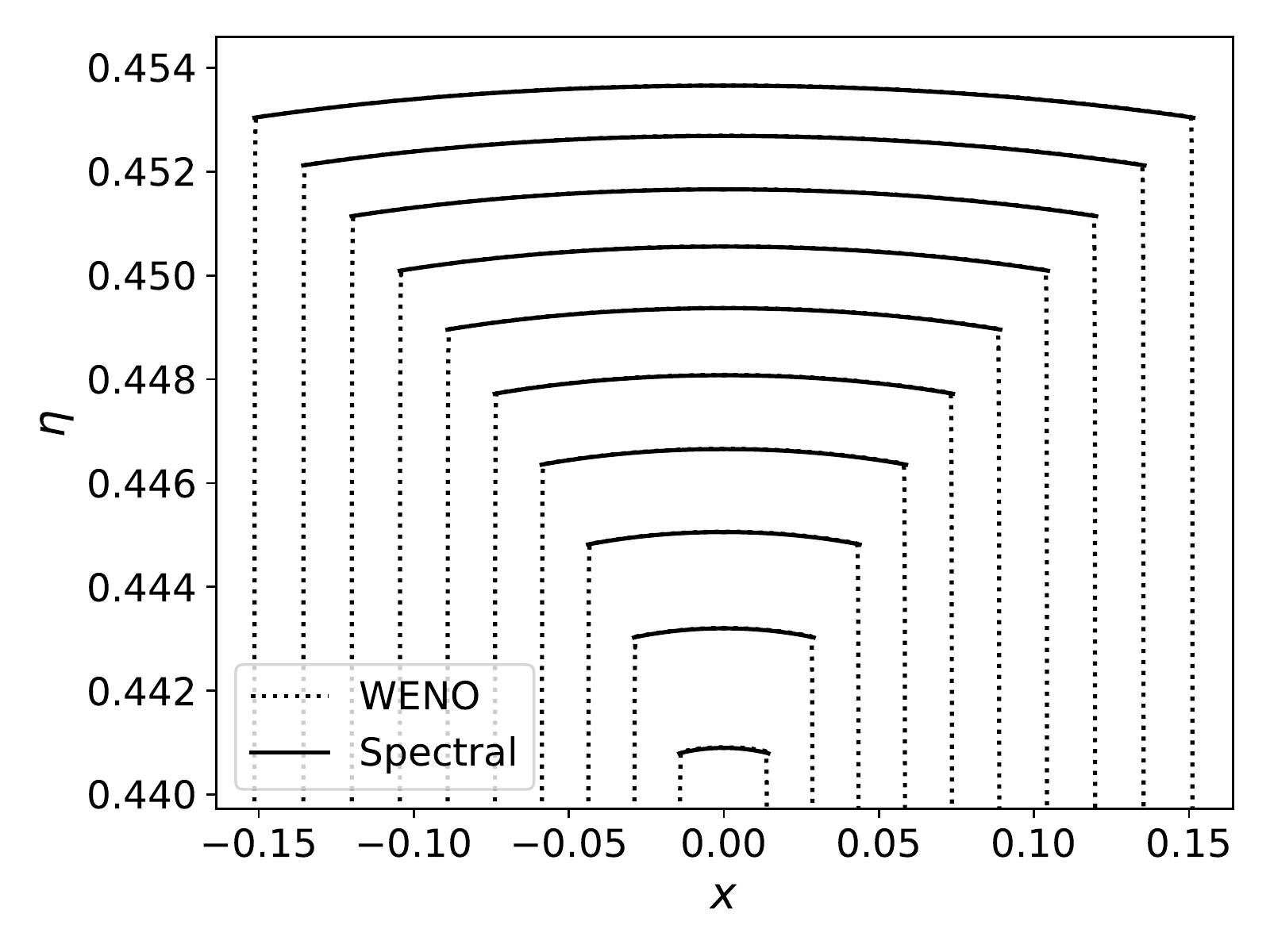}
\includegraphics[width=0.4\textwidth]{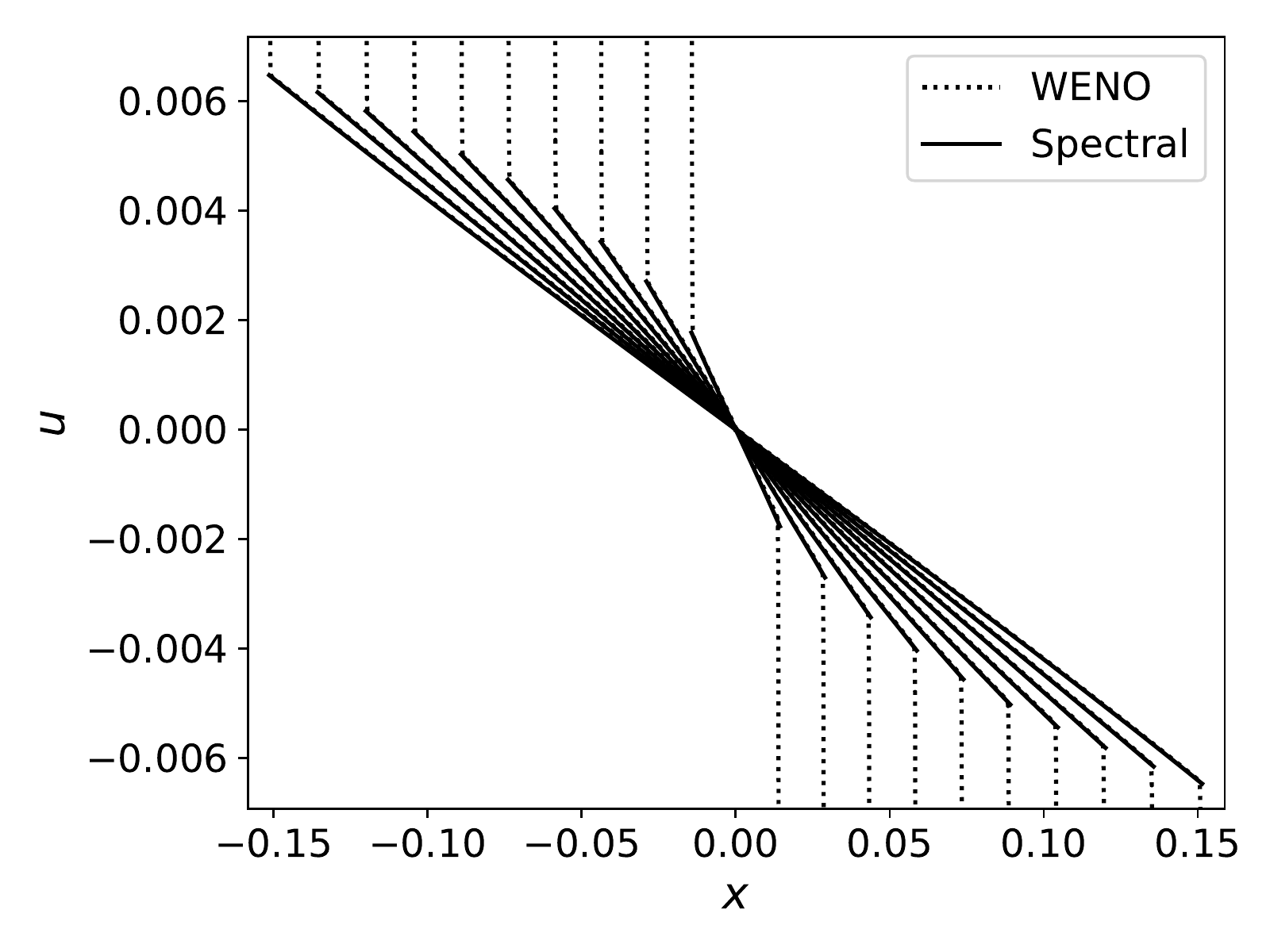}
\caption{Evolution of the initial datum defined implicitly in equation~\eqref{xnv} by the WENO scheme with $M=2^{18}$ grid points in the interval $[-\sqrt{3/2},\sqrt{3/2}\,]$ and timestep  $h=2\cdot 10^{-7}$. Top row, left panel: snapshots of $\eta$ as a function of $x$; right panel: $u$ as a function of $x$. Snapshots are taken at $0.1$ time units apart, from~$t=0.5$ to $t=1.4$. Bottom
row shows corresponding blowups of the region between shocks, compared with the inner solution from the shock-fitted coordinate evolution, numerically solved by the spectral code and converted to $(x,t)$ coordinates, with reflection across the $x=0$ axis. Snapshots are taken at $0.05$ time units apart, from $t=0.05$ to $t=0.5$.}
\label{fig:plots_weno}
\end{figure}

\begin{figure}[htbp]
\centering
\includegraphics[width=0.4\textwidth]{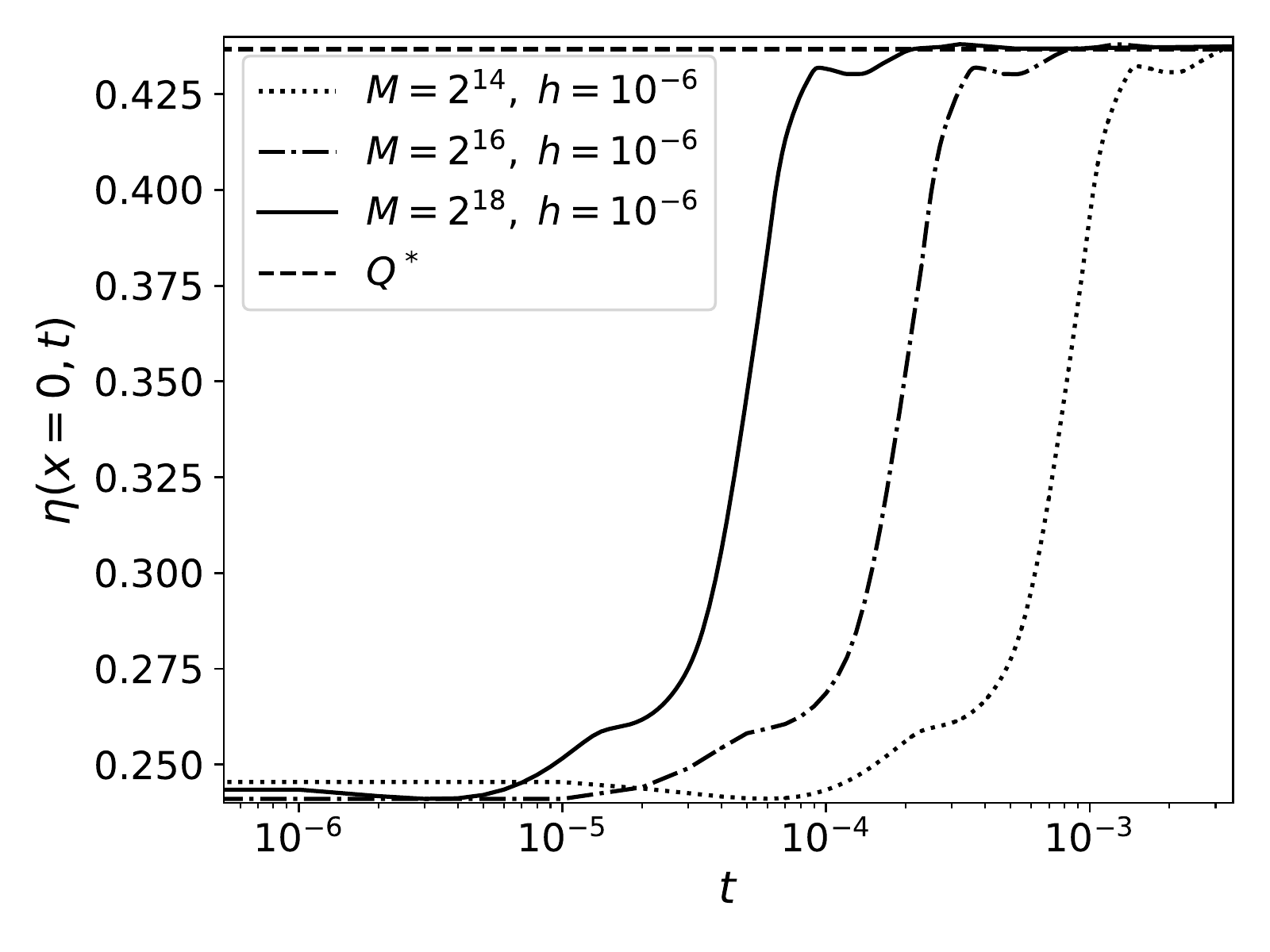}
\caption{Computed evolution with the WENO code of $\eta(0,t)$ for the full case  around the  collapse time~$t_c$; parameters are $Q=1/2$, $\gamma_0=1/16$, time step $h=10^{-6}$ and the spatial resolution is increased from $M=2^{14}$ to $M=2^{18}$ points in the interval $x\in [-\displaystyle \sqrt{3/2},\displaystyle \sqrt{3/2}\,]$. The oscillations in reaching the theoretical value $Q^*\simeq0.87349\,Q =0.43674$ similar to those exhibited in the Riemann case of figure~\ref{fig:riemann_refinement} can be clearly seen.}
\label{etabol}
\end{figure}

We conclude the presentation of the numerical computations by showing the results of the shock-capturing (WENO) scheme.
The shock-capturing scheme is set on the interval $[-\sqrt{3/2},\sqrt{3/2}]$, with $M=2^{18}$ grid points, and a time step $h=2\cdot 10^{-7}$. The initial datum in this case is obtained by solving for the implicit expression~\eqref{xnv} at the grid point coordinates using the Mathematica package.
In figure~\ref{fig:plots_weno} we show an example of the numerical evolution using this approach and also compare the WENO evolution with the shock-fitted spectral code, 
mapped into $(x,t)$ coordinates.
The profiles of $\eta$ and $u$ by the algorithms coincide up to three decimal digits, and indeed the results obtained with the two methods can hardly be distinguished. However, it should be stressed that the WENO evolution is affected by numerical uncertainty, similar to that occurring for the Riemann problem shown in figure~\ref{fig:riemann_refinement} and displayed here in figure~\ref{etabol},  for the time of shock formation.
There is also a slight difference in location of the endpoints of the inner solution between shocks in each snapshot, which follows from the relative lack of accuracy of the shock-capturing WENO code in detecting the shock position.
This latter quantity is smeared over a few grid points in the shock-capturing method, making it less precise as far as the shock position and its velocity are concerned.

\section{Discussion and conclusions}
In this work we have investigated the evolution of a class of exact solutions of the Airy shallow water model. These correspond to a particular choice of initial data with zero initial velocity, and surface elevation in the form of a downward parabola sandwiched between, and joined continuously to, constant flat states $\eta=Q$.  When the parabola contacts the bottom in a ``dry point," the evolution of the elevation $\eta$ and velocity $u$ is markedly different than their counterpart in the ``wet" case, where the interface is at a finite, however small, distance from the bottom. In fact, the dry point persists until the time when strong shocks develop, corresponding to the collapse of the initial parabola to a segment of height $Q/4$. In contrast, for the wet case, the parabola's minimum eventually lifts up to the background elevation $Q$, and a zero velocity, constant elevation state $\eta=Q$ develops in a neighbourhood of where the minimum was originally located. The dry case parabola collapse can be viewed as a mechanism for the fluid to ``fill the hole" that goes all the way to bottom, thereby allowing the reconnection of the fluid domain. 

It is remarkable that both the dynamics prior to reconnection, and that ensuing afterwards, can be followed entirely by analytical means, either in exact form, for pre-collapse times, or asymptotically, for short times after collapse.  
Until the collapse time, all the evolution details can be followed  with closed form exact solutions. The self-similar form of the parabolic part can be followed by solving a set of ODEs for its coefficients (for the dry point case, these are simply the  curvature  $\eta_{xx}$ at the minimum $x=0$ and the slope $u_x$ at the same point), while the connection to the background rest state is achieved by simple waves whose analytic form is expressed implicitly through an appropriate characteristic parameter.  The continuation past collapse and strong shock formation is more delicate and requires proper shock fitting. We have achieved this by introducing stretched coordinates to magnify the region where the shock first appears. This tool borrows ideas from singular perturbation theory,  and allows the analysis to proceed in the form of asymptotics for $\eta$ and $u$  between shocks, thereby continuing the solution in the limit of short times beyond that of collapse.

In order to illustrate, and provide intuition on, how this can be implemented, we constructed  two ad hoc  classes of initial data that capture the essential mechanisms of shock formation and propagation, but avoid the technical difficulties posed by the original case's branch singularities at the collapse location. The first of these two classes, which we dubbed ``double-Riemann,"  is a variant of the classic shock problem for piecewise constant data. The second case, dubbed ``double-Stoker," is obtained by splicing together two mirroring simple wave solutions of the dam-break type illustrated in Stoker's book~\cite{Stoker}. This increases the level of complexity with respect to the double-Riemann case, and gets qualitatively closer to the ``full" case by adding the fundamental component  of a time- and space-dependent medium through which the shocks must propagate. 

In all cases, the elevation $\eta$ at the origin jumps from the level $Q/4$ (of course, this particular choice is dictated by the pre-collapse evolution for the full case, and it is not arbitrary as in the ad-hoc constructions) at $t=t_c^-$ to a new level $Q^*\simeq 0.87\, Q$ at $t=t_c^+$, in the form of a thin jet between shock locations at $x=0$. The shocks propagate outward, with nontrivial dynamics, unlike the double-Riemann problem where the shock speed is constant, thickening the jet while its maximum grows with developing  curvature. These are signature features of the dry point case with respect to its wet counterpart, where the background state with a flat interface  $\eta=Q$ and $u=0$ is always fully recovered, for a finite time dependent on the initial minimum thickness of the fluid layer, in a neighbourhood of the origin.  
In this respect, a question that deserves to be investigated is whether this full recovery is a consequence of discontinuities in the derivatives of our initial conditions, and whether  the possible generation of  more generic singularities emerging at earlier times, and travelling towards  the vacuum regions, can significantly alter the qualitative features of the evolution we have identified.  However, these details would presumably be heavily dependent on the nature of the initial conditions; for instance, the corners at $\eta=Q$ could in principle be smoothed out with $C^\infty$ compactly supported bump functions, so that the core parabolic solution and the splicing point dynamics $a(t)$ could be maintained with the resulting smooth initial data. A study along these lines is undergoing and will be reported in the future.

With the analytical results at hand, we have implemented algorithms for 
the numerical solution of the Airy model, including the continuation past shock formation, and gauged their performance with respect to the theory. Not unexpectedly, the WENO scheme is able to give a qualitatively accurate global perspective in the physical coordinates. However, our study shows that a quantitatively accurate solution is more problematic  for this scheme, at least on short time scales after collapse. Thus, for instance, the scheme is plagued at short times by oscillations that cannot be eliminated even pushing grid refinement to levels where computational costs become unreasonable. As a result of this, the time-scaling of monitoring quantities such slope and curvature of $u$ and $\eta$ at the origin can not be pinned down accurately, making the validation against the short time theory impractical. Further, while these oscillations die out, presumably due to dissipation at the shocks, the longer time evolution seems to be affected by the initial errors accumulated by the scheme, though for the double-Riemann case, where the shock propagates in a constant medium, the correct (theoretical) values are eventually recovered.  However, even in this simplest case the shock elevation $Q^*$ is reached after an initial spiky overshoot that persists in magnitude while its time-support decreases as the space grid is refined and the time step is reduced. We remark  that these oscillations might be of different nature than the numerical ones the WENO scheme is designed to suppress. The implementation of the unfolding coordinate scheme, and its perturbation formulation,  in fact reveals that left- and right-propagating waves in the interval $\xi\in [-1,1]$ can be caused by a (small) compatibility mismatch between the initial (as $\tau \to -\infty$) and boundary conditions. This bouncing back and forth modes in the interval $\xi \in [-1,1]$ can persist until dissipated at the shocks, and, at least initially, can appear as high frequency oscillations in the compressed interval~$[-x_s,x_s]$ of the physical space. 
It must be noted, however, that while the numerical algorithm we have developed for the unfolding coordinate is far more accurate than WENO's, as attested by comparison with the theory (also a consequence of the local solution remaining regular in these coordinates, allowing our implementation to be spectrally accurate), numerical round-off errors are nonetheless magnified when converting back to physical coordinates, owing to the multiplication of small floating point numbers for the solution by large, $O(e^{-\tau})$ for negative $\tau$, values.
A thorough investigation to identify optimal algorithms able to handle the kind of evolutions that we have studied analytically is  interesting  from a numerical perspective, but would need a dedicated study outside of the scope of this work. 

The theoretical framework developed here for shocks emanating from the class of initial conditions we have focussed on may be extended to more general set-ups, and possibly be generalized to a broad class of hyperbolic conservation laws. We plan to pursue this in future work, where we will also  examine the interplay between dissipation at the shock location vs. its realization via the boundary condition in the shock fitted domain. 
A relevant  goal, which we touched upon above and will be reported more in depth elsewhere, is that of extending  our results for the class of initial data we studied to more general, and in particular smooth, initial data. Furthermore, a thorough investigation of the differences between initial data that are limiting sequences to those with dry spots is being carried out and will be reported in the future. We remark here that such a study would bear some similarity with the investigation on the formation of ``splash" type singularities in models of free surface flows~\cite{CFG}, and it would be interesting to explore this direction further. 

We also remark that the reflection symmetry of the solutions we have studied, together with the assumption of negligible viscosity for the fluid applications, allows us to interpret our results as equivalent to the case of a vertical impermeable boundary condition, a ``wall," at $x=0$. Thus, for the full case, the initial bottom contact of the interface at $x=0$ would imply that the upward jet resulting from the impact of the water layer with the vertical wall would be initially more than $10\%$ lower than it would  be if a wetting liquid film were maintained over the bottom. Of course, for such a simplified model this estimate cannot be expected to be quantitatively accurate for a real physical situation, nonetheless it isolates a trend that could be significant enough not to be swamped by other real-fluid properties, such as viscosity and surface tension, and possibly be observable in experiments.

Finally, we stress that the solution features uncovered by our analysis and simulations cannot obviously be expected to represent in all details the physics of fluids modelled by the Airy's system, based as this is on a long-wave asymptotics of the Euler equations,  which would be clearly violated by shock formation. However, by enforcing the fundamental conservation laws of mass and momentum, this system is known to be quite robust in capturing general features of the dynamics of shallow water, with shocks representing hydraulic jumps. Initial conditions of the kind studied here, 
while hard to implement in actual experiments, can be studied in silico, and in future work we plan to present results for the parent Euler equations  governing ideal fluids. Generalizations of our class of initial conditions and a thorough study of the energy dissipation due to shock formation are also interesting new developments, which will be reported  in future work.

\section*{Acknowledgments}

RC thanks the hospitality of the ICERM's program ``Singularities and Waves in Incompressible Fluids" in the Spring of 2017, and the support by the National Science Foundation under grants RTG DMS-0943851, CMG ARC-1025523, DMS-1009750, DMS-1517879, and by the Office of Naval Research under grants N00014-18-1-2490 and  DURIP N00014-12-1-0749. All authors gratefully acknowledge the auspices of the GNFM Section of INdAM under which part of this work was carried out. R.C., M.P. and G.P. thank  the {Dipartimento di Matematica e Applicazioni\/} of Universit\`a Milano-Bicocca  for its hospitality,
G.F., G.O.,  M.P. and G.P. thank the Carolina Center for Interdisciplinary Applied Mathematics at the University of North Carolina for hosting their visits in~2018. Support by grant H2020-MSCA-RISE-2017 Project No. 778010 IPaDEGAN is also gratefully acknowledged.

\subsection*{Appendix: Shock continuation via local Taylor series}
\renewcommand{\theequation}{A\arabic{equation}}

We seek a local solution of system~(\ref{Airymodel}) as $|x|\to 0$ by an expansion in powers of 
$x$, 
\begin{equation}\label{Ssolnear0}
\eta(x,t)=\sum_{n=0}^\infty \eta_n(t)x^{2n}\, , \qquad
u(x,t)=\sum_{n=0}^\infty u_n(t)x^{2n+1}\, ,
\end{equation}
where the choice of even and odd powers respectively for $\eta$ and $u$ 
enforces the symmetry $\eta(-x,\cdot)=\eta(x,\cdot)$, $u(-x,\cdot)=-u(x,\cdot)$.
In particular, as well known, when the initial data are analytic functions of $x$, the Cauchy-Kovalevskaya (C-K) Theorem (see, e.g.,~\cite{Evans}) assures that these power series have a finite radius of convergence, so that all coefficients are analytic in a neighbourhood of $t=0$. 
The exact solutions~(\ref{exact}) are of course in this class, and are special in that they derive from a hierarchy that truncates at the leading orders, $n=0,1$, so that $\eta$ and $u$ are second and first degree polynomials, respectively. Of course, this property is special, and 
the series~(\ref{Ssolnear0}) cannot in general be expected to truncate to a polynomial form; once substituted in~(\ref{Airymodel}) the series  generate a recursive infinite hierarchy of ODE's for the coefficients 
$\eta_n(t)$ and $u_n(t)$.  The first equations in the hierarchy are (cf.~(\ref{coeffODEs})) 
\begin{equation}
\label{paraeq}
\dot{u}_0+u_0^2+2\eta_1=0\,,\qquad \dot{\eta}_0+u_0\eta_0=0\,, \qquad \dot{\eta_1}+3(\eta_1 u_0+\eta_0 u_1)=0\,. 
\end{equation}
Note that, when applied to zero velocity smooth initial data with a dry point~$\eta_0(0)=0$, then $\eta_0(t)=0$  for all times and these equations reduce to a closed system for $u_0$ and $\eta_1$ which coincides with~(\ref{coeffODEsr}). In fact, the presence of the dry point has important implications, 
since when $\eta_0(t)=0$ the equations in the hierarchy for the pairs $\eta_{n+1}$ and $u_n$ are {\it linear} , with coefficients and inhomogeneous terms that depend solely on the ($\eta_{k+1},u_k$) pairs, $k<n$. The recursive solution for all coefficients thus shows that time singularities are determined locally in an neighbourhood of $x=0$ solely by the nonlinear system for the first pair $(\eta_1,u_0)$, i.e., equations~(\ref{paraeq}) with $\eta_0=0$.

While the C-K Theorem refers to smooth initial data, we assume next that a local solution in a neighbourhood of the origin away from shocks can still be expressed in the form~(\ref{Ssolnear0}), for suitable initial conditions such as the double-Stoker case~\S\ref{doubleSt}, and  that equations~(\ref{paraeq})  govern the leading order approximation to the solution between shocks for short times $t>0$.
Accordingly, we seek the coefficients $u_0(t), \eta_0(t)$, and $\eta_1(t)$ as 
 Taylor expansions in time,
\begin{equation}\label{expamg}
u_0(t)=\nu^{(0)}+ \nu^{(1)}t+O(t^2),\quad \eta_0(t)=\mu^{(0)}+\mu^{(1)} t+O(t^2),\quad \eta_1(t)=\gamma^{(0)}+\gamma^{(1)} t+O(t^2)\,, 
\end{equation}
where, as mentioned in~\S\ref{2stoksec}, we use this notation to remind the reader of the role played by the analogous coefficients in the exact counterpart from initial data~(\ref{inidata-up})  for system~(\ref{Airymodel}). 
With the above assumptions, the shock position can be similarly expanded as $t\to 0^+$, 
\begin{equation}
x_s(t)={s}_0\,t+O(t^2)\,.
\end{equation}
As in section~\ref{sec:beyond_shock}, the shock speed and the consistency condition are given by~(\ref{shockspeed}) and~(\ref{shockcc}), respectively. 
In this case $f_+$ corresponds to the rarefaction (simple wave) solution~(\ref{2stok}) while $f_-$ is determined by the local solution~(\ref{Ssolnear0}).

Carrying out the above expansions through the recursive steps, we obtain for the first few coefficients the following decimal values 
\begin{equation}\label{anresbS}\begin{split}
&\mu^{(0)}= Q^*\, ,\quad {s}^{(0)}\simeq0.40097\sqrt{Q} \,, \quad \nu^{(0)}\simeq-0.23769\sqrt{g_0}\,, \quad \mu^{(1)}\simeq 0.20762 \sqrt{g_0}\, Q\, , 
\end{split}
\end{equation}
where the ordering in this list reflects that of the recursion method for finding the asymptotic series coefficients, beginning with the solution of the cubic equation that sets the shock amplitude, $\mu^{(0)}$, at time $t=0^+$. These are identical to the corresponding values computed in~\S\ref{2stoksec}, however we remark that higher order coefficients lead to calculations that can be even  lengthier than their counterpart in the unfolding coordinate approach.


\end{document}